\documentclass[12pt]{uofsthesis-cs} 
\usepackage{hyperref} 
\usepackage{amsmath,amssymb} 
\usepackage{graphicx} 
\usepackage{setspace} 
\usepackage{url} 
\usepackage{simplewick} 
\usepackage{slashed} 
\usepackage{tabulary} 

\title{QCD sum rule studies of Heavy Quarkonium-like states}

\author{Robin Thomas Kleiv}

\degree{\PhD}         

\convocationdate{September 2013}

\department{Physics and Engineering Physics}

\ptuaddress{
Head of the Department of Physics and Engineering Physics\\
116 Science Place\\
University of Saskatchewan\\
Saskatoon, Saskatchewan\\
Canada\\
S7N 5E2\\
}

\abstract{
In 2003 the Belle collaboration announced the discovery of the $X(3872)$ particle. This was confirmed shortly thereafter by the CDF, D0 and BaBar collaborations, and later by the LHCb collaboration. Based on the decay modes that have been observed to date, it is clear that this particle is a hadron, that is, a composite particle that experiences the strong nuclear force. The $X(3872)$ was found within a family of well understood hadrons called charmonia. Interestingly, it is quite difficult to interpret the $X(3872)$ as a charmonium state. For this reason it has been widely speculated that the $X(3872)$ cannot be understood in terms of the quark model, unlike the vast majority of hadrons observed to date. Such hitherto unobserved particles are called exotic hadrons. Since the discovery of the $X(3872)$, many similarly anomalous charmonium-like particles have been discovered. As would be expected, some unanticipated hadrons have also been found in the closely related bottomonium spectrum. These particles are 
collectively referred to as heavy quarkonium-like. Evidence is growing that at least some of these particles are exotic hadrons. If confirmed, this would have dramatic implications for our understanding of the strong nuclear force. 

A major experimental and theoretical effort is now underway in the field of hadron spectroscopy to determine the identities of the heavy quarkonium-like states. In order to investigate the possibility that some of these states could be exotic hadrons, theoretical calculations are needed to firmly establish their properties. One of the main arguments for the existence of exotic hadrons is that they are predicted by the fundamental theory of the strong interaction, Quantum Chromodynamics (QCD). Therefore it is desirable to predict the properties of exotic hadrons using a theoretical approach that is firmly based in QCD. One such method is QCD sum rules (QSR). 

The research presented here uses the QSR technique to study exotic hadrons. There are several themes in this work. First is the use of QSR to predict the masses of exotic hadrons that may exist among the heavy quarkonium-like states. The second theme is the application of sophisticated loop integration methods in order to obtain more complete theoretical results. These in turn can be extended to higher orders in the perturbative expansion in order to predict the properties of exotic hadrons more accurately. The third theme involves developing a renormalization methodology for these higher order calculations. This research has implications for the $Y(3940)$, $X(3872)$, $Z_c^\pm\left(3895\right)$, $Y_b\left(10890\right)$, $Z_b^{\pm}(10610)$ and $Z_b^{\pm}(10650)$ particles, thereby contributing to the ongoing effort to understand these and other heavy quarkonium-like states.
}

\acknowledgements{
First, I wish to express my deepest gratitude to my supervisor, Dr.~Tom Steele. His skilled guidance, thoughtful advice, constant support and deep knowledge of physics have benefited me enormously. I would also like to thank the members of my advisory committee, Dr.~Murray Bremner, Dr.~Rainer Dick, Dr.~Glenn Hussey, and Dr.~Kaori Tanaka, as well as my external examiner, Dr.~Marcelo Loewe. I am grateful for their advice, support and encouragement. Thanks also to my collaborators on the articles included in this thesis: Ryan Berg, Dr.~Ian Blokland, Dr.~Derek Harnett, Dr.~Hong-Ying Jin, Dr.~Ken Moats, and Dr.~Ailin Zhang. Part of this work was completed at Shanghai University, which I thank for its hospitality. I am also grateful to the National Science and Engineering Research Council (NSERC) and the University of Saskatchewan for financial support. 

I am grateful to my family and friends for their unwavering support and encouragement. Without them, this thesis would not have been possible. Thanks to Ryan Berg, Brian Bewer, Yasmin Carter, Katie Knorr, Keith Kotyk, Martin Lepage, John McLeod, Denin Nienaber, Eric Nienaber, Kurt Nienaber, Percy Paul, Gareth Perry, Fred Sage, Zhi-Wei Wang, Mark Wurtz and Melissa Wurtz. {\it Mange Takk}\, to Margaret and Thor Kleiv for their kind hospitality when I first arrived in Saskatoon and for their constant encouragement since then. Special thanks are due to my teacher, mentor, colleague and friend Dr. Derek Harnett for setting me on this path in the first place. Above all I thank my family for their love and support.
} 

\dedication{To my family.}

\loa{
\abbrev{MS}{Minimal Subtraction}
\abbrev{$\overline{\rm MS}$}{Modified-Minimal Subtraction}
\abbrev{OPE}{Operator Product Expansion}
\abbrev{QCD}{Quantum Chromodynamics}
\abbrev{QED}{Quantum Electrodynamics}
\abbrev{QSR}{QCD Sum Rules}
\abbrev{SM}{Standard Model}
}

\begin{document}

\maketitle

\frontmatter


%
% ...
%

\chapter{Introduction}
\label{chapter_1_intro}


\section{Motivation for Research}

In 2003, the $X(3872)$ was discovered by the Belle collaboration~\cite{Choi_2003_a}. The discovery was subsequently confirmed by the Babar~\cite{Aubert_2004_a}, CDF~\cite{Acosta_2003_a}, D0~\cite{Abazov_2004_a} and LHCb collaborations~\cite{Aaij_2011_a}. The decays of this particle that have been observed to date clearly indicate that it is a hadron, that is, a composite particle composed of quarks that experiences the strong nuclear force. This particle was found within the mass region occupied by a well understood family of hadrons known as charmonia. However, the properties of the $X(3872)$ make it difficult to interpret it as a member of the charmonium spectrum~\cite{Swanson_2006_a}. Since 2003, more hadrons have been discovered in the charmonium mass region that are difficult to interpret as charmonium states. A few anomalous hadrons have also been found within the closely related bottomonium spectrum. These anomalous particles are called heavy quarkonium-like, or XYZ states. Table~\ref{XYZ_experiments_info} includes basic information for experiments that have discovered XYZ states and Table~\ref{XYZ_confirmed_table} lists the XYZ states that have been confirmed by more than one experiment at a high level of statistical significance. Ref.~\cite{Beringer_2012_a} provides a more complete list that includes particles that have only been observed by a single experiment and particles that have been observed at a lower level of statistical significance.

\begin{table}[hbt]
\centering
\begin{tabular}{lll}
\hline
Experiment & Facility & Process \\
\hline
 & \\
Babar & SLAC, Stanford, USA & Electron-Positron Collider \\
 & \\
Belle & KEK, Tsukuba, Japan & Electron-Positron Collider \\
 & \\
BES-III & BES, Beijing, China & Electron-Positron Collider \\
 & \\
CDF & Fermilab, Chicago, USA & Proton-Antiproton Collider \\
 & \\ 
CLEO & CESR, Ithaca, USA & Electron-Positron Collider \\
 & \\  
D0 & Fermilab, Chicago, USA & Proton-Antiproton Collider \\
 & \\  
LHCb & CERN, Geneva, Switzerland & Proton-Proton Collider \\
 & \\   
\hline
\end{tabular}
\caption{Experiments that have detected heavy quarkonium-like states.}
\label{XYZ_experiments_info}
\end{table}

\begin{table}[hbt]
\centering
\begin{tabular}{ll}
\hline
Particle & Experiments \\
\hline
 & \\
$X(3872)$ & Babar, Belle, CDF, D0, LHCb \\
 & \\
$G(3900)$ & Babar, Belle \\
 & \\
$Y(4260)$ & Babar, Belle, CLEO \\
 & \\
$Y(4360)$ & Babar, Belle \\
 & \\
$Z_c^{\pm}(3895)$ & Belle, BES-III, CLEO \\
 & \\
\hline
\end{tabular}
\caption{Experimentally confirmed heavy quarkonium-like states. In all cases the number in parentheses indicates the mass in units of ${\rm MeV}$. This system of units is called natural units and is discussed in Appendix~\ref{appendix_a_conventions}.}
\label{XYZ_confirmed_table}
\end{table}

Nearly all hadrons that have been observed to date can be classified according to the quark model. The quark model was introduced in Refs.~\cite{Gell_Mann_1964_a,Zweig_1964_a} to bring some order to the already large number of hadrons that were known at the time. The model introduces two families of hadrons: baryons such as the neutron and proton that are fermions, and mesons such as the pion that are bosons. Both baryons and mesons are composite particles composed of fundamental particles, called quarks. Baryons are composed of three quarks and mesons are composed of a quark and an antiquark. It should be emphasized that the quark model does not describe the dynamics of quarks. Rather, it is a classification scheme that successfully explains the large variety of hadrons as various combinations of a small number of quarks.

Quantum Chromodynamics (QCD) successfully describes the interactions of quarks, and as such, it is a fundamental theory of the strong nuclear force. However, it is not clear how the quark model of hadrons emerges from QCD. Interestingly, QCD seems to suggest that a much richer spectrum of hadrons is possible than the simple baryons and mesons of the quark model. These hadrons that exist outside the quark model are called exotic hadrons. To date there is no unambiguous proof for the existence of any exotic hadron, although the $Z_c^{\pm}(3895)$ is a very strong candidate. There are experimentally established hadrons that are difficult to interpret within the quark model and are often speculated to be exotic hadrons. It has been widely speculated that some of the heavy quarkonium-like states may be exotic hadrons. In order to investigate this possibility, theoretical calculations are needed to firmly establish the expected properties of exotic hadrons. The methods of QCD sum rules (QSR) can be used to predict 
the physical properties of exotic hadrons that may exist in the same mass region as heavy quarkonia. This is the main motivation for the research presented in this thesis.

\section{Hadronic Physics}

\subsection{The Standard Model}

\begin{figure}[htb]
\centering
\includegraphics[scale=0.5]{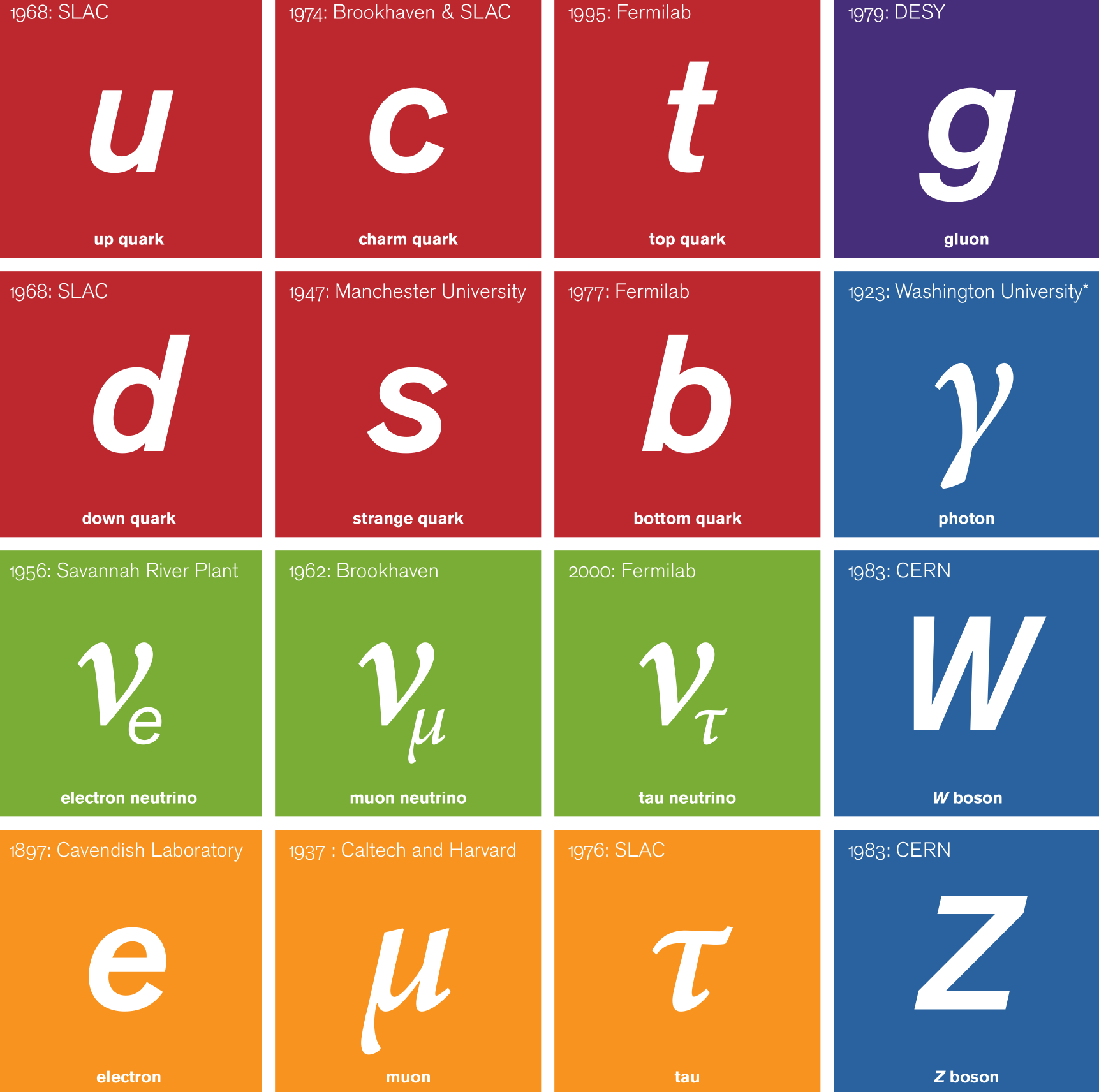}
\caption{
The particle content of the Standard Model, excluding the Higgs boson. Figure from Ref.~\cite{Harris_2009_a}.}
\label{SM_fig}
\end{figure}

The Standard Model (SM) of particle physics is an extremely successful theoretical framework that describes all fundamental interactions in nature at the quantum level, apart from gravity. The particle content of the SM is shown in Fig.~\ref{SM_fig}. There are three main categories of particles: spin-1 gauge bosons ($g$, $\gamma$, $W$, $Z$), spin-1/2 leptons ($e$, $\mu$, $\tau$, $\nu_e$, $\nu_\mu$, $\nu_\tau$), and spin-1/2 quarks ($u$, $d$, $s$, $c$, $b$, $t$). Because they mediate interactions between particles in quantum field theory, gauge bosons are often referred to as force carriers. Leptons are particles that do not experience the strong nuclear force, such as the electron. All of the particles in Fig.~\ref{SM_fig} have been confirmed experimentally. However, the SM also predicts the existence of an additional particle known as the Higgs boson. On July 4, 2012, the ATLAS~\cite{Aad_2012_a} and CMS~\cite{Chatrchyan_2012_a} collaborations announced the discovery of a particle that is likely to be the 
Higgs boson. Observation of the Higgs boson is a crucial test of the Higgs mechanism, which is essential to the SM. Experimental work to precisely determine the properties of this particle is ongoing.

This thesis will focus on quarks and gluons, which are the only particles in the SM that directly experience the strong nuclear force. There are six types, or flavours of quarks: up $(u)$, down $(d)$, strange $(s)$, charm $(c)$, bottom $(b)$ and top $(t)$. These can be divided into light quarks $\left(u\,,d\,,s\right)$ and heavy quarks $\left(c\,,b\,,t\right)$, which have much greater masses than the light quarks (masses are given in Table~\ref{quark_masses}). Gluons are massless and serve as the mediators of the strong interaction. Interestingly, quarks and gluons only occur within hadrons, and cannot be isolated or otherwise removed from hadrons. This peculiar feature of the strong interaction is known as confinement, and understanding how it emerges from QCD is one of the great problems of modern physics. All approaches to this problem must invariably deal with hadrons, which are how quarks and gluons manifest themselves in nature.

\begin{table}[hbt]
\centering
\begin{tabular}{cccc}
\hline
Flavour & Mass $\left({\rm MeV}\right)$ & Flavour & Mass $\left({\rm GeV}\right)$ \\
\hline
& & & \\
u & 2.3 & c & 1.28 \\
& & & \\
d & 4.8 & b & 4.18 \\
& & & \\
s & 95 & t & 173.07 \\
& & & \\
\hline
\end{tabular}
\caption{Phenomenological values of the light and heavy quark masses as given in Ref.~\cite{Beringer_2012_a}.}
\label{quark_masses}
\end{table}

\subsection{The Quark Model}

\begin{figure}[htb]
\centering
\includegraphics[scale=1.5]{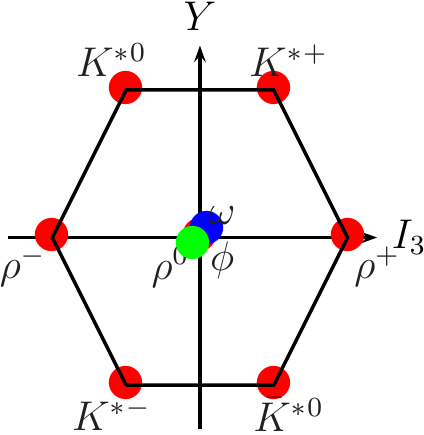}
\includegraphics[scale=1.5]{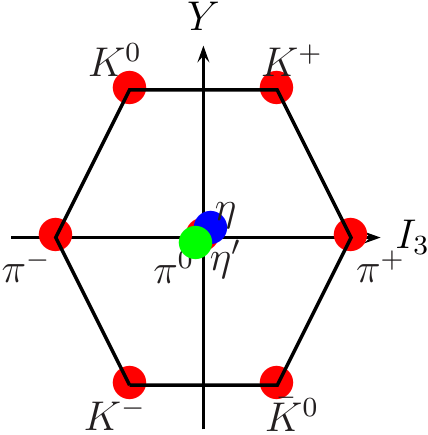}
\caption{
The vector (left) and pseudoscalar (right) meson nonets. The labels $Y$ and $I_3$ indicate the hypercharge and isospin quantum numbers, respectively. The masses of all of these particles are less than about $1\,{\rm GeV}$, and typically vary by a few hundred ${\rm MeV}$ within each nonet. However, the pions $\left(\pi^- \,,\, \pi^0\,,\, \pi^+\right)$ are anomalously light. The reason for this emerges naturally from QCD. The figures are from Ref.~\cite{Richard_2012_a}.}
\label{meson_nonets}
\end{figure}

In 1964, Gell-Mann~\cite{Gell_Mann_1964_a} and Zweig~\cite{Zweig_1964_a} independently introduced the quark model, which proposes that hadrons are not fundamental particles. Rather, they are composite objects composed of more fundamental particles, which Gell-Mann called quarks. At the time, all known hadrons could be explained in terms of just three types of quarks $\left(u\,,\,d\,,\,s\right)$, and their corresponding antimatter counterparts, antiquarks. The quark model suggests that there are only two kinds of hadrons: baryons that contain three quarks $\left(qqq\right)$, and mesons that contain a quark and an antiquark $\left(q\bar q\right)$. Baryons and mesons naturally arrange themselves into multiplets containing hadrons with similar properties and masses that are roughly degenerate. This approximate flavour symmetry is the origin of the multiplets. The vector and pseudoscalar meson nonets are shown in Fig.~\ref{meson_nonets}.

In many ways, the quark model is analogous to the periodic table. Initially the periodic table served to classify the elements based on their physical and chemical properties, without attempting to explain the underlying reasons for these properties. Of course, we now know that these properties ultimately derive from the electron shell structure of the elements as dictated by quantum mechanics. Similarly, the utility of the quark model lies in its explanation of the large number of hadrons in terms of a small number quarks. 

The quark model also led to an important insight into the nature of hadrons. Consider for instance the $\Delta^{++}$ baryon, which has spin-$3/2$. In the quark model, it is composed of three identical spin-$1/2$ up quarks that are not orbitally excited with respect to one another. This means that the spin, flavour and spatial wave functions are symmetric under particle interchange, meaning that the total wave function is also. Because the $\Delta^{++}$ is a fermion, this violates the spin-statistics theorem. This observation led to the introduction of the colour quantum number for quarks~\cite{Greenberg_1964_a}, which in the case of the $\Delta^{++}$ has an anti-symmetric wave function, thus ensuring that the spin-statistics theorem is upheld. The name colour was chosen in order to emphasize a key feature of hadrons, and is only used as an analogy. The number of colours can be inferred from experimental data, such as from the decay rate of the neutral pion~\cite{Narison_2007_a}. An individual quark may have 
one of three colours: red, green or blue. Each of the three quarks within a baryon must have a unique colour, thus the combination of red, green and blue is considered to have no net colour charge. All baryons and mesons are colourless, or colour singlets. 

\subsection{Exotic Hadrons}

In QCD, the colour quantum number of quarks is understood as a kind of generalization of electric charge. Quantum Electrodynamics (QED) describes electromagnetic interactions between electrically charged particles that are mediated by electrically neutral photons. In QCD, quarks with colour charge interact via gluons, which also carry colour charge. This fact means that QCD is radically different from QED, and it is also responsible for many of the interesting features of QCD. Unlike QED, the fundamental degrees of freedom in QCD are not directly manifested in nature. Instead, quarks and gluons are realized in terms of hadrons. 

QCD suggests the possibility of a far richer hadronic spectrum than the quark model. Exotic hadrons are colour singlet hadrons that are neither baryons nor mesons (see, {\it e.g.}\, Ref.~\cite{Klempt_2007_a} for a review). One such possibility is a hadrons with four quarks $\left(qq\bar{q}\bar{q}\right)$. Four-quark hadrons can be realized in two distinct ways. The first is as a weakly bound state of two colour singlet mesons $\left[\left(q\bar{q}\right)\left(q\bar{q}\right)\right]$, which is called a molecular state. The second is as a tetraquark, which is composed of diquark clusters that have a net colour charge $\left[\left(qq\right)\left(\bar{q}\bar{q}\right)\right]$ and hence is more strongly bound than a molecular state. Diquarks are best thought of as a kind of strong correlation between two quarks within a hadron~\cite{Anselmino_1992_a}. Because gluons also carry colour charge, colour singlet hadrons with explicit gluonic content are also possible. Hybrids are hadrons that can be thought of as a 
conventional meson with an excited gluon $\left(qG\bar{q}\right)$. Perhaps the most exotic of all exotic hadrons are glueballs, which are composed entirely of gluons $\left(GG\,\text{or}\,GGG\right)$. Note that four-quark states, hybrids and glueballs are all bosons. It should be noted that fermionic exotic hadrons are also possible, an example of which is a pentaquark $\left(qqqq\bar{q}\right)$. However, these will not be discussed in this thesis. The majority of the candidates for exotic hadrons exist among heavy quarkonia, all of which are bosons.

\subsection{Heavy Quarkonium-like States}

A meson that is composed of two heavy quarks of the same flavour is called heavy quarkonium. Those that are composed of charm quarks $\left(c\bar c\right)$ are called charmonia, while those that are composed of bottom quarks $\left(b\bar b\right)$ are called bottomonia. The top quark decays very rapidly via the weak interaction and does not form bound states. Because of the large masses of the charm and bottom quarks, relativistic effects are small, and hence heavy quarkonia can be approximated reasonably well using non-relativistic quantum mechanics. It is important to note that this approach does not derive directly from QCD. Rather, a potential is chosen that is inspired by QCD. The potential includes a short distance Coulombic term and a long distance term that models the effects of confinement. Spin dependent terms are crucial and relativistic corrections can also be included. The energy levels of the quarkonium system can be calculated using potential models. Each energy level, {\it i.e.}\ each 
charmonium or bottomonium state, is interpreted as a distinct meson. Ref.~\cite{Kwong_1987_a} provides a review of potential model methods.

\begin{figure}[htb]
\centering
\includegraphics[scale=0.7]{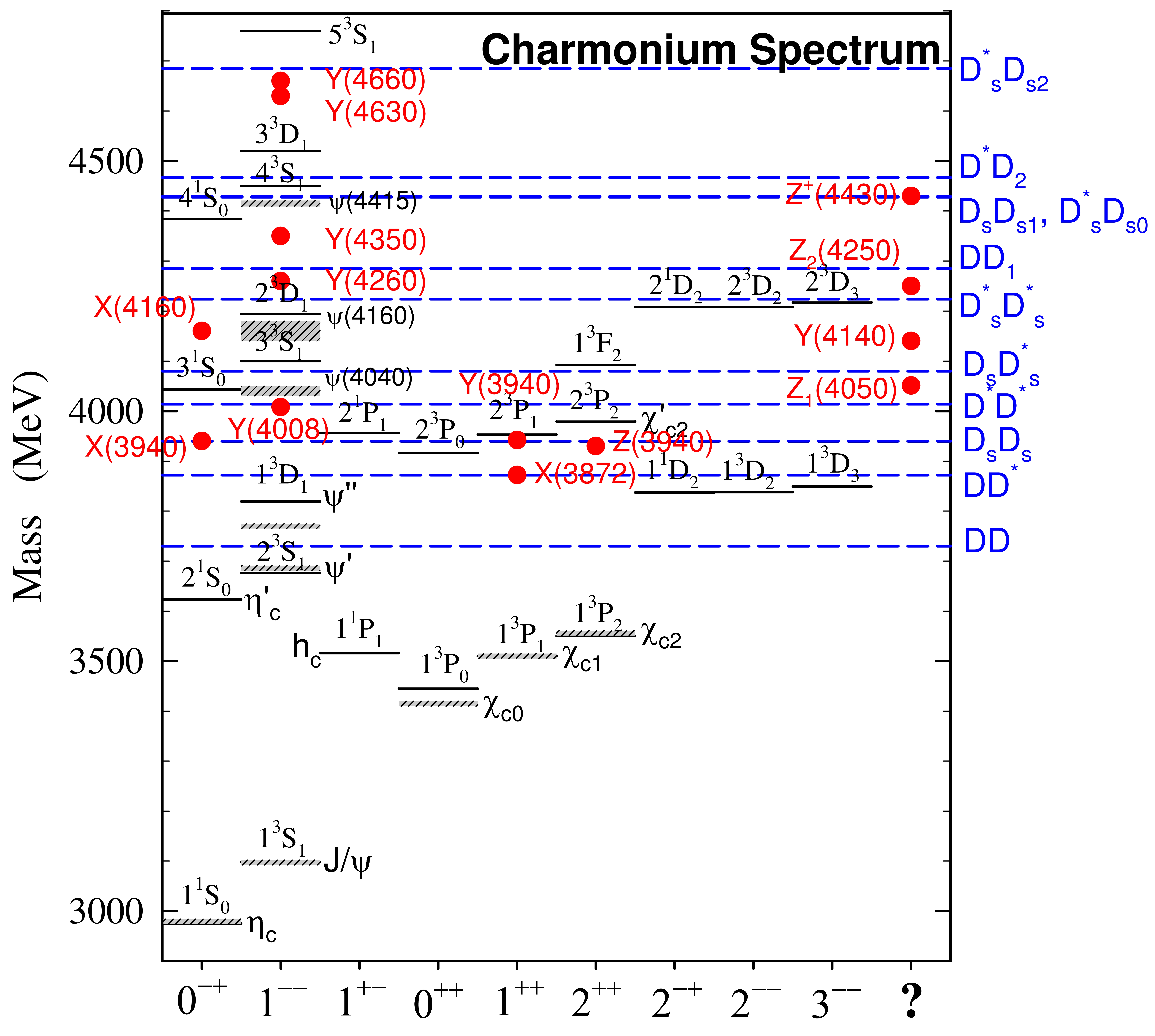}
\caption{
The charmonium spectrum. Black lines denote charmonium states, and red dots indicate charmonium-like states. Blue lines indicate the thresholds at which states can decay into a pair of $D$ mesons, which contain a charm quark and a light quark. Note that most of the $J^{PC}$ quantum numbers assigned to the XYZ states here are speculative (Ref.~\cite{Beringer_2012_a} lists the possible $J^{PC}$ for each XYZ state). Figure taken from Ref.~\cite{Godfrey_2009_a}.}
\label{charmonium_spectrum}
\end{figure}

Potential model predictions for the low-lying members of the charmonium and bottomonium spectra are in excellent agreement with experiment. However, in recent years experiments have begun to probe the mass region that higher mass charmonium states are expected to occupy. The results of these experiments have been quite surprising: numerous states that were not predicted by potential models have been found in the $3.8-4.7\,{\rm GeV}$ mass region, and a few unanticipated states have been found within the bottomonium mass region as well~\cite{Brambilla_2010_a}. These anomalous states are called heavy quarkonium-like, or XYZ states. The current experimental situation is summarized in detail in Ref.~\cite{Beringer_2012_a}. Fig.~\ref{charmonium_spectrum} shows the charmonium spectrum, including many of the charmonium-like states.

Charmonium states are labeled using spectroscopic notation $n^{2s+1}L_J$, where $n$ is the principal quantum number $\left(n=1\,,2\,,\ldots\right)$, $s=s_1+s_2$ is the total spin $\left(s=0\,,1\right)$, $L$ is the relative orbital angular momentum, and $J=L+S$ is the total angular momentum of the quark-antiquark pair. Following standard conventions, states with $L=0\,,1\,,2$ are denoted as $S\,,P\,,D$, and so on. In addition, all hadrons can be classified according to their $J^{PC}$ quantum numbers, where $P=(+,-)$ denotes parity and $C=(+,-)$ denotes charge conjugation (which is relevant to electrically neutral states). For quarkonia, it can be shown that the parity and charge conjugation quantum numbers are related to the spin and orbital angular momentum by $P=\left(-\right)^{L+1}$ and $C=\left(-\right)^{L+S}$, respectively~\cite{Griffiths_1987_a}. Accordingly, heavy quarkonium states can have $J^{PC}=0^{-+}$, $1^{--}$, $1^{+-}$, $0^{++}$, $1^{++}$, $2^{++}$, for example. However, it is impossible for a 
heavy quarkonium state to have the quantum numbers $J^{PC}=0^{--}$, $0^{+-}$, $1^{-+}$, or $2^{+-}$. Quantum numbers that are forbidden for heavy quarkonia are called exotic quantum numbers.

It has been widely speculated that some of the heavy quarkonium-like states could be exotic hadrons (see \cite{Brambilla_2010_a,Swanson_2006_a} for comprehensive reviews). This would explain why the XYZ states were unanticipated by potential models that consider only quark-antiquark hadrons. Experimentally, there are some simple signals for the existence of exotic hadrons. The first is that potential models predict a certain number of states for each $J^{PC}$ channel, and any supernumerary states could be exotic hadrons. A more obvious signal would be the observation of a state with exotic $J^{PC}$ quantum numbers, which cannot be realized by quark-antiquark bound states. To date, no hadrons with exotic quantum numbers have been definitively observed. Hadrons with unusual decay modes could also be exotic. For instance, states that are above open flavour thresholds in Fig.~\ref{charmonium_spectrum} are kinematically allowed to decay into pairs of $D$ mesons. Hadrons that are kinematically allowed to undergo 
such decays but fail to do so could be exotic.

An overview of the exotic interpretations of the heavy quarkonium-like states is given in Ref.~\cite{Brambilla_2010_a}. The most well known exotic hadron candidate is the $X(3872)$, whose quantum numbers have been confirmed to be $J^{PC}=1^{++}$ by the LHCb collaboration~\cite{Aaij_2013_a}. Several of its decay modes involve a $J/\psi$, which is the lightest spin-1 charmonium state~\cite{Beringer_2012_a}. Therefore the quark content of the $X(3872)$ must be at least $\bar{c}c$. However, the properties of $X(3872)$ are incompatible with a charmonium interpretation~\cite{Swanson_2006_a}. Shortly after its discovery, it was soon recognized that its mass is very close to the combined mass of the $D^{0}$ and $\bar{D}^{0*}$ mesons. For this reason the $X(3872)$ has been widely interpreted as a loosely bound $D^{0}\bar{D}^{0*}$ molecular state~\cite{Close_2003_a,Voloshin_2003_a,Swanson_2003_a,Tornqvist_2004_a,AlFiky_2005_a,Thomas_2008_a,Liu_2008_a,Lee_2009_a}. Another interpretation is that the $X(3872)$ is a 
tetraquark, and is expected to be only one member of a nonet of tetraquarks~\cite{Maiani_2004_a,Ebert_2005_a,Matheus_2006_a,Terasaki_2007_a,Dubnicka_2010_a}. Quite recently, the $Z_c^{\pm}(3895)$ was discovered by the BES-III collaboration~\cite{Ablikim_2013_a} and confirmed by the Belle~\cite{Liu_2013_a} and CLEO collaborations~\cite{Xiao_2013_a}. Because charmonium states cannot be electrically charged, this state cannot be a charmonium state. Like the $X(3872)$, the $Z_c^{\pm}(3895)$ decays to $J/\psi$, hence it must contain $\bar{c}c$. However, this combination cannot produce an electric charge. The simplest explanation for this state is that it is a four-quark state of the form $c\bar{c}q_i\bar{q}_j$, where $q_i$ and $\bar{q}_j$ are light quarks with different flavours. In fact, Ref.~\cite{Maiani_2004_a} predicted the existence of the $Z_c^{\pm}(3895)$ on the basis of a tetraquark model of the $X(3872)$. The $X(3872)$ and $Z_c^{\pm}(3895)$ are discussed in Chapter~\ref{chapter_4_Qq_diquark}.

\section{Quantum Chromodynamics}

Quantum Chromodynamics (QCD) is the fundamental theory of strong interactions. It is a quantum field theory, which is a generalization of quantum mechanics to describe physical processes involving particle creation or annihilation. It is important to stress that quantum mechanics is incapable of this: the wave function of a particle that has not yet been created or has been annihilated cannot be normalized, and thus is incompatible with the statistical interpretation of quantum mechanics. In quantum field theory particles are understood as being excitations, or quanta, of quantum fields. There are two distinct approaches that are used to construct a quantum field theory. Canonical quantization involves reinterpreting classical fields as operators that satisfy a certain algebra. A second approach utilizes the path integral formulation of quantum mechanics (see Ref.~\cite{Feynman_1965_a} for a review). Both methods will be utilized in this chapter to formulate QCD.

\subsection{Canonical Quantization of Quark Fields}
\label{CanonicalQuantizationOfQuarkFields}

QCD begins with quantizing spin-$1/2$ fermion fields that represent quarks. These satisfy the Dirac equation,
\begin{equation}
\left[i\slashed{\partial} - m \right] Q\left(x\right) = 0 \,, \quad \slashed{\partial}=\gamma^\mu \frac{\partial}{\partial x^\mu} \,,
\label{dirac_eqn} 
\end{equation}
where the quark field $Q\left(x\right)$ is a complex four-component spinor field and we are using natural units (see Appendix~\ref{appendix_a_conventions}). The set of four matrices $\gamma^\mu$  satisfy the algebra $\{\gamma^\mu \,,\, \gamma^\nu \} = 2\,g^{\mu\nu}$. A peculiarity of the Dirac equation is that it permits both positive and negative energy solutions for free particles. Negative energy solutions represent antiparticles, which are identical in every way to their particle counterparts, except that they have opposite electric charge. Particles and antiparticles can interact to annihilate one another and particle-antiparticle pairs can be created spontaneously. When interactions are included, the statistical interpretation of non-relativistic quantum theory cannot be applied to the Dirac equation.

The solution to this problem is to reinterpret the Dirac equation as a field equation, rather than a single particle wave equation, and then quantize the field. In this way, a consistent quantum field theory that incorporates interactions can be constructed. The dynamics of fields are governed by the principle of least action, where the action is defined as
\begin{equation}
 S = \int \, d^4x \, \mathcal{L}\left(Q, \, \partial_\mu Q \right) \,,
\label{action}
\end{equation}
where $\mathcal{L}\left(Q, \, \partial_\mu Q \right)$ is the Lagrangian density, which is commonly referred to as the Lagrangian. The principle of least action states that as the field $Q\left(x\right)$ evolves in spacetime it does so in a way that minimizes the action (\ref{action}). It can be shown that in order to satisfy the principle of least action, the field must satisfy the Euler-Lagrange equation,
\begin{equation}
\frac{\partial\mathcal{L}}{\partial Q} - \partial_\mu \left(\frac{\partial\mathcal{L}}{\partial\left(\partial_\mu Q \right)}\right) =0 \,.
\label{euler_lagrange}
\end{equation}
Note that this must be satisfied by each distinct field in a given Lagrangian. Given the Lagrangian for a field, the equations of motion for the field can be determined using (\ref{euler_lagrange}). It is important to emphasize that at this stage the fields are still classical quantities. Only when the fields have been reinterpreted as operators that satisfy an appropriate algebra will we pass to a quantum field theory.

Let us now consider canonical quantization of Dirac fields. Quantum theory uses the Hamiltonian to determine the time evolution of a system. In the Heisenberg picture of quantum theory, time-dependence is carried by operators governed by the Heisenberg equation of motion,
\begin{equation}
i \partial_0 Q\left(x\right) = \left[Q\left(x\right) \,,\,H\right] \,.
\label{H_eqn_of_motion}
\end{equation}
In order to quantize the Dirac fields, we must first know what Hamiltonian operator to use in (\ref{H_eqn_of_motion}). Since the Hamiltonian and Lagrangian are related, we may determine a suitable Lagrangian for the Dirac fields and use this to find the corresponding Hamiltonian. The simplest form of the Lagrangian can be written as
\begin{equation}
\mathcal{L} = \bar{Q}\left[i\slashed{\partial} - m \right] Q \,,
\label{Ldirac}
\end{equation}
where $\bar{Q}=Q^\dag\gamma^0$. When this Lagrangian is substituted into the Euler-Lagrange equation (\ref{euler_lagrange}) with $Q$ and $\bar{Q}$ treated as dynamical fields, the correct equations of motion for $Q$ and $\bar{Q}$ result, so this is a suitable Lagrangian for the quark fields. Using the relationship between the Hamiltonian and the Lagrangian along with (\ref{Ldirac}), the Hamiltonian for Dirac fields can be shown to be
\begin{equation}
H = \int d^3x \, \mathcal{H} = \int d^3x 
\left[ \frac{\partial\mathcal{L}}{\partial\left[\partial_0\,Q\left(x\right)\right]}\partial_0 Q\,\left(x\right) -  \mathcal{L} \right] 
= i \int d^3x \, \bar{Q}\left(x\right) \gamma^0 \partial_0 Q\left(x\right)
\,.
\label{Hdirac}
\end{equation}
We may now use the Heisenberg equation of motion (\ref{H_eqn_of_motion}) to quantize Dirac fields. Using the Hamiltonian (\ref{Hdirac}) and the identity $\left[A\,,\,BC\right]= \{A\,,\,B\}C-B\{A\,,\,C\}$, 
\begin{gather}
\begin{split}
\left[Q\left(x\right) \,,\, H \right] &= i \int d^3y \left[Q\left(x\right) \,,\,  \bar{Q}\left(y\right) \gamma^0 \frac{\partial}{\partial y^0} Q\left(y\right) \right] 
\\
&= i \int d^3y 
\left( 
\{ Q\left(x\right) \,,\, \bar{Q}\left(y\right) \} \gamma^0 \frac{\partial}{\partial y^0} Q\left(y\right) - 
\bar{Q}\left(y\right) \{ Q\left(x\right) \,,\, \gamma^0 \frac{\partial}{\partial y^0} Q\left(y\right) \}
\right) 
\\
&= i \int d^3y 
\left( 
\{ Q\left(x\right) \,,\, \bar{Q}\left(y\right) \} \gamma^0 \frac{\partial}{\partial y^0} Q\left(y\right) - 
\bar{Q}\left(y\right) \gamma^0 \frac{\partial}{\partial y^0} \{ Q\left(x\right) \,,\, Q\left(y\right) \}
\right) 
\\
&= i \frac{\partial}{\partial x^0} Q\left(x\right) \,.
\end{split}
\end{gather}
In order to satisfy the Heisenberg equation of motion, the quark fields must satisfy an equal time anticommutator algebra where
\begin{equation}
\{ Q^\alpha_j\left(x\right) \,,\, \bar{Q}^\beta_k\left(y\right) \} = \delta^3\left(\, \mathbf{x}-\mathbf{y}\,\right)\,\gamma^0_{jk} \,,
\label{dirac_algebra}
\end{equation}
and all other anticommutators are zero. Note that we have restored implicit spinor indices $j\,,k$ and used the property $\left(\gamma^0\right)^2=1$ as well as the Heisenberg equation of motion for $\bar{Q}$. It is important to note that the Heisenberg equation of motion can also be satisfied by operators that have a commutator algebra. However, the spin-statistics theorem requires that fermions satisfy an anticommutator algebra. Since Dirac fields are fermions, we must use the algebra (\ref{dirac_algebra}). This issue is discussed in many standard texts, see for instance Ref.~\cite{Peskin_1995_a}.

The Dirac fields can be expanded in a basis of plane wave states,
\begin{gather}
\begin{split}
& Q_j\left(x\right) = \int \frac{d^3p}{\left(2\pi\right)^3} \frac{1}{\sqrt{2E_p}} \sum_s 
\left( a^s\left(\mathbf{p}\right)u_j^s\left(p\right)e^{-ip \cdot x} + b^{s\dag}\left(\mathbf{p}\right)v_j^s\left(p\right)e^{ip \cdot x} \right) 
\\
& \bar{Q}_k\left(x\right) = \int \frac{d^3p}{\left(2\pi\right)^3} \frac{1}{\sqrt{2E_p}} \sum_s
\left( b^s\left(\mathbf{p}\right)\bar{v}_k^s\left(p\right)e^{-ip \cdot x} + a^{s\dag}\left(\mathbf{p}\right)\bar{u}_k^s\left(p\right)e^{ip \cdot x} \right)
\end{split}
\label{quark_field_mode_expansion}
\end{gather}
where $p^0 = E_p$ and $s$ denotes the spin state. Using these expressions and the algebra (\ref{dirac_algebra}) it is easy to show that 
\begin{equation}
\{ a^r_j\left(\mathbf{p}\right) \,,\, a^{s\dag}_k\left(\mathbf{q}\right) \} = \{ b^r_j\left(\mathbf{p}\right) \,,\, b^{s\dag}_k\left(\mathbf{q}\right) \} = \left(2\pi\right)^3  \delta^3\left(\mathbf{p}-\mathbf{q}\right) \, \delta^{rs} \, \delta_{jk}
\end{equation}
and all other anticommutators involving these operators are zero. The solutions (\ref{quark_field_mode_expansion}) are linear combinations of the basis vectors of the Hamiltonian (\ref{Hdirac}). In quantum field theory, the operators (\ref{quark_field_mode_expansion}) are interpreted as creating and annihilating field quanta by acting on the vacuum (ground) state $|0\rangle$, which contains no field quanta. This requires that $a^r_j\left(\mathbf{p}\right) |0\rangle = 0$. Particles and antiparticles are understood as field quanta and are represented by momentum eigenstates with an associated spin state $s$. For later convenience it is useful to define the normal-ordering operator. When acting on a product of creation and annihilation operators, the normal-ordering operator moves all creation operators $a^{s\dag}_k\left(\mathbf{q}\right)$ to the left. For example,
\begin{gather}
:a^r_j\left(\mathbf{p}\right) a^{s\dag}_k\left(\mathbf{q}\right): \,
= - a^{s\dag}_k\left(\mathbf{q}\right) a^r_j\left(\mathbf{p}\right) \,.
\label{normal_ordering_operator} 
\end{gather}
This can also be applied to the quark field operators~\eqref{quark_field_mode_expansion}. A crucial property is that 
\begin{gather}
\langle 0 | : \ldots : | 0 \rangle = 0 \,, 
\end{gather}
where the dots denote any combination of quantum fields.

The quark field operators can be used to determine the amplitude for a quark to propagate between two distinct locations in spacetime. In order to calculate this amplitude we must first define the time-ordering operator, 
\begin{gather}
T\left[Q\left(x\right)\bar{Q}\left(y\right)\right] 
= \left\{ 
\begin{array}{rr}
Q\left(x\right)\bar{Q}\left(y\right) \,, & x^0 > y^0  \\
-\bar{Q}\left(y\right)Q\left(x\right) \,, & x^0 < y^0  
\end{array}
\right.
\label{time_ordering_operator} 
\end{gather}
which anticommutes a product of Dirac fields so that the field with the earliest time is the furthest right and the field with the latest time is the furthest left. The time-ordering operator ensures that particles only propagate forward in time. Explicitly calculating Eqn.~\eqref{time_ordering_operator} using the expressions for the quark fields~\eqref{quark_field_mode_expansion} and their algebra~\eqref{dirac_algebra}, it can be shown that
\begin{gather}
\langle 0 | \, T\left[ \right. Q\left(x\right)\bar{Q}\left(y\right) \left.\right] |0\rangle
= S\left(x-y\right)
= i \int \frac{d^4p}{\left(2\pi\right)^4} \frac{\slashed{p}+m}{p^2-m^2+i\eta} e^{-ip \cdot \left(x-y\right)} \,,
\label{quark_propagator}
\end{gather}
where $\slashed{p}=\gamma^\mu p_\mu$ and $\eta \rightarrow 0^{+}$. Equation (\ref{quark_propagator}) is the Feynman quark propagator, which is also a Green's function of the Dirac equation~\eqref{dirac_eqn}. The $i\eta$ pole prescription ensures that time-ordering is respected. The Feynman propagator $S\left(x-y\right)$ is the quantum mechanical amplitude for a quark to travel between the spacetime points $y$ and $x$, if $y^0<x^0$.

\subsection{Perturbation Theory}

The $S$-matrix formalism relates physical quantities such as scattering cross sections and decay rates to correlation functions, which are also referred to as Green's functions or $n$-point functions. Therefore correlation functions are of paramount importance in quantum field theory. As an example, consider the four-point function
\begin{gather}
\langle 0 | T\left[Q\left(x_1\right)\bar{Q}\left(x_2\right)Q\left(x_3\right)\bar{Q}\left(x_4\right)\right] | 0 \rangle \,. 
\label{example_correlation_function}
\end{gather}
Correlation functions of this form can be evaluated using Wick's theorem, which can be used to express any time ordered product in terms of Feynman propagators and normal ordered products. For the time ordered product in~\eqref{example_correlation_function}, Wick's theorem yields
\begin{gather}
\begin{split}
T\left[Q\left(x_1\right)\bar{Q}\left(x_2\right)Q\left(x_3\right)\bar{Q}\left(x_4\right)\right] 
=& \,:Q\left(x_1\right)\bar{Q}\left(x_2\right)Q\left(x_3\right)\bar{Q}\left(x_4\right):
\\
&\!+\contraction{}{Q}{\left(x_1\right)}{\bar{Q}}Q\left(x_1\right)\bar{Q}\left(x_2\right):Q\left(x_3\right)\bar{Q}\left(x_4\right):
\\
&+:Q\left(x_1\right)\bar{Q}\left(x_2\right):\contraction{}{Q}{\left(x_3\right)}{\bar{Q}}Q\left(x_3\right)\bar{Q}\left(x_4\right)
\\
&\!+\contraction{}{Q}{\left(x_1\right)}{\bar{Q}}\contraction{Q\left(x_1\right)\bar{Q}\left(x_2\right)}{Q}{\left(x_3\right)}{\bar{Q}} Q\left(x_1\right)\bar{Q}\left(x_2\right)Q\left(x_3\right)\bar{Q}\left(x_4\right)
\\
&\!+\contraction[2ex]{}{Q}{\left(x_1\right)\bar{Q}\left(x_2\right)Q\left(x_3\right)}{\bar{Q}}\contraction{Q\left(x_1\right)}{\bar{Q}}{\left(x_2\right)}{Q} Q\left(x_1\right)\bar{Q}\left(x_2\right)Q\left(x_3\right)\bar{Q}\left(x_4\right) \,.
\end{split}
\label{wicks_theorem}
\end{gather}
The contraction of the quark fields is defined as
\begin{gather}
\contraction{}{Q}{\left(x\right)}{\bar{Q}} Q\left(x\right) \bar{Q}\left(y\right) = S\left(x-y\right) \,,
\label{wick_contraction}
\end{gather}
where $S\left(x-y\right)$ is the quark propagator~\eqref{quark_propagator}. Note that in the last line of Eq.~\eqref{wicks_theorem} there are contractions where the fields are not adjacent and are not in the same order as those in~\eqref{wick_contraction}. The quark fields can be moved so that they are adjacent and in the proper order using the anticommutator algebra~\eqref{dirac_algebra}. Wick's theorem yields the following for the correlation function:
\begin{gather}
\begin{split}
\langle 0 | T\left[Q\left(x_1\right)\bar{Q}\left(x_2\right)Q\left(x_3\right)\bar{Q}\left(x_4\right)\right] | 0 \rangle  &=  S\left(x_1-x_2\right)S\left(x_3-x_4\right)
\\
&-S\left(x_3-x_2\right)S\left(x_1-x_4\right) \,,
\end{split}
\label{example_wicks_theorem_correlation_function} 
\end{gather}
where we have used the fact that vacuum expectation values of normal ordered products are identically zero. Wick's theorem is valid for all quantum fields, and can be generalized to time ordered products involving any number of fields. 

All physical theories involve interactions between quantum fields. When interactions are included correlation functions can be calculated via perturbation theory. It can be shown that correlation functions in the interacting theory are related to those in the non-interacting theory by
\begin{equation}
\langle \Omega | \, T\left[ Q\left(x\right)\bar{Q}\left(y\right) \right] | \Omega \rangle = 
\lim_{t \to \infty\left(1-i\epsilon\right)} 
\frac{ \langle0|\, T\left[ Q\left(x\right)\bar{Q}\left(y\right)e^{iS_{\rm int}} \right] |0\rangle}{\langle0|\, T\left[e^{iS_{\rm int}}\right]|0\rangle} \,,
\label{perturbative_expansion_of_correlation_function}
\end{equation}
where $| \Omega \rangle$ and $|0\rangle$ denote the vacua of the interacting and free (non-interacting) theories, respectively~\cite{Peskin_1995_a}. The limit is needed in order to define $| \Omega \rangle$ as a perturbation of $|0\rangle$. The exponential is defined as 
\begin{gather}
S_{\rm int} = \int d^4 x \, \mathcal{L}_{\rm int} \,,
\label{defintion_of_interaction_lagrangian} 
\end{gather}
where $\mathcal{L}_{\rm int}$ is the part of the interacting theory Lagrangian that defines an interaction between quantum fields. One of the fundamental assumptions of quantum field theory is that interactions between quantum fields are local, that is, fields interact at a single point in spacetime. For instance, in the next section we shall see that the interaction between quark and gluon fields is given by
\begin{gather}
\mathcal{L} = \frac{g}{2} \bar{Q}\left(x\right) \lambda^a \gamma^\mu A_\mu^a\left(x\right) Q\left(x\right) \,,
\label{quark_gluon_interaction_lagrangian}
\end{gather}
where $A_\mu^a$ denotes a gluon field and the coupling $g$ characterizes the strength of the interaction. The interacting theory correlation function~\eqref{perturbative_expansion_of_correlation_function} can be calculated as a power series in the coupling $g$. Equation~\eqref{perturbative_expansion_of_correlation_function} can be generalized to calculate correlation functions involving any number of quark or gluon fields by simply adding these fields to both sides of the equation. Wick's theorem remains valid and can be used to calculate correlation functions in the interacting theory in terms of the propagators of the non-interacting theory~\eqref{quark_propagator}. However, in the QCD vacuum $|\Omega\rangle$ there are some normal ordered products whose expectation values are non-zero. These are called condensates and will be discussed in Section~\ref{QCDLaplaceSumRules}.

\begin{figure}[htb]
\centering
\includegraphics[scale=1]{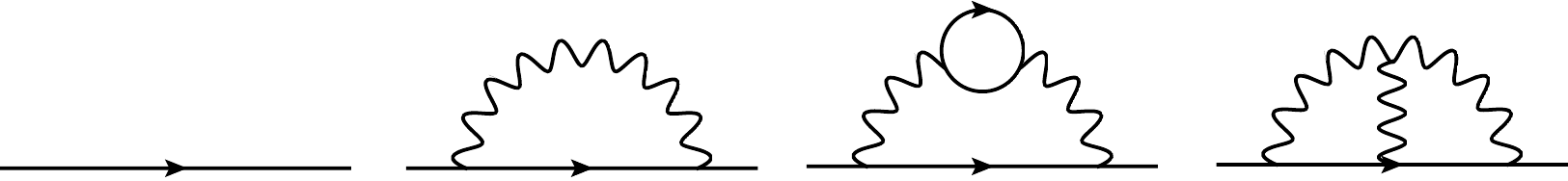}
\caption{Examples of Feynman diagrams that contribution to the perturbative expansion of the correlation function in Eq.~\eqref{perturbative_expansion_of_correlation_function}. Straight and wavy lines represent quark and gluon propagators, respectively. Each vertex represents the local interaction between quark and gluon fields given by~\eqref{quark_gluon_interaction_lagrangian}, and hence introduces a factor of the coupling $g$. Diagrams with more vertices correspond to higher order terms in the perturbative expansion. Note that the first three diagrams can also occur in QED, where the gluons are replaced with photons. However, the fourth diagram includes a direct interaction between gluon fields and has no equivalent in QED. These Feynman diagrams were produced using JaxoDraw~\cite{Binosi_2003_a}.}
\label{examples_of_feynman_diagrams}
\end{figure}

Correlation functions can be represented in terms of Feynman diagrams. For instance, the perturbative expansion of the correlation function~\eqref{perturbative_expansion_of_correlation_function} involves quark and gluon propagators, as well as interactions between quarks and gluons due to the interaction term~\eqref{quark_gluon_interaction_lagrangian}. Some of these terms are shown in Fig.~\ref{examples_of_feynman_diagrams}.

Note that the Feynman diagrams in Fig.~\eqref{examples_of_feynman_diagrams} are connected, that is, all of the propagators are linked to one another. However, the perturbative expansion of the numerator in Eq.~\eqref{perturbative_expansion_of_correlation_function} includes disconnected diagrams. Diagrams of this type represent vacuum processes. It can be shown that the denominator of Eq.~\eqref{perturbative_expansion_of_correlation_function} serves to cancel all disconnected diagrams that arise in the perturbative expansion. In practice this cancellation can be implemented by simply ignoring terms in the perturbative expansion that correspond to disconnected diagrams.

Feynman diagrams can be used as mnemonics to keep track of terms in the perturbative expansion of a correlation function. Using Wick's theorem and the expression for the perturbative expansion~\eqref{perturbative_expansion_of_correlation_function}, it is possible to relate each diagram component to a certain mathematical expression. These are called Feynman rules. One of the Feynman rules for QCD is that every quark line in a Feynman diagram mathematically corresponds to a quark propagator~\eqref{quark_propagator}. Another is that every quark-gluon vertex is associated with a factor of $i\frac{g}{2}\lambda^a \gamma^\mu$. This vertex rule can be derived easily using Wick's theorem and the perturbative expansion. The Feynman rules for QCD are given in Ref.~\cite{Peskin_1995_a}. Note, however, that any correlation function can be calculated using Wick's theorem. It is important to emphasize that Wick's theorem is more fundamental than the Feynman rules.

\subsection{Non-Abelian Gauge Theory}

Correlation functions can be calculated perturbatively once the complete Lagrangian for a theory is known. In the SM, interactions are introduced through gauge symmetries. For instance, we have seen that the Lagrangian for free quark fields is given by
\begin{equation}
\mathcal{L}_{\rm quarks} = \bar{Q}_j\left[i\slashed{\partial} - m \right] Q_j \,,
\label{quark_lagrangian}
\end{equation}
where we have introduced the index $j\in\left\{1\,,2\,,3\right\}$ to denote the colour degree of freedom of the quark fields. The Lagrangian~\eqref{quark_lagrangian} is invariant under the global gauge transformation
\begin{gather}
Q_i \to Q_i^\prime = \exp{\left[-igT^a_{ij}\theta^a\right]} Q_j \,, \quad \bar{Q}_i \to \bar{Q}_i^\prime = \exp{\left[igT^a_{ij}\theta^a\right]} \bar{Q}_j \,,
\label{global_gauge_transformations} 
\end{gather}
where $T^a$ is a generator of the non-Abelian group $SU\left(3\right)$. The generators satisfy the Lie algebra $\left[T^a\,,T^b\right]=if^{abc}T^c$ where $f^{abc}$ are the structure constants of $SU\left(3\right)$. The generators are related to the Gell-Mann matrices $\lambda^a$ via $T^a=\frac{\lambda^a}{2}$. Now, suppose that we alter the gauge transformation~\eqref{global_gauge_transformations} so that $\theta^a\to\theta^a\left(x\right)$. Clearly the Lagrangian~\eqref{quark_lagrangian} is not invariant under this local gauge transformation. However, it can be made so by introducing a gauge field. This can be done by replacing the derivative in~\eqref{quark_lagrangian} with a covariant derivative
\begin{gather}
D_\mu = \partial_\mu - igT^a A^a_\mu \,,
\label{covariant_derivative} 
\end{gather}
where $A^a_\mu$ is the gauge field. In fact, this is the gluon field, which has its own Lagrangian
\begin{gather}
 \mathcal{L}_{\rm YM} = - \frac{1}{4} G^a_{\mu\nu} G_a^{\mu\nu} \,, \quad G^c_{\mu\nu} = \partial_\mu A^c_\nu - \partial_\nu A^c_\mu + g f^{abc} A^a_\mu A^b_\nu \,,
\label{yang_mills_lagrangian}
\end{gather}
where $G^a_{\mu\nu}$ is the gluon field strength tensor and the gluon field is massless. This is called the Yang-Mills Lagrangian. It can be shown that the following Lagrangian is invariant under local $SU(3)$ gauge transformations~\cite{Srednicki_2007_a}:
\begin{gather}
\mathcal{L} = \bar{Q}\left[i\slashed{D} - m \right] Q - \frac{1}{4} G^a_{\mu\nu} G_a^{\mu\nu} \,.
\label{classical_QCD_lagrangian} 
\end{gather}
The covariant derivative~\eqref{covariant_derivative} leads to the quark-gluon interaction term~\eqref{quark_gluon_interaction_lagrangian} discussed earlier. Also, notice that the definition of the gluon field strength tensor~\eqref{yang_mills_lagrangian} leads to self-interactions among gluon fields. Interactions between gauge fields with a universal coupling are a distinguishing feature of non-Abelian gauge theories.

\subsection{Path Integral Quantization of Gluon Fields}

Although the Lagrangian~\eqref{yang_mills_lagrangian} contains interactions, the gluon field still has to be quantized. The methods of canonical quantization that was used in Section~\ref{CanonicalQuantizationOfQuarkFields} to quantize quark fields are ill-suited for this purpose. Instead, we will utilize the path integral to quantize the gluon field. The discussion in this section closely follows that of Ref.~\cite{Srednicki_2007_a}.

The generating functional for a quantum field $\phi$ is defined as 
\begin{gather}
Z\left[J\right] = \int D\phi \exp\left[i\int d^4 x \left[ \mathcal{L}+J\left(x\right)\phi\left(x\right) \right] \right] \,,
\label{generating_functional} 
\end{gather}
where $\mathcal{L}=\mathcal{L}\left(\phi\,,\partial_\mu\phi\right)$ is the Lagrangian for the field $\phi$. The integration in~\eqref{generating_functional} is over the space of configurations of the field $\phi$. An integral of this form is called a path integral. The path integral is a functional, that is, a function that acts upon functions and returns numbers. The term $J\left(x\right)$ in the exponential is known as a source term. It is useful to define the functional derivative
\begin{gather}
\frac{\delta}{\delta J\left(x\right)} J\left(y\right) = \delta^4\left(x-y\right) \,.
\label{functional_derivative} 
\end{gather}
It can be shown that correlation functions involving the field $\phi$ can be calculated as functional derivatives of the generating functional~\cite{Peskin_1995_a}. For example,
\begin{gather}
\langle 0 | T\left[\right. \phi\left(x_1\right)\phi\left(x_2\right) \left.\right] 0 \rangle = \frac{1}{Z_0} \left. \left(-i \frac{\delta}{\delta J\left(x_1\right)} \right) \left(-i \frac{\delta}{\delta J\left(x_2\right)} \right) Z\left[J\right] \right|_{J=0} \,,
\label{example_of_correlation_function_from_path_integral} 
\end{gather}
where $Z_0=Z\left[J=0\right]$. Correlation functions involving more $\phi$ fields can be calculated simply by calculating more functional derivatives of the generating functional. Note that the field $\phi$ has been quantized: the path integral can be used to calculate correlation functions of the field $\phi$, which are quantum mechanical amplitudes. This procedure generalizes to any quantum field, provided that the statistics of the field are incorporated. For instance, path integrals involving fermion fields require the use of Grassmann variables~\cite{Peskin_1995_a}. Once the generating function for a quantum field has been defined, the field has been quantized.

Now we will construct the generating functional for the gluon field. By analogy with the generating functional for the field $\phi$~\eqref{generating_functional}, we might guess that the generating functional for the gluon field is given by
\begin{gather}
 Z\left[J\right] = \int D A \exp\left[i\int d^4 x \left[ \mathcal{L}_{\rm YM} +J^\mu_a A^a_\mu \right] \right] \,,
\label{guess_for_gluon_generating_functional}
\end{gather}
where $\mathcal{L}_{\rm YM}$ denotes the Yang-Mills Lagrangian~\eqref{yang_mills_lagrangian}. Unfortunately, the integration over the configurations of the gluon field is ill-defined. This is due to the gauge symmetry of $\mathcal{L}_{\rm YM}$. It can be shown that under an infinitesimal gauge transformation, the gluon field transforms as 
\begin{gather}
A_\mu^a \left(x\right) \to \tilde{A}_\mu^a \left(x\right) = A_\mu^a \left(x\right) - D_\mu^{ab}\theta^b\left(x\right) \,, \quad D_\mu^{ab} = \delta^{ab}\partial_\mu - gf^{abc}A_\mu^c \,.
\label{gluon_field_gauge_transformation} 
\end{gather}
This reflects a redundancy among the configurations of the field $A_\mu^a$, which spoils the definition of the generating functional~\eqref{guess_for_gluon_generating_functional}. 

The generating functional given in Eq.~\eqref{guess_for_gluon_generating_functional} cannot be used to quantize the gluon fields in its present form. We will use a method introduced by Faddeev and Popov~\cite{Faddeev_1967_a} to modify the generating functional so that quantization is possible. The redundancy in the integration over the field $A_\mu^a$ can be removed by introducing the gauge-fixing function
\begin{gather}
 Z\left[J\right] = \int D A \, {\rm det}\left(\frac{\delta G}{\delta \theta}\right) \delta\left(G\right) \exp\left[i\int d^4 x \left[ \mathcal{L}_{\rm YM} +J^\mu_a A^a_\mu \right] \right] \,,
\label{constrained_gluon_generating_functional}
\end{gather}
where $G^a\left(x\right)=\partial^\mu A^a_\mu - \omega^a\left(x\right)$ for some arbitrary function $\omega^a$. Using Eq.~\eqref{gluon_field_gauge_transformation}, it can be shown that the gauge-fixing function transforms as 
\begin{gather}
G^a \left(x\right) \to \tilde{G}^a \left(x\right) = G^a \left(x\right) - \partial^\mu D_\mu^{ab}\theta^b\left(x\right) \,.
\label{gauge_fixing_function_gauge_transformation} 
\end{gather}
Using this, the functional derivative in Eq.~\eqref{constrained_gluon_generating_functional} is
\begin{gather}
\frac{\delta G^a\left(x\right)}{\delta \theta^b\left(y\right)} = - \partial^\mu D_\mu^{ab} \delta^4\left(x-y\right) \,.
\label{functional_derivative_of_G}
\end{gather}
Note that the Faddeev-Popov method can be used to quantize QED, but there Eq.~\eqref{functional_derivative_of_G} does not depend on the photon field and hence it cannot introduce any new dynamics into the theory. However, in QCD the functional determinant explicitly depends on the gluon field $A^a_\mu$, because Eq.~\eqref{functional_derivative_of_G} contains the covariant derivative. The functional determinant that appears in~\eqref{constrained_gluon_generating_functional} can be expressed in terms of a path integral involving Faddeev-Popov ghosts:
\begin{gather}
\begin{split}
{\rm det}\left(\frac{\delta G}{\delta \theta}\right) = \int Dc \, D\bar{c} \, \exp{\left[i\int d^4 x \, \mathcal{L}_{\rm gh} \right]} \,, 
\\
\mathcal{L}_{\rm gh} = \bar{c}^a \partial^\mu D^{ab}_\mu c^b = -\partial^\mu \bar{c}^a \partial_\mu c^a + g f^{abc}A_\mu^c\partial^\mu \bar{c}^a c^b \,.
\end{split}
\label{ghost_path_integral}
\end{gather}
The ghost fields $c^a$ and $\bar{c}^a$ are unphysical, but are needed to defined the generating functional for the gluon field. The first term in $\mathcal{L}_{\rm gh}$ can be used to calculated the ghost propagator (given in Ref.~\cite{Peskin_1995_a}, for instance) and the second term in $\mathcal{L}_{\rm gh}$ denotes an interaction between the ghost and gluon field. The delta functional appearing can be dealt with by multiplying the generating functional~\eqref{constrained_gluon_generating_functional} by 
\begin{gather}
 \exp{\left[-\frac{i}{2a} \int d^4 x \, w^a\left(x\right) w^a\left(x\right) \right]} \,.
\label{omega_function_integral}
\end{gather}
This is permitted because $\omega^a\left(x\right)$ does not depend on the gluon field $A^a_\mu$, and hence multiplying the generating functional~\eqref{constrained_gluon_generating_functional} by Eq.~\eqref{omega_function_integral} can only alter the overall normalization of the generating functional. The delta function in~\eqref{constrained_gluon_generating_functional} can be used to evaluate the integral~\eqref{omega_function_integral}. This effectively introduces a new term into the generating functional that has the form
\begin{gather}
\mathcal{L}_{\rm gf} = -\frac{1}{2a} \partial^\mu A_\mu^a \partial^\nu A_\nu^a \,.
\label{gauge_fixing_lagrangian} 
\end{gather}
This is called the gauge-fixing Lagrangian, and $a$ is the gauge parameter. Finally, the generating functional for the gluon field is 
\begin{gather}
Z\left[J\right] = \int DA \, Dc \, D\bar{c} \, \exp{\left[i\int d^4 x \left[\mathcal{L}_{\rm YM} + J^\mu_a A^a_\mu  + \mathcal{L}_{\rm gf} + \mathcal{L}_{\rm gh} \right]\right]} \,,
\label{complete_gluon_generating_functional} 
\end{gather}
where $\mathcal{L}_{\rm YM}$ is the Yang-Mills Lagrangian~\eqref{yang_mills_lagrangian}, $\mathcal{L}_{\rm gf}$ is the gauge-fixing Lagrangian~\eqref{gauge_fixing_lagrangian} and $\mathcal{L}_{\rm gh}$ is the ghost Lagrangian~\eqref{ghost_path_integral}. Using~\eqref{complete_gluon_generating_functional}, it can be shown that the gluon propagator is given by
\begin{gather}
\langle 0 | \, T\left[ \right. A_\mu^a\left(x\right)A_\nu^b\left(y\right) \left. \right] |0\rangle
= D_{\mu\nu}^{ab}\left(x-y\right)
= -i \delta^{ab} \int \frac{d^4p}{\left(2\pi\right)^4} \left[g_{\mu\nu}-\left(1-a\right)\frac{p_\mu p_\nu}{p^2+i\eta}\right]\frac{e^{-ip \cdot \left(x-y\right)}}{p^2+i\eta}  \,.
\label{gluon_propagator}
\end{gather}

\subsection{Regularization and Renormalization}

Now that the quark and gluon fields have been quantized, the complete QCD Lagrangian is given by
\begin{gather}
\begin{split}
\mathcal{L}_{\rm QCD} &=  \bar{Q}\left[i\slashed{\partial} - m \right] Q 
- \frac{1}{4}\left[\partial_\mu A_\nu^b - \partial_\nu A_\mu^b\right]\left[\partial^\mu A^\nu_b - \partial^\nu A^\mu_b\right]
- \frac{1}{2a} \partial^\mu A_\mu^b \partial^\nu A_\nu^b
\\
&+\frac{g}{2} \bar{Q}\lambda^a \gamma^\mu A_\mu^a Q - \frac{g}{4}\left[\partial_\mu A_\nu^a - \partial_\nu A_\mu^a\right]f^{abc}A_b^\mu A_c^\nu
-\frac{g^2}{4}f^{abc}f^{ade} A^a_\mu A^b_\nu A_d^\mu A_e^\nu
\\
&-\partial^\mu \bar{c}^a \partial_\mu c^a + g f^{abc}A_\mu^c \partial^\mu \bar{c}^a c^b \,.
\label{bare_qcd_lagrangian}
\end{split}
\end{gather}
The terms in Eq.~\eqref{bare_qcd_lagrangian} can be interpreted as follows: the first term can be used to derive the quark propagator, the second and third terms can be used to derive the gluon propagator, the fourth term represents an interaction between quark and gluon fields, the fifth term represents an interaction between three gluon fields, the sixth term represents an interaction between four gluon fields, the seventh term can be used to derive the ghost propagator, while the eighth term represents an interaction between ghost and gluon fields. It is important to note that the gauge-fixing Lagrangian in Eq.~\eqref{gauge_fixing_lagrangian} is not gauge invariant. Although the QCD Lagrangian~\eqref{bare_qcd_lagrangian} is not gauge invariant, it is invariant under a generalized form of gauge symmetry known as BRST symmetry~\cite{Becchi_1976_a,Iofa_1976_a}. This can be used to prove the Slavnov-Taylor identities which relate various correlation functions in QCD~\cite{Slavnov_1975_a,Taylor_1971_a}.

\begin{figure}[htb]
\centering
\includegraphics[scale=0.75]{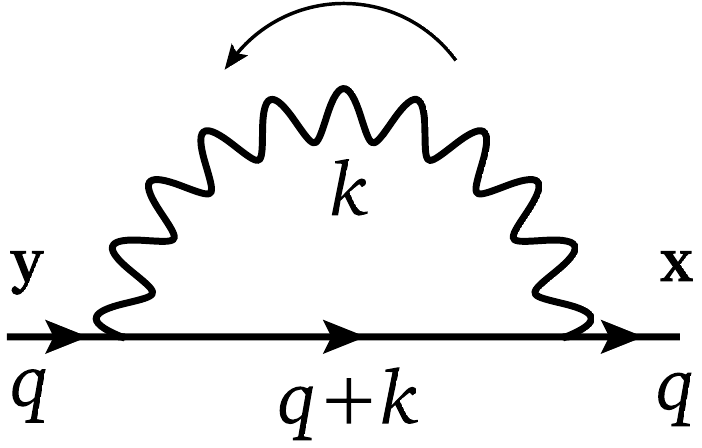}
\caption{Feynman diagram representing the quark self-energy. The quark propagates between the spacetime points $y$ and $x$, and has momentum $q$ at these locations. The quark interacts with a gluon that has momentum $k$, which flows from right to left in the diagram. Momentum is conserved at each vertex in the diagram.}
\label{quark_self_energy_diagram}
\end{figure}

Any QCD correlation function can be calculated to any order in $g$ using the perturbative expansion~\eqref{perturbative_expansion_of_correlation_function}, the interaction terms in the QCD Lagrangian~\eqref{bare_qcd_lagrangian}, as well as the quark, gluon and ghost propagators. In practice, this can be done via Wick's theorem or using the Feynman rules for QCD, which can be derived from the QCD Lagrangian~\eqref{bare_qcd_lagrangian}. Higher order terms in the expansion can be represented by Feynman diagrams that contain loops. For instance, consider the correlation function $\langle \Omega |T\left[\right.Q\left(x\right)\bar{Q}\left(y\right)\left.\right]| \Omega \rangle$, which is related to the amplitude for a quark to propagate between the spacetime points $y$ and $x$ in the presence of interactions. We will consider the next-to-leading order term in the perturbative expansion of this correlation function, which is $\mathcal{O}\left(g^2\right)$. This is called the quark self-energy and can be represented 
by the Feynman diagram shown in Fig.~\ref{quark_self_energy_diagram}. It is conventional to depict Feynman diagrams in momentum space, with the four-momentum of each propagator uniquely labeled. For brevity we will refer to four-momenta as momenta in what follows. 

The quark self-energy is represented by the Feynman diagram in Fig.~\ref{quark_self_energy_diagram} and is proportional to an integral over the momentum of the gluon. Schematically, the quark self-energy is given by
\begin{gather}
\Sigma\left(q\right) \simeq g^2 \int \frac{d^4 k}{(2\pi)^4} \frac{1}{k^2 \left[\right. \left(k+q\right)^2 - m^2 \left.\right]} \,.
\label{example_of_loop_integral}
\end{gather}
Momentum integrals such as this are called loop integrals, because they emerge naturally from Feynman diagrams that contain loops. It should be understood that we are integrating over the entire infinite range of each integration variable, that is, the integration in \eqref{example_of_loop_integral} is over the entire volume of the four-dimensional momentum space. In what follows we will suppress the limits of integration in loop integrals. For brevity we have also omitted the $i\eta$ pole prescription in the propagators in Eq.~\eqref{example_of_loop_integral}. In addition, we have ignored the leftmost and rightmost quark propagators with momentum $q$ in Fig.~\ref{quark_self_energy_diagram}. It is customary to remove (or amputate) external propagators in Feynman diagrams. The integral in Eq.~\eqref{example_of_loop_integral} can be evaluated in spherical coordinates~\cite{Peskin_1995_a}. However, the result is surprising: the integral diverges at large values of the gluon momentum $k$. Integrals that diverge 
in this way are called ultraviolet divergent. 

In order to extract meaningful physical information from the integral~\eqref{example_of_loop_integral}, the ultraviolet divergence must first be brought under control, or regulated. In order to do this, we will utilize dimensional regularization~\cite{tHooft_1972_a,Bollini_1972_a}. With this method, integrals in four dimensional Minkowski space are reinterpreted as integrals in $d$-dimensions. For example, the integral above is reinterpreted as 
\begin{gather}
\Sigma\left(q\right) \simeq \left(g^2\mu^{d-4}\right) \frac{1}{\mu^{d-4}} \int \frac{d^d k}{(2\pi)^d} \frac{1}{k^2 \left[\right. \left(k+q\right)^2 - m^2 \left.\right]} \,.
\label{example_of_dim_reg_int}
\end{gather}
In dimensional regularization the number of dimensions, $d$, is best thought of as a parameter that can be adjusted such that the integral~\eqref{example_of_dim_reg_int} converges. Integrals that are formally divergent in a certain number of dimensions can be uniquely defined through analytic continuation in the parameter $d$. In four dimensions, the coupling $g$ is dimensionless and hence is suitable to be used as an expansion parameter. However, in $d$-dimensions the combination $g^2\mu^{d-4}$ is dimensionless, where $\mu$ is the renormalization scale. This can be used to define the $d$-dimensional expansion parameter
\begin{gather}
\alpha = \frac{g^2 \mu^{d-4}}{4\pi} \,.
\label{defintion_of_alpha_in_d_dim}
\end{gather}
The remaining factor of the renormalization scale in the denominator of Eq.~\eqref{example_of_dim_reg_int} ensures that the $d$-dimensional integral has the same dimensions as the original four dimensional integral~\eqref{example_of_loop_integral}. The integral~\eqref{example_of_dim_reg_int} can be evaluated in $d$-dimensions using the methods described in Chapter~\ref{chapter_2_loops}. The result naturally depends on $d$, and we may examine the behaviour of the integral near four dimensions by setting $d=4+2\epsilon$ and expanding around $\epsilon=0$. The methods used to perform this expansion are discussed in Section~\ref{EpsilonExpansion}. For the integral~\eqref{example_of_dim_reg_int}, the result is
\begin{gather}
\Sigma\left(q\right) \simeq -\frac{i}{4} \left(\frac{\alpha}{\pi}\right) \left[ \frac{1}{\epsilon} - \log{(4\pi)} + \gamma_E -\log{\left[\frac{Q^2}{\mu^2}\right]} + f\left(\frac{Q^2}{m^2}\right)\right] \,,
\label{example_of_epsilon_expansion}
\end{gather}
where $\gamma_E$ is the Euler-Mascheroni constant (see Appendix~\ref{appendix_b_math}), $Q^2=-q^2$ is the Euclidean momentum and $f$ is a function of the dimensionless ratio $Q^2/m^2$. The divergence has been regulated and appears as a simple pole at $\epsilon=0$.

Theories in which divergences can be removed systematically order by order in perturbation theory are called renormalizable. The proof that QCD is renormalizable was given in Refs.~\cite{tHooft_1972_a,tHooft_1972_b}. Renormalization is the process of canceling these divergences. Formally, this can be achieved by rescaling the parameters of the QCD Lagrangian~\eqref{bare_qcd_lagrangian} as follows:

\begin{gather}
\begin{split}
\begin{array}{lll}
\left[ A^\mu_a \right]_B = Z_{\rm 3YM}^{1/2} \left[ A^\mu_a \right]_R	& 
\left[ Q \right]_B  = Z_{\rm 2F}^{1/2} \left[ Q \right]_R & 
\left[ c_a \right]_B =\tilde{Z}_3^{1/2} \left[ c_a \right]_R 
\\
\left[ g_{\rm 1YM} \right]_B = Z_{\rm 1YM} Z_{\rm 3YM}^{-3/2} \left[ g \right]_R &
\left[ \tilde{g} \right]_B = \tilde{Z}_1 \tilde{Z}_{3}^{-1} Z_{\rm 3YM}^{-1/2} \left[ g \right]_R & 
\left[ g_{\rm F} \right]_B = Z_{\rm 1F} Z_{\rm 3YM}^{-1/2} Z_{\rm 2F}^{-1} \left[ g \right]_R 
\\
\left[ g_5 \right]_B = Z_5^{1/2} Z_{\rm 3YM}^{-1} \left[ g \right]_R & 
\left[ m \right]_B = Z_4 Z_{\rm 2F}^{-1} \left[ m \right]_R & 
\left[ a \right]_B = Z_6^{-1} Z_{\rm 3YM} \left[ a \right]_R \,, 
\\
\end{array}
\end{split}
\label{renorm_coeffs}
\end{gather}
where we have used the notations of Ref.~\cite{Pascual_1984_a}. The constants $Z_i$ are called renormalization factors, and the subscripts $B$ and $R$ denote bare and renormalized quantities, respectively. The bare couplings $\left[ g_{\rm YM} \right]_B$, $\left[ \tilde{g} \right]_B$, $\left[ g_{\rm F} \right]_B$, and $\left[ g_5 \right]_B$ are associated with the three-gluon, ghost-gluon, quark-gluon and four-gluon interaction terms in the bare QCD Lagrangian~\eqref{bare_qcd_lagrangian}. However, all of these couplings must be identical in order for the QCD Lagrangian to be BRST invariant. This implies the that renormalization factors satisfy
\begin{equation}
\frac{Z_{\rm 3YM}}{Z_{\rm 1YM}}=\frac{\tilde{Z}_3}{\tilde{Z}_1} \,, \quad  \frac{Z_{\rm 3YM}}{Z_{\rm 1YM}}=\frac{Z_{\rm 2F}}{Z_{\rm 1F}}
\,, \quad  Z_5 =  \frac{Z_{\rm 1YM}^2}{Z_{\rm 3YM}} \,.
\label{slavnov} 
\end{equation}
These identities are closely related to the Slavnov-Taylor identities that are needed in order to prove that QCD is renormalizable.

The Lagrangian given above in Eq.~\eqref{bare_qcd_lagrangian} is implicitly in terms of bare parameters, hence it is called the bare Lagrangian. The renormalized Lagrangian has the same form as the bare Lagrangian, and can be obtained using the relations given in Eq.~\eqref{renorm_coeffs}. Because QCD is a renormalizable theory, all correlation functions calculated with the renormalized Lagrangian must be free of divergences. The renormalization factors are of the form
\begin{gather}
Z = 1 + \frac{\alpha}{\pi} \frac{A}{\epsilon} + \mathcal{O}\left(\alpha^2\right) \,,
\label{example_renormalization_factor}
\end{gather}
where $A$ is a constant and dimensional regularization is used with $d=4+2\epsilon$. In practice, a correlation function can be calculated using the bare Lagrangian~\eqref{bare_qcd_lagrangian}, and the result in terms of the bare parameters will contain divergences. The bare parameters can be rewritten in terms of the renormalization factors and renormalized parameters using~\eqref{renorm_coeffs}. When this is done, the divergences in the bare correlation function are canceled by compensating divergences in the renormalization factors. The resulting renormalized correlation function is purely in terms of the renormalized parameters and is free of divergences. This approach is called bare perturbation theory, because it involves calculating correlation functions in terms of bare parameters which are then renormalized. A complementary approach is renormalized perturbation theory, where correlation functions are calculated in terms of renormalized parameters. This procedure involves inverting the 
relations~\eqref{renorm_coeffs}, leading to the introduction of counterterms. Renormalized perturbation theory is described in Ref.~\cite{Peskin_1995_a}.

The renormalization process is somewhat arbitrary because there are many ways that it can be implemented. The minimal subtraction $({\rm MS})$ scheme is often used in conjunction with dimensional regularization. In the ${\rm MS}$ scheme the renormalization factors $Z_i$ are defined such that only the poles at $\epsilon=0$ are canceled, while the $\gamma_E$ and $\log{(4\pi)}$ terms appearing in Eq.~\eqref{example_of_epsilon_expansion} remain. However, these terms are merely artifacts of dimensional regularization and are unphysical. In the modified minimal subtraction ($\overline{\rm MS}$) scheme the renormalization constants are defined so that these terms are also canceled. A convenient method of partially implementing the $\overline{\rm MS}$ renormalization scheme is discussed in Chapter~\ref{chapter_2_loops}.

Renormalization factors can be calculated as an expansion in the coupling $\alpha$. This can be done to one-loop order as follows. First, all possible one-loop connected correlation functions are calculated in terms of bare parameters. Second, the bare parameters are eliminated in favor of the renormalized parameters using Eq.~\eqref{renorm_coeffs}. Finally, the requirement that the renormalized correlation functions must be finite can be used to determine the renormalization factors. As an example, consider the quark self-energy represented by Fig.~\ref{quark_self_energy_diagram}. The renormalized and bare quark self-energy are related by
\begin{gather}
\Sigma_R\left(q\,;m_R\,,a_R\,,\alpha_R\right) = \lim_{\epsilon\to0}\left[Z^{-1}_{\rm 2F} \Sigma_B\left(q\,;m_B\,,a_B\,,\alpha_B\right) \right] \,,
\label{renormalized_quark_self_energy}
\end{gather}
where $\Sigma_B$ denotes the bare quark self-energy~\eqref{example_of_epsilon_expansion} which is in terms of bare parameters. Using the relationships between the bare and renormalized parameters given in Eq.~\eqref{renorm_coeffs}, the bare parameters can be expressed in terms of renormalized parameters along with the corresponding renormalization factors. The renormalization factors are defined such that the limit in Eq.~\eqref{renormalized_quark_self_energy} can be taken order by order in $\alpha$. Ref.~\cite{Pascual_1984_a} provides expressions for the renormalization factors in Eq.~\eqref{renorm_coeffs} to two-loop order. 

So far, we have only considered correlation functions that involve multiple quark or gluon fields at distinct spacetime locations. However, in QSR calculations correlation functions involving composite local operators are needed. For instance, the following correlation function can be used to study a heavy-light pseudoscalar meson:
\begin{gather}
\Pi\left(Q^2\right) = i\int d^4 x \, e^{iq\cdot x} \langle \Omega | \, T\left[ J\left(x\right) J^\dag\left(0\right) \right] | \Omega \rangle \,, \quad Q^2 = -q^2 \,.
\label{example_of_composite_operator_correlation_function}
\end{gather}
The current $J\left(x\right)=i\bar{q}\left(x\right) \gamma_5 Q\left(x\right)$, where $q$ and $Q$ respectively denote light and heavy quark fields, is a composite local operator that couples to the heavy-light pseudoscalar mesons~\cite{Jamin_2001_a}. In order to extend QSR calculations to higher orders, we must consider the renormalization of correlation functions that involve composite operators. 

The renormalization of composite operators is complicated by the fact that multiple composite operators may share the same quantum numbers. Because the fields in the QCD Lagrangian~\eqref{bare_qcd_lagrangian} have distinct quantum numbers, they must renormalize separately. However, this is not the case with composite operators: those with the same quantum numbers can mix under renormalization. The renormalization of composite operators is discussed in detail in Ref.~\cite{Collins_1984_a}. In general, in order to study the renormalization of an operator $\mathcal{O}_a$ with dimension $a$, one must also consider operators $\mathcal{O}_b$ with the same quantum numbers and dimension $b<a$ (the dimensions of quark and gluon fields are given in Appendix~\ref{appendix_a_conventions}). The renormalization factors of these operators are formally defined as
\begin{gather}
\left[\mathcal{O}_{i} \right]_R = Z_{ij} \left[\mathcal{O}_{j} \right]_B
\label{operator_mixing_renormalization_matrix}
\end{gather}
where the vector $\mathcal{O}_{i} = \left( \mathcal{O}_a \,, \Lambda^{a-b} \mathcal{O}_b \,, \, \ldots \, \right)$ and the parameters $\Lambda$ ensure that all elements of the vector have the same dimension. The matrix $Z_{ij}$ is an upper diagonal matrix containing the renormalization factors. The renormalization factors $Z_{ij}$ can be determined by calculating correlation functions composed of the operators $\mathcal{O}_{i}$. 

In Chapter~\ref{chapter_6_mixing} mixing between scalar $\left(J^{PC}=0^{++}\right)$ glueballs and quark mesons is studied using the currents
\begin{gather}
J_g = \alpha G^2 \,, \quad G^2=G^a_{\mu\nu} G_a^{\mu\nu} \,, \quad J_q = m_q \left(\bar{u}u + \bar{d}d\right) \,.
\label{scalar_glueball_and_quark_meson_currents}
\end{gather}
The scalar glueball operator $J_g$ mixes under renormalization with the scalar quark meson operator $J_q$, which has the same dimension and quantum numbers. In Refs.~\cite{Pascual_1984_a,Narison_2007_a} the renormalization of the scalar glueball operator is studied using background field techniques. The resulting renormalized scalar glueball operator is given by
\begin{gather}
G^2_R = \left[1+\frac{1}{\epsilon}\frac{\alpha}{\pi}\left(\frac{11}{4}-\frac{n_f}{6}\right)\right] G_B^2 - \frac{4}{\epsilon}\frac{\alpha}{\pi} \left[m_u \bar{u}u + m_d \bar{d}d\right]_B \,, 
\label{renormalized_gluonic_current}
\end{gather}
where $n_f$ is the number of active quark flavours. The mixing between the operators $J_g$ and $J_q$ under renormalization is signaled by the second term in Eq.~\eqref{renormalized_gluonic_current}. This second term leads to a crucial renormalization-induced contribution in the mixing analysis in Chapter~\ref{chapter_6_mixing}. 

A somewhat simpler example of composite operator renormalization is considered in Chapter~\ref{chapter_5_diquark_renorm}. There the renormalization of the scalar diquark current is considered, which is given by
\begin{gather}
J^d_\alpha\left(x\right) = \epsilon_{\alpha\beta\gamma} Q_{\beta}\left(x\right) C\gamma_5 q_\gamma\left(x\right) \,,
\label{scalar_diquark_current}
\end{gather}
where $\alpha\,,\,\beta\,,\,\gamma$ are colour indices, $C$ is the charge conjugation operator and $\gamma_5$ is a Dirac matrix (both are defined in Appendix~\ref{appendix_a_conventions}). There are no composite operators of lower dimension with the same quantum numbers as the scalar diquark current, hence it cannot mix under renormalization with any other operators. This greatly simplifies the task of determining the renormalization factor of the scalar diquark current. Composite operators typically require an additional renormalization beyond that of their component fields and parameters. In Chapter~\ref{chapter_5_diquark_renorm} the scalar diquark operator renormalization factor is determined to two-loop order by considering the correlation function
\begin{gather}
\Gamma^d = \langle \Omega | \, T\left[ Q\left(x\right) J^d\left(0\right) q\left(y\right) \right] | \Omega \rangle \,,
\label{scalar_diquark_renorm_correlation_function}
\end{gather}
where $J^d$ is the scalar diquark current~\eqref{scalar_diquark_current} and colour indices have been omitted for brevity. Conventionally the correlation in Eq.~\eqref{scalar_diquark_renorm_correlation_function} is calculated in momentum space and the external quark propagators are amputated, in an identical fashion to the quark self-energy~\eqref{example_of_epsilon_expansion} However, in the case of Eq.~\eqref{scalar_diquark_renorm_correlation_function} the scalar diquark operator is inserted with zero momentum. This is justified because renormalization factors are momentum independent. Then the renormalized correlation function is related to the bare correlation function by
\begin{gather}
\Gamma^d_R\left(q\,;m_R\,,a_R\,,\alpha_R\right) = \lim_{\epsilon\to0}\left[Z_{\rm d} \, Z^{-1}_{\rm 2F}  \Gamma^d_B\left(q\,;m_B\,,a_B\,,\alpha_B\right) \right] \,.
\label{scalar_diquark_renorm_bare_and_renormalized_correlation_functions}
\end{gather}
Notice that this expression is identical to Eq.~\eqref{renormalized_quark_self_energy}, apart from the factor of $Z_{\rm d}$. This extra factor is the additional renormalization that is required in order to evaluate the limit in Eq.~\eqref{scalar_diquark_renorm_bare_and_renormalized_correlation_functions}. This extra factor is precisely the scalar diquark current renormalization factor. In Chapter~\ref{chapter_5_diquark_renorm} the scalar diquark operator renormalization factor is calculated to two-loop order using Eq.~\eqref{scalar_diquark_renorm_bare_and_renormalized_correlation_functions}.

Renormalized correlation functions explicitly depend on the renormalization scale $\mu$. However, bare correlation functions which are calculated prior to renormalization do not. For instance, consider an amputated bare correlation function with $n=n_{\rm YM}+\tilde{n}+n_{\rm F}$ external gluon, ghost and quark propagators. Then the bare correlation function must satisfy
\begin{gather}
\mu \frac{d}{d\mu} \Gamma_B\left(q_1\,,q_2\,,\ldots\,, q_n\,; \alpha_B \,, a_B\,, m^i_B \,; \epsilon \right) = 0 \,,
\label{bare_correlation_function_mu_dependence} 
\end{gather}
where $q_i$ denote the momenta of each external propagator and $m_i$ is to distinguish distinct quark flavours. The bare and renormalized correlation functions are related by
\begin{gather}
\Gamma_R\left(q_1\,,q_2\,,\ldots\,, q_n\,; \alpha_R\,,a_R\,, m^i_R \,; \mu \right) = \lim_{\epsilon\to0} \left[ Z_\Gamma\left(\mu\,,\epsilon\right) \Gamma_B\left(q_1\,,q_2\,,\ldots\,, q_n\,; \alpha_B \,, a_B\,, m^i_B \,; \epsilon \right) \right] \,,
\\
Z_\Gamma\left(\mu\,,\epsilon\right) = Z_{\rm 3YM}^{-n_{\rm YM}/2}\left(\mu\,,\epsilon\right) \tilde{Z}_{\rm 3}^{-\tilde{n}/2}\left(\mu\,,\epsilon\right) Z_{\rm 2F}^{-n_{\rm F}/2}\left(\mu\,,\epsilon\right) \,.
\label{bare_and_renorm_correlators} 
\end{gather}
Using Eqs.~\eqref{bare_correlation_function_mu_dependence} and~\eqref{bare_and_renorm_correlators} it can be shown that the renormalized correlation function must satisfy the differential equation
\begin{gather}
\left[ \mu \frac{\partial}{\partial\mu} + \beta\left(\alpha\right)\alpha \frac{\partial}{\partial\alpha} + \delta\left(\alpha\right) a \frac{\partial}{\partial a} - \gamma_i\left(\alpha\right)x_i \frac{\partial}{\partial x_i} - \gamma_\Gamma\left(\alpha\right)  \right] \Gamma_R\left(\alpha\,,a\,, m_i \,; \mu \right) = 0 \,,
\\
x_i = \frac{m_i}{\mu} \,, \quad \gamma_\Gamma\left(\alpha\right) = -\frac{1}{2} \left[ n_{\rm YM} \gamma_{\rm YM}\left(\alpha\right) + n_{\rm F} \gamma_{\rm F}\left(\alpha\right) +\tilde{n} \tilde{\gamma}\left(\alpha\right)  \right] \,,
\label{renormalization_group_equation} 
\end{gather}
where all parameters should be interpreted as renormalized parameters and we have omitted the momentum dependence of the renormalized correlation function. The parameter $x_i$ is implicitly summed over all quark flavours. The differential equation above is called the renormalization group equation. The renormalization group functions in Eq.~\eqref{renormalization_group_equation} are defined as
\begin{gather}
\mu \frac{d\alpha}{d\mu} = \alpha \beta\left(\alpha\,,a\,,x_i\right)  \,, \quad \frac{\mu}{m_i} \frac{dm_i}{d\mu} = -\gamma_i\left(\alpha\,,a\,,x_i\right) \,, 
\quad \mu \frac{da}{d\mu} = a \delta\left(\alpha\,,a\,,x_i\right) \,,
\\
\frac{\mu}{Z_{\rm 3YM}} \frac{d Z_{\rm 3YM}}{d\mu} = \gamma_{\rm YM}\left(\alpha\,,a\,,x_i\right) \,, \quad \frac{\mu}{Z_{\rm 2F}} \frac{d Z_{\rm 2F}}{d\mu} = \gamma_{\rm F}\left(\alpha\,,a\,,x_i\right) \,,
\quad \frac{\mu}{\tilde{Z}_3} \frac{d\tilde{Z}_3}{d\mu} = \tilde{\gamma}\left(\alpha\,,a\,,x_i\right) \,.
\label{renormalization_group_functions}
\end{gather}
Note that the $a$ and $x_i$ dependence of these functions was suppressed in Eq.~\eqref{renormalization_group_equation}. In the ${\rm MS}$ and $\overline{\rm MS}$ renormalization schemes all renormalization group functions are independent of the mass parameter $x_i$, and the $\beta$ function is also independent of that gauge parameter $a$~\cite{Pascual_1984_a}.

An important consequence of the renormalization process is that parameters of the renormalized QCD Lagrangian depend on the renormalization scale, and hence are called running parameters. The renormalization group equation can be used to determine how the running parameters vary with the renormalization scale. First, however, the renormalization group functions must be calculated. For instance, the $\beta$ function is calculated to $\mathcal{O}\left(\alpha^3\right)$ in Ref.~\cite{Pascual_1984_a}. We will now outline the calculation of the leading order term in the expansion. First, the Slavnov-Taylor identities~\eqref{slavnov} allow us to write
\begin{gather}
\alpha = \frac{\left(g\mu\right)^{2\epsilon}}{4\pi} \,, \quad \alpha_R = Z_\alpha^{-1} \alpha_B \,, \quad  Z_\alpha = \tilde{Z}_1^2 \tilde{Z}_3^{-2} Z_{\rm 3YM}^{-1} \,.
\end{gather}
The renormalization factor $Z_\alpha$ can be determined using the methods described previously. To one-loop order, 
\begin{gather}
Z_\alpha = 1 + \frac{\alpha}{\pi}\left[\frac{11}{4}-\frac{n_f}{6}\right]\frac{1}{\epsilon} \,,
\label{Z_alpha}
\end{gather}
where $n_f$ denotes the number of quark flavours. In Ref.~\cite{Pascual_1984_a} it is shown that to lowest order in $\alpha$,
\begin{gather}
\beta\left(\alpha\right) = -2\alpha \frac{\partial Z^{(1)}_\alpha}{\partial \alpha} \,, \quad \beta\left(\alpha\right) = \frac{\alpha}{\pi} \beta_1 \,, \quad \beta_1 = -\frac{11}{2}+\frac{n_f}{3} \,,
\label{QCD_beta_function}
\end{gather}
where $Z^{(1)}_\alpha$ denotes the divergent term in Eq.~\eqref{Z_alpha}. In a similar fashion it can be shown that to lowest order
\begin{gather}
\gamma \left(\alpha\right) = \frac{\alpha}{\pi} \gamma_1 \,, \quad \gamma_1 = 2 \,.
\label{QCD_gamma_function}
\end{gather}
The differential equations defining the renormalization group functions in Eq~\eqref{renormalization_group_functions} can be solved to determine how the QCD Lagrangian parameters depend on the renormalization scale. For instance, using the one-loop expression for the $\beta$ function~\eqref{QCD_beta_function}, we find
\begin{gather}
\mu \frac{d\alpha}{d\mu} = \alpha^2 \frac{\beta_1}{\pi} \quad \to \quad \alpha \left(\mu \right) = \frac{\alpha\left(M\right)}{1-\frac{\beta_1 \alpha\left(M\right)}{2\pi} \log{\left[\frac{\mu^2}{M^2}\right]}} \,,
\label{one_loop_running_coupling} 
\end{gather}
The value of $\beta_1$ depends on the number of active quark flavours $n_f$, as can be seen from Eq.~\eqref{QCD_beta_function}. In QSR calculations $n_f$ is chosen to encompass the heaviest quark in the hadron being studied. For instance, if the heaviest quark is the charm quark $n_f=4$, whereas if it is the bottom quark $n_f=5$. This is justified by the decoupling theorem, which states that contributions from quarks that are much heavier than the characteristic scale of the problem are suppressed by the heavy quark mass~\cite{Appelquist_1975_a}. In the $\overline{\rm MS}$ scheme the value of the coupling at the reference scale is taken to be
\begin{gather}
\begin{array}{lll}
 n_f=4\,: & \alpha\left(M\right) = \alpha\left(M_\tau\right) = 0.33\pm0.01 \,, & M_\tau = 1.77\,{\rm GeV} \,, \\
 n_f=5\,: & \alpha\left(M\right) = \alpha\left(M_Z\right) = 0.1184\pm0.0007 \,, & M_Z = 91.118\,{\rm GeV} \,,
\end{array}
\label{coupling_at_reference_scales}
\end{gather}
where all numerical values have been taken from Ref.~\cite{Beringer_2012_a}. Similarly, using the one-loop expression for the $\gamma$ function~\eqref{QCD_gamma_function}, it can be shown that
\begin{gather}
\frac{\mu}{m} \frac{dm}{d\mu} =  - \alpha \frac{\gamma_1}{\pi}  \quad \to \quad m\left(\mu\right) = \overline{m} \left[\frac{\alpha\left(\mu\right)}{\alpha\left(\overline{m}\right)}\right]^{-\frac{\gamma_1}{\beta_1}} \,, \quad \overline{m} = m\left(\mu=m\right) \,.
\label{one_loop_running_mass}
\end{gather}
 In the $\overline{\rm MS}$ scheme the value of the quark mass at the reference scale is taken to be
\begin{gather}
\begin{array}{ll}
 n_f=4 \,: & \overline{m}=\overline{m}_c = m\left(\mu=m_c\right) = 1.28 \pm 0.03 \,{\rm GeV} \,, \\
 n_f=5 \,: & \overline{m}=\overline{m}_b = m\left(\mu=m_b\right) = 4.18 \pm 0.03 \,{\rm GeV} \,,
\end{array}
\label{mass_at_reference_scales}
\end{gather}
where the numerical values are taken again from Ref.~\cite{Beringer_2012_a} and $\alpha\left(\overline{m}\right)$ can be determined using Eq.~\eqref{one_loop_running_coupling}.

\begin{figure}[htb]
\centering
\includegraphics[scale=1]{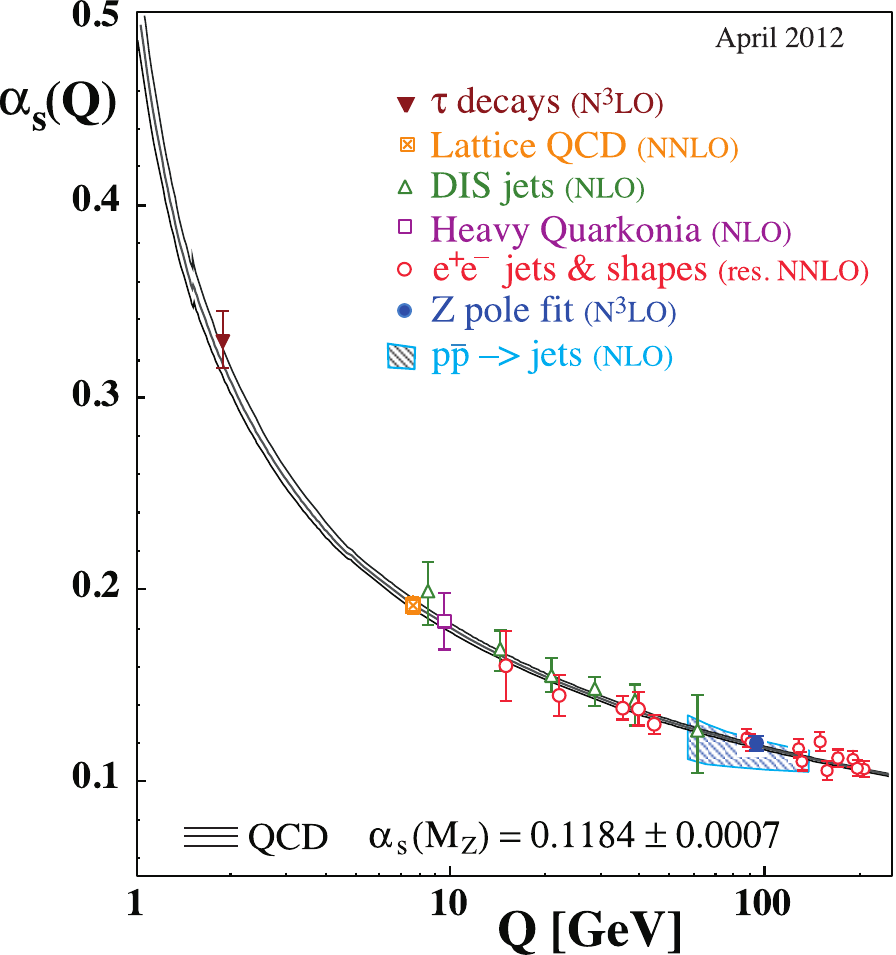}
\caption{Experimental and theoretical predictions for the running QCD coupling $\alpha\left(\mu\right)$. Figure taken from Ref.~\cite{Beringer_2012_a}.}
\label{plot_of_qcd_coupling}
\end{figure}

The $\beta$ function signals an essential feature of QCD. Because $\beta_1 < 0$, the one-loop QCD coupling decreases with increasing energy scale. This defining characteristic of QCD is called asymptotic freedom. Fig.~\ref{plot_of_qcd_coupling} compares theoretical predictions and experimental measurements of $\alpha\left(\mu\right)$ at several different energy scales $\mu$. The predicted and measured values are in excellent agreement. Because asymptotic freedom is a prediction of QCD, this agreement is a strong experimental confirmation of QCD~\cite{Bethke_2009_a}.

\section{QCD Laplace sum rules}
\label{QCDLaplaceSumRules}

The QCD coupling is small at high energies due to asymptotic freedom. This means that perturbative expansions in QCD converge rapidly at high energies. However, at low energies the coupling increases, and the convergence of the perturbative expansion suffers. In practical terms this means that perturbative techniques alone are insufficient to describe QCD at low energies.

There are two broad classes of theoretical techniques that are used to study hadrons: those that are inspired by QCD and those that are based in QCD. The key distinction between these two is that the latter utilize the QCD Lagrangian~\eqref{bare_qcd_lagrangian} while the former do not. Most QCD-inspired techniques are based on effective field theory methods, such as chiral perturbation theory~\cite{Ecker_1994_a} or heavy quark effective theory~\cite{Neubert_1993_a}. Additional QCD-inspired methods include potential models~\cite{Kwong_1987_a} and techniques based on the AdS/CFT correspondence in string theory~\cite{Kim_2012_a}. Methods that are based in QCD typically augment perturbation theory in some way or avoid it entirely. In lattice QCD the path integral is calculated numerically in a discretized Euclidean space~\cite{Kronfeld_2012_a}. The Dyson-Schwinger equations are an infinite set of coupled integral equations relating various correlation functions in the interacting theory. When truncated, the 
equations can be solved and used to determine hadronic parameters~\cite{Maris_2003_a}. Another QCD-based approach is QCD sum rules (QSR).

The QSR method is based upon the concept of quark-hadron duality and on the operator product expansion (OPE). Refs.~\cite{Shifman_1978_a,Shifman_1978_b} are the original papers outlining the QSR technique and reviews of its methodology are given in Refs.~\cite{Reinders_1984_a,Colangelo_2000_a,Narison_2007_a}. QSR depends critically on the concept of quark-hadron duality, which asserts that hadrons can be described equally well in terms resonances or in terms of bound states composed of quarks and gluons. This duality is realized globally rather than locally, in the sense that the two descriptions agree when suitably averaged. Calculations on the QCD side of the duality relation can be performed using the OPE, which naturally includes both perturbative and non-perturbative effects. The hadron side of the duality can be invoked using an experimentally known hadronic spectral function, or a suitable resonance model. Ultimately there are two main applications of QSR that utilize this duality in opposite 
directions. The first uses experimentally known hadronic parameters to determine unknown QCD parameters, such as quark masses (see {\it e.g.}\ Ref.~\cite{Narison_2011_a}). The second involves determining unknown hadronic parameters in terms of known QCD parameters, using an appropriate model for the hadronic spectral function. The research presented in Chapters \ref{chapter_3_hybrids}, \ref{chapter_4_Qq_diquark} and \ref{chapter_6_mixing} uses the second approach to predict the properties of exotic hadrons.

\subsection{Dispersion Relation}

All QSR calculations begin with a QCD correlation function of the form
\begin{gather}
\Pi\left(Q^2\right) = i \int d^4 x \, e^{iq\cdot x} \, \langle \Omega | \left[\right. J\left(x\right) J^\dag\left(0\right) \left. \right] | \Omega \rangle \,, \quad Q^2 = -q^2 \,,
\label{qsr_correlation_function} 
\end{gather}
where the current $J$ is a composite operator that couples to the hadron being studied. Techniques for calculating the correlation function will be discussed in Section~\ref{OperatorProductExpansion}. The analytic properties of the correlation function~\eqref{qsr_correlation_function} can be used to show that $\Pi\left(Q^2\right)$ and its imaginary part ${\rm Im}\Pi\left(Q^2\right)$ are related by a dispersion relation. In turn, ${\rm Im}\Pi\left(Q^2\right)$ is related to a hadronic spectral function. Quark-hadron duality is therefore encoded through this dispersion relation.

\begin{figure}[htb]
\centering
\includegraphics[scale=0.5]{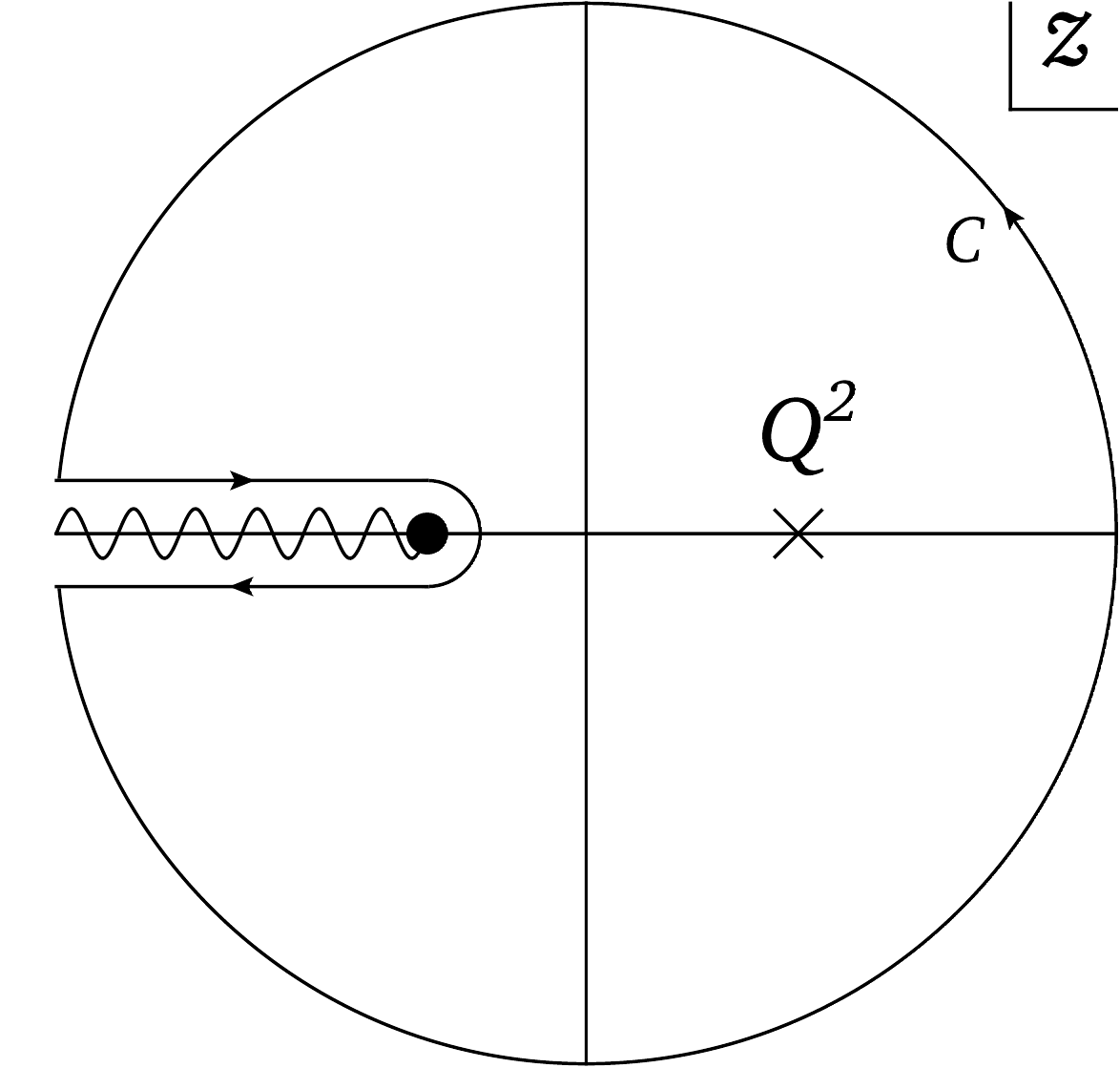}
\caption{Integration contour used to derive the dispersion relation between the correlation function and its imaginary part. The correlation function has a branch cut on the interval $z\in\left(-\infty\,,-t_0\right]$, where $t_0$ is the hadronic threshold.}
\label{dispersion_relation_contour}
\end{figure}

We will now demonstrate how the dispersion relation can be derived by appealing to the analytic properties of the correlation function. To do so, we will calculate the contour integral
\begin{gather}
I = \oint_C \frac{dz}{2\pi i} \frac{\Pi\left(z\right)}{z^n\left(z-Q^2\right)} \,,
\label{dispersion_relation_contour_integral} 
\end{gather}
which is depicted in Fig.~\ref{dispersion_relation_contour}. The branch cut singularity is due to the correlation function and the integrand has poles at $z=0$ and $z=Q^2$ as depicted in the figure. The value of $n$ is chosen to ensure that the contribution of the radial contour vanishes as its radius is taken to infinity. As an example, we will derive the dispersion relation used in Chapter~\ref{chapter_4_Qq_diquark}, where the correlation function satisfies
\begin{gather}
\lim_{z\to\infty} \Pi\left(z\right) \sim z \log^2{\left(z\right)} \,.
\label{Qq_diquark_correlator_asymptotic_behaviour} 
\end{gather}
The dispersion relation can be derived by evaluating the contour integral~\eqref{dispersion_relation_contour_integral} in two ways and equating the results. First, we will evaluate the contribution from each portion of the contour $C$. The radial portion of the contour can be bounded using~\eqref{Qq_diquark_correlator_asymptotic_behaviour}:
\begin{gather}
|I_R| \leq \lim_{R\to\infty} \frac{\Pi\left(R\right)}{R^n} = \lim_{R\to\infty} \frac{R \log^2{\left(R\right)}}{R^n} = \lim_{R\to\infty} \frac{\log^2{\left(R\right)}}{R} = 0 \,,
\label{dispersion_relation_radial_contour} 
\end{gather}
where we have set $n=2$ to ensure that the contribution of the radial contour is zero. In order to determine the contribution of the portion of $C$ that circles the branch point, we must know the behaviour of the correlation function near the hadronic threshold $t_0$. The correlation function in Chapter~\ref{chapter_4_Qq_diquark} is regular at this point, therefore the contribution of the portion of the contour that circles the branch point is zero. The only remaining portions of the contour $C$ are those that are above and below the branch cut in Fig.~\ref{dispersion_relation_contour}. For these, we find
\begin{gather}
I = \int_{t_0}^\infty \frac{dt}{2\pi i} \frac{ \Pi\left(te^{-i\pi}\right) - \Pi\left(te^{i\pi}\right) }{ t^2\left(t+Q^2\right) } \,,
\label{dispersion_relation_discontinuity}
\end{gather}
where $\Pi\left(te^{-i\pi}\right)$ and $\Pi\left(te^{i\pi}\right)$ are the values of the correlation function at points below and above the branch cut, respectively. Therefore Eq.~\eqref{dispersion_relation_discontinuity} effectively requires the discontinuity of the correlation function across the branch cut. However, the correlation function satisfies Schwarz reflection~\cite{Polya_1974_a}, which implies that 
\begin{gather}
\left[\Pi \left(z^*\right)\right]^* = \Pi\left(z\right) \quad \to \quad \Pi\left(te^{-i\pi}\right) - \Pi\left(te^{i\pi}\right) = 2i\, {\rm Im}\Pi\left(te^{-i\pi}\right) \,,
\label{dispersion_relation_schwarz_reflection}
\end{gather}
where $z^*$ denotes the complex conjugate of $z$ and ${\rm Im}\Pi\left(te^{-i\pi}\right)$ is the imaginary part of the correlation function evaluated at a point below the branch cut. The imaginary part of the correlation function is equivalent to the hadronic spectral function $\rho^{\rm had}\left(t\right)$ (see Ref.~\cite{Narison_2007_a} for a proof of this). Using this and substituting Eq.~\eqref{dispersion_relation_schwarz_reflection} into Eq.~\eqref{dispersion_relation_discontinuity} yields
\begin{gather}
I = \frac{1}{\pi} \int_{t_0}^\infty dt \, \frac{\rho^{\rm had}\left(t\right)}{t^2 \left(t+Q^2\right)} \,.
\label{dispersion_relation_result_via_contours} 
\end{gather}
The contour integral~\eqref{dispersion_relation_contour_integral} can also be evaluated using the residue theorem, with the result
\begin{gather}
I = \frac{1}{Q^4} \left[ \Pi\left(Q^2\right) - \Pi\left(0\right) - Q^2 \Pi^\prime\left(0\right) \right] \,, \quad \Pi^\prime\left(0\right) = \left. \frac{d}{dQ^2} \Pi\left(Q^2\right) \right|_{Q^2=0} \,.
\label{dispersion_relation_result_via_residue_theorem}  
\end{gather}
Equating the results for the contour integral given in Eq.~\eqref{dispersion_relation_result_via_contours} and Eq.~\eqref{dispersion_relation_result_via_residue_theorem}, the following dispersion relation results:
\begin{gather}
\Pi\left(Q^2\right) = \Pi\left(0\right) + Q^2 \Pi^\prime\left(0\right) + \frac{Q^4}{\pi} \int_{t_0}^\infty dt \, \frac{\rho^{\rm had}\left(t\right)}{t^2 \left(t+Q^2\right)} \,.
\label{Qq_diquark_dispersion_relation}
\end{gather}

\subsection{Borel Transform}
\label{BorelTransform}

The dispersion relation~\eqref{Qq_diquark_dispersion_relation} relates the correlation function $\Pi\left(Q^2\right)$ that can be calculated in QCD to the hadronic spectral function $\rho^{\rm had}\left(t\right)$ which can be parametrized in terms of the hadronic parameters. In principle this can be used to calculate hadronic parameters, such as masses, in terms of QCD parameters. However, in practice this approach fails. In general we are interested in the ground state hadron in a certain $J^{PC}$ channel. The spectral function will include this state, along with excited states and the continuum. Hence it is difficult to isolate the ground state contribution when such a dispersion relation is used. In addition, the correlation function $\Pi\left(Q^2\right)$ often contains field theoretical divergences and its value at $Q^2=0$ is usually unknown.

The critical insight of Refs.~\cite{Shifman_1978_a,Shifman_1978_b} is that these difficulties can be overcome by applying the Borel transform to the dispersion relation~\eqref{Qq_diquark_dispersion_relation}, which is defined as
\begin{gather}
\hat B\equiv 
\lim_{\stackrel{N,~Q^2\rightarrow \infty}{N/Q^2\equiv \tau}}
\frac{\left(-Q^2\right)^N}{\Gamma(N)}\left(\frac{\mathrm{d}}{\mathrm{d}Q^2}\right)^N\quad \,.
\label{borel_transform}
\end{gather}
The Borel transform has the following properties:
\begin{gather}
\hat B \left[ Q^{2n} \right] = 0 \,, \quad \hat B \left[ \frac{Q^{2n}}{t+Q^2} \right] = \tau \left(-1\right)^n e^{-t\tau} \,,
\label{properties_of_borel_transform}
\end{gather}
where $n>0$. In Ref.~\cite{Bertlmann_1984_a} it was shown that the Borel transform is related to the inverse Laplace transform via
\begin{gather}
\frac{\hat B}{\tau} \left[ f\left(Q^2\right) \right] = \mathcal{L}^{-1}\left[f\left(Q^2\right) \,; \tau \right] = \frac{1}{2\pi i} \int\limits_{b-i\infty}^{b+i\infty} dQ^2 f\left(Q^2\right) e^{Q^2 \tau} \,,
\label{borel_and_inverse_laplace_transform} 
\end{gather}
where $b$ is defined such that $f\left(Q^2\right)$ is analytic to the right of the integration contour. Multiplying both sides of Eq.~\eqref{Qq_diquark_dispersion_relation} by $\left(-Q^2\right)^k$ and taking the Borel transform using Eq.~\eqref{properties_of_borel_transform}, the dispersion relation becomes
\begin{gather}
\frac{\hat B}{\tau} \left[ \left(-Q^2\right)^k \Pi\left(Q^2\right) \right] = \frac{1}{\pi} \int_{t_0}^\infty dt \, t^k \, e^{-t\tau} \, \rho^{\rm had}\left(t\right)  \,.
\label{borel_transformed_dispersion_relation} 
\end{gather}
Note that the Borel transform has removed the $\Pi\left(0\right)$ and $\Pi^\prime\left(0\right)$ terms. In addition, any terms in the explicit field-theoretic expression for $\Pi\left(Q^2\right)$ that are polynomials in $Q^2$ will be removed by the Borel transform. Note that this includes any divergences of the form $\epsilon^{-n} f\left(Q^2\right)$ where $f\left(Q^2\right)$ is a polynomial in $Q^2$. However, such terms will not be eliminated by the Borel transform when the function $f\left(Q^2\right)$ is not a polynomial in $Q^2$. These are called non-local divergences and must be dealt with through renormalization. Furthermore, the Borel transform has introduced an exponential factor which serves to suppress excited state contributions to the hadronic spectral function. In order to isolate the ground state contribution, it is conventional to parametrize the hadronic spectral function in terms of a resonance and continuum:
\begin{gather}
\rho^{\rm had}\left(t\right) = \rho^{\rm res}\left(t\right) + \theta\left(t-s_0\right)\rho^{\rm cont}\left(t\right) \,, \quad \rho^{\rm cont}\left(t\right) = {\rm Im}\Pi\left(te^{-i\pi}\right) \,,
\label{resonance_and_continuum_model_of_hadronic_spectral_function} 
\end{gather}
where $\theta\left(t-s_0\right)$ is the Heaviside step function and $s_0$ is the continuum threshold $(s_0>t_0)$. The continuum contribution is related to the imaginary part of the QCD correlation function through the optical theorem~\cite{Peskin_1995_a}. Inserting this into Eq.~\eqref{borel_transformed_dispersion_relation} yields
\begin{gather}
\mathcal{R}_k\left(\tau\,,s_0\right) = \frac{1}{\pi} \int_{t_0}^\infty dt \, t^k \, e^{-t\tau} \, \rho^{\rm res}\left(t\right) 
\,, 
\\
\mathcal{R}_k\left(\tau\,,s_0\right) = \frac{\hat B}{\tau} \left[ \left(-Q^2\right)^k \Pi\left(Q^2\right) \right] 
- \frac{1}{\pi} \int_{s_0}^\infty dt \, t^k \, e^{-t\tau} \, {\rm Im}\Pi\left(te^{-i\pi}\right)  \,.
\label{laplace_sum_rules} 
\end{gather}
The quantity $\mathcal{R}_k\left(\tau\,,s_0\right)$ can be calculated in QCD, and is related to the spectral function $\rho^{\rm res}\left(t\right)$. The spectral function $\rho^{\rm res}\left(t\right)$ can be measured experimentally, or it can be modeled in terms of the physical properties of the hadron being studied. Therefore, Eq.~\eqref{laplace_sum_rules} provides a direct relationship between QCD calculations and hadronic parameters. This is the central identity of QCD Laplace sum rules.

Before proceeding it is useful to consider possible forms that $\mathcal{R}_k\left(\tau\,,s_0\right)$ can take. Typically, the correlation function $\Pi\left(Q^2\right)$ involves functions that have a branch cut on the interval $Q^2\in\left(-\infty\,,-t_0\right]$ and functions that have a pole at $Q^2=-t_0$. Those that have a pole generally have the form
\begin{gather}
\Pi^{\rm pole} \left(Q^2\right) \sim \frac{1}{\left(Q^2+t_0\right)^n} \,,
\label{example_of_correlator_with_a_pole}
\end{gather}
where $n$ is a positive integer. Because Eq.~\eqref{example_of_correlator_with_a_pole} has no imaginary part, the contribution of such a function to the sum rule is given by
\begin{gather}
\mathcal{R}^{\rm pole}_k\left(\tau\right) = \frac{\hat B}{\tau} \left[ \left(-Q^2\right)^k \Pi^{\rm pole}\left(Q^2\right) \right] \,,
\label{sum_rule_contribution_of_pole}
\end{gather}
which is independent of the continuum threshold $s_0$. The following result is useful in order to calculate the Borel transform~\cite{Pascual_1984_a}:
\begin{gather}
\frac{\hat B}{\tau} \left[ \frac{\left(-Q^2\right)^k}{Q^2+t_0}  \right] = t_0^{2k} e^{-t_0\tau} \,.
\label{identity_for_borel_transform_of_pole}
\end{gather}
Note that Eq.~\eqref{identity_for_borel_transform_of_pole} can be extended to cases where the denominator is raised to a higher power by differentiating with respect to $t_0$. 

\begin{figure}[htb]
\centering
\includegraphics[scale=0.5]{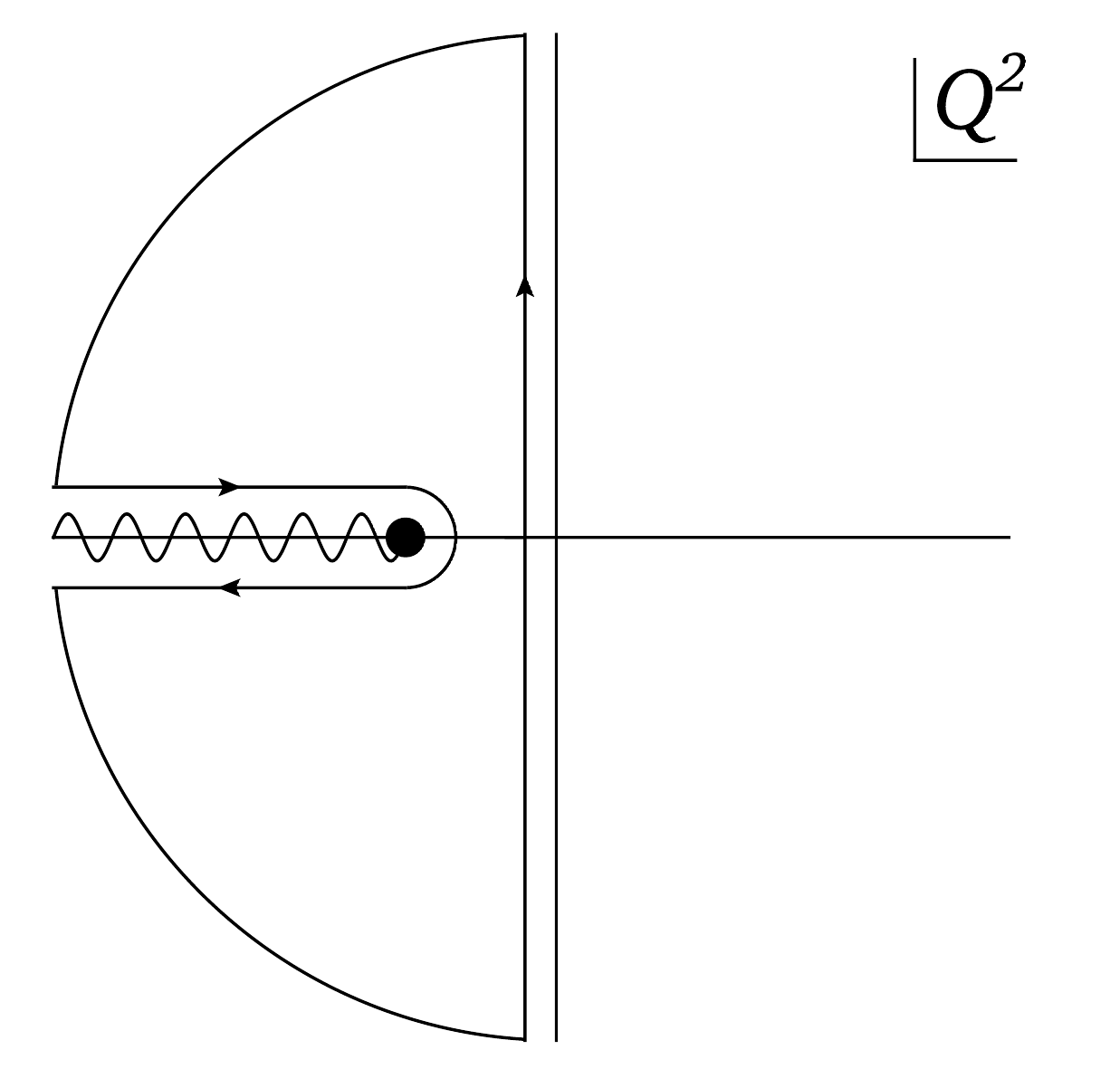}
\caption{Contour integral used to calculate the inverse Laplace transform in Eq.~\eqref{inverse_laplace_transform_of_branch}. The function has a branch cut on the interval $Q^2\in\left(-\infty\,,-t_0\right]$, where $t_0$ is the hadronic threshold.}
\label{inverse_laplace_transform_contour}
\end{figure}

\noindent Functions that have a branch cut can be dealt with using the relationship between the Borel transform and the inverse Laplace transform~\eqref{borel_and_inverse_laplace_transform}. The contribution to the sum rule is given by
\begin{gather}
\mathcal{R}^{\rm branch}_k\left(\tau\,,s_0\right) = \frac{\hat B}{\tau} \left[ \left(-Q^2\right)^k \Pi^{\rm branch}\left(Q^2\right) \right] - \frac{1}{\pi} \int_{s_0}^\infty dt \, t^k \, e^{-t\tau} \, {\rm Im}\Pi^{\rm branch}\left(te^{-i\pi}\right)  \,,
\label{sum_rule_contribution_of_branch}
\end{gather}
The first term in Eq.~\eqref{sum_rule_contribution_of_branch} is an inverse Laplace transform
\begin{gather}
\mathcal{L}^{-1}\left[\left(-Q^2\right)^k \Pi^{\rm branch}\left(Q^2\right) \,; \tau \right] = \frac{1}{2\pi i} \int\limits_{b-i\infty}^{b+i\infty} dQ^2 \left(-Q^2\right)^k \Pi^{\rm branch}\left(Q^2\right) e^{Q^2 \tau} \,.
\label{inverse_laplace_transform_of_branch} 
\end{gather}
This can be calculated using the residue theorem. For instance, consider the contour integral
\begin{gather}
I = \frac{1}{2\pi i} \oint_C dQ^2 \left(-Q^2\right)^k \Pi^{\rm branch}\left(Q^2\right) e^{Q^2 \tau} \,,
\label{inverse_laplace_contour_integral}
\end{gather}
where the integration contour $C$ is depicted in Fig.~\ref{inverse_laplace_transform_contour}. By the residue theorem $I=0$, and hence
\begin{gather}
I_\mathcal{L} = - I_R - I_{\rm top} - I_{\eta} - I_{\rm bottom} \,,
\label{inverse_laplace_contours}
\end{gather}
where $I_\mathcal{L}$, $I_R$, $I_{\rm top}$, $I_{\eta}$ and $I_{\rm bottom}$ denote the portion of contour $C$ in Fig~\ref{inverse_laplace_transform_contour} parallel to the imaginary axis, the radial contour, the portion above the branch cut, the portion that circles the branch cut, and the portion below the branch cut, respectively. The exponential factor in Eq.~\eqref{inverse_laplace_transform_of_branch} ensures that $I_R=0$ for $R\to\infty$ and $I_{\eta}=0$ provided that 
$\Pi^{\rm branch}\left(-t_0\right)$ is regular. It can be shown that the remaining portions of the contour give
\begin{gather}
\begin{split}
 I_\mathcal{L} &= \frac{1}{2\pi i} \int\limits_{t_0}^{\infty} dt \, t^k \, \left[ \Pi^{\rm branch}\left(te^{-i\pi}\right) - \Pi^{\rm branch}\left(te^{i\pi}\right)  \right] e^{-t\tau}
= \frac{1}{\pi} \int\limits_{t_0}^{\infty} dt \, t^k {\rm Im}\Pi\left(te^{-i\pi}\right) e^{-t\tau} \,,
\end{split}
\end{gather}
where we have used the fact that $\Pi^{\rm branch}\left(Q^2\right)$ must satisfy Schwarz reflection. Inserting this into Eq.~\eqref{sum_rule_contribution_of_branch}, we find
\begin{gather}
\mathcal{R}^{\rm branch}_k\left(\tau\,,s_0\right) = \frac{1}{\pi} \int\limits_{t_0}^{s_0} dt \, t^k \, e^{-t\tau} \, {\rm Im}\Pi^{\rm branch}\left(te^{-i\pi}\right)  \,.
\label{final_sum_rule_contribution_of_branch}
\end{gather}
In general, the field theoretic correlation function $\Pi\left(Q^2\right)$ will contain functions that have a branch cut, and in order to formulate the contribution of these to the sum rule we must evaluate the imaginary part of these functions below the branch cut. In principle, only the imaginary part of the correlation function is needed in order to use Eq.~\eqref{final_sum_rule_contribution_of_branch}. However, there are some situations in which the entire correlation function is needed in order to properly formulate the sum rule. For example, some terms in the correlation function may be singular at the branch point. This occurs in Chapter~\ref{chapter_3_hybrids}, for instance. In this case, the integrand in Eq.~\eqref{final_sum_rule_contribution_of_branch} is singular at the lower limit of integration. However, this difficulty can be overcome by noting that contribution to the inverse Laplace transform from the integration contour $I_{\eta}$ in Eq.~\eqref{inverse_laplace_contours} is singular as the 
radius of the contour is taken to zero. This compensates for the integration divergence in Eq.~\eqref{final_sum_rule_contribution_of_branch}. In this way a limiting procedure can be developed such that the integration in Eq.~\eqref{final_sum_rule_contribution_of_branch} is well-defined. In order to do so the entire correlation function $\Pi\left(Q^2\right)$ must be known, however.

\subsection{Operator Product Expansion}
\label{OperatorProductExpansion}

In QSR analyses we typically wish to study a hadronic state $|h\rangle$ with certain $J^{PC}$ quantum numbers. To do so, we define a current $J$ with the same quantum numbers that couples to the hadronic state
\begin{gather}
\langle \Omega | J | h \rangle = \Lambda f_h 
\end{gather}
where $\Lambda$ is a dimensionful constant and $f_h$ is a dimensionless factor that measures how strongly the hadronic state $|h\rangle$ couples to the current $J$. The current is a local composite operator composed of quark and gluon fields that approximate the valence quark and gluon content of the hadronic state $|h\rangle$. However, it is important to note that more than one current $J$ may couple to a single hadronic state. Chapter~\ref{chapter_6_mixing} explores such a scenario. Once the current $J$ has been constructed, we form the correlation function
\begin{gather}
\Pi\left(Q^2\right) = i\int d^4 x \, e^{iq\cdot x} \langle \Omega | \, T\left[ J\left(x\right) J^\dag\left(0\right) \right] | \Omega \rangle \,, \quad Q^2 = -q^2 \,,
\label{example_of_qsr_correlation_function}
\end{gather}
which can be calculated using the perturbative expansion~\eqref{perturbative_expansion_of_correlation_function}. However, as mentioned previously the QCD coupling becomes large at hadronic energy scales and hence a purely perturbative approach cannot adequately describe low energy phenomena. 

In QSR, confinement is assumed to exist, and its effects are parametrized through the operator product expansion (OPE)~\cite{Wilson_1969_a}:
\begin{gather}
\lim_{x \to 0} J\left(x\right)J^\dag\left(0\right) = \sum_n C_n\left(x\right) :\mathcal{O}_n\left(0\right): \,,
\label{position_space_ope}
\end{gather}
where the Wilson coefficients $C_n\left(x\right)$ are functions of $x$ and $:\mathcal{O}_n\left(0\right):$ are normal ordered composite operators of dimension $n$ (the dimensions of quark and gluon fields are discussed in Appendix~\ref{appendix_a_conventions}). Taking the vacuum expectation value and moving to momentum space, the OPE reads
\begin{gather}
\lim_{Q^2 \to \infty} \int d^4 x \, e^{iq\cdot x}  J\left(x\right)J^\dag\left(0\right) = \sum_n \langle \Omega | \,  :\mathcal{O}_n\left(0\right):  | \Omega \rangle \int d^4 x \, e^{iq\cdot x} C_n\left(x\right)  \,.
\label{momentum_space_ope}
\end{gather}
The lowest dimensional operator in the OPE is the identity operator, which corresponds to purely perturbative contributions. Each higher dimensional operator $\mathcal{O}_n\left(0\right)$ is a normal ordered combination of quark and gluon fields whose vacuum expectation value does not vanish. These are called condensates and represent non-trivial features of the QCD vacuum $|\Omega\rangle$. Through the OPE, QSR analyses naturally include both perturbative and non-perturbative effects. The OPE involves an implicit separation of scales: the condensates and Wilson coefficients represent low and high energy phenomena, respectively. As such, the Wilson coefficients can be calculated perturbatively. The condensates are gauge invariant and Lorentz invariant combinations of quark and gluon fields. The two most important condensates are the quark and gluon condensates
\begin{gather}
m_q \langle \bar{q}q \rangle = m_q \langle \Omega | : \bar{q}\left(0\right)q\left(0\right) : | \Omega \rangle \,, \quad
\alpha \langle G^2 \rangle = \alpha \langle \Omega | : G^a_{\mu\nu}\left(0\right)G_a^{\mu\nu}\left(0\right) : | \Omega \rangle \,, 
\label{quark_and_gluon_condensate_definition}
\end{gather}
both of which have dimension four. Note that the quark condensate does not include heavy flavours because the heavy quark condensate can be related to the gluon condensate. It is important to stress that the numerical values of condensates cannot be calculated directly within QCD. Rather, they must be determined empirically. One such method involves using QSR duality relations to relate condensates to experimental data, for instance. The quark condensate can be defined in terms of the pion mass and decay constant via the Gell-Mann-Oakes-Renner relation~\cite{Gell_Mann_1968_a}:
\begin{gather}
 m_q \langle \bar{q} q \rangle = -\frac{1}{2} f_\pi^2 m_\pi^2 \,, \quad f_\pi = 0.093\,{\rm GeV} \,, \quad m_\pi = 0.139\,{\rm GeV} \,,
\label{gmor_relation} 
\end{gather}
where the numerical values have been taken from Ref.~\cite{Beringer_2012_a}. The gluon condensate can be extracted from a QSR analysis of charmonium~\cite{Narison_2010_a}, which yields
\begin{gather}
\langle \alpha G^2 \rangle = \left(7.5\pm2.0\right) \times 10^{-2} \,{\rm GeV}^4 \,.
\label{gluon_condensate}
\end{gather}
Higher-dimensional condensates involving more quark and gluon fields also exist. For instance, the mixed condensate has dimension-five and is given by
\begin{gather}
\langle \Omega | : g \, \bar{q}\left(0\right) \frac{\lambda^a}{2} \sigma^{\mu\nu} G^a_{\mu\nu}\left(0\right) q\left(0\right) : | \Omega \rangle 
= \langle \bar{q}\sigma G q \rangle = M_0^2 \langle \bar{q}q \rangle \,, 
\label{mixed_condensate}
\end{gather}
where $M_0^2=\left(0.8\pm0.1\right)\,{\rm GeV}^2$, which was determined from baryon sum rules~\cite{Dosch_1988_b}. The dimension-six gluon condensate is given by
\begin{gather}
 \langle \Omega | : g^3 \, f_{abc} \, G^a_{\alpha\beta}\left(0\right)G^b_{\beta\gamma}\left(0\right)G^c_{\gamma\alpha}\left(0\right) : | \Omega \rangle 
 = \langle g^3 G^3 \rangle = \left(8.2\pm1.0\right) \,{\rm GeV^2} \, \langle \alpha G^2 \rangle \,,
\label{dimension_six_gluon_condensate} 
\end{gather}
which was also determined in Ref.~\cite{Narison_2010_a}. Additional condensates include the dimension-six quark condensate and the dimension-eight gluon condensate, which are given in Ref.~\cite{Narison_2007_a}.

In QSR calculations the correlation function~\eqref{example_of_qsr_correlation_function} is evaluated using the OPE~\eqref{momentum_space_ope}. Contributions that are proportional to the identity operator correspond to purely perturbative effects, while contributions from higher dimensional operators in the OPE correspond to non-perturbative effects that are represented through condensates. In practice, the simplest way to calculate these contributions is with the aid of Wick's theorem~\eqref{wicks_theorem}. For instance, consider   the calculation of a correlation function of the form
\begin{gather}
\Pi\left(Q^2\right) = i\int d^4 x \, e^{iq\cdot x} \langle \Omega | \, T\left[ J\left(x\right) J^\dag\left(0\right) \right] | \Omega \rangle \,, \quad Q^2 = -q^2 \,,
\label{scalar_current_correlation_function}
\end{gather}
where $J\left(x\right)=\bar{q}\left(x\right)q\left(x\right)$. The correlation function can be calculated using the perturbative expansion~\eqref{perturbative_expansion_of_correlation_function}. The lowest order term in the perturbative expansion involves the following time ordered product, which can be evaluated using Wick's theorem~\eqref{wicks_theorem}:
\begin{gather}
\begin{split}
T\left[q\left(x\right)\bar{q}\left(x\right)\bar{q}\left(0\right)q\left(0\right)\right] 
=&\,:q\left(x\right)\bar{q}\left(x\right)\bar{q}\left(0\right)q\left(0\right):
\\
&\!-\contraction{}{q}{\left(x\right)}{\bar{q}}\contraction{q\left(x\right)\bar{q}\left(0\right)}{q}{\left(0\right)}{\bar{q}} 
q\left(x\right)\bar{q}\left(0\right)q\left(0\right)\bar{q}\left(x\right)
\\
&\!+\contraction{}{q}{\left(0\right)}{\bar{q}}q\left(0\right)\bar{q}\left(x\right) :\bar{q}\left(0\right)q\left(x\right): 
\\
&\!+:\bar{q}\left(x\right)q\left(0\right):\contraction{}{q}{\left(x\right)}{\bar{q}}q\left(x\right)\bar{q}\left(0\right) \,.
\end{split}
\label{example_of_wicks_theorem}
\end{gather}
Note that the first term in Eq.~\eqref{example_of_wicks_theorem} corresponds to a disconnected diagram and can be ignored. The second term in Eq.~\eqref{example_of_wicks_theorem} is the $\mathcal{O}\left(g^0\right)$ contribution to the $n=0$ Wilson coefficient in the OPE. That is, the second term represents the leading-order perturbative contribution to the correlation function~\eqref{scalar_current_correlation_function}. The third and fourth terms in Eq.~\eqref{example_of_wicks_theorem} involve the normal ordered product of two quark fields at distinct locations. Ultimately, the normal ordered products will be related to condensate contributions, {\it i.e.}\, terms in the OPE with $n>0$. The propagators that multiply these terms will lead to the corresponding Wilson coefficients. 

Note that because of the limit the definition of the perturbative expansion~\eqref{perturbative_expansion_of_correlation_function}, the vacuum expectation value of these terms is taken using $|\Omega\rangle$. For instance, the fourth term in Eq.~\eqref{example_of_wicks_theorem} involves the vacuum expectation value
\begin{gather}
\langle \Omega | : \bar{q}\left(x\right)q\left(0\right) : | \Omega \rangle = \langle \Omega | : \bar{q}\left(0\right)q\left(0\right) : | \Omega \rangle
+ x^\mu \partial_\mu \left. \langle \Omega | : \bar{q}\left(x\right)q\left(0\right) : | \Omega \rangle \right|_{x=0} + \mathcal{O}\left(x^2\right)
\label{expansion_of_non_local_vev}
\end{gather}
The first term in this expansion can be identified with the quark condensate $\langle \bar{q}q \rangle$~\eqref{gmor_relation}. The second term is problematic because it involves the derivative $\partial_\mu$ and hence is not gauge invariant. Ultimately the higher order terms in this expansion will be related to higher dimensional condensates, which are gauge invariant by definition. Therefore the expansion in Eq.~\eqref{expansion_of_non_local_vev} must be performed in a gauge invariant fashion. This can be achieved using fixed-point gauge techniques~\cite{Novikov_1983_a}, or equivalently using plane wave methods~\cite{Bagan_1992_a}. Here we will use fixed-point gauge, where the gluon field satisfies 
\begin{gather}
x^\mu A^a_\mu\left(x\right) = 0 \,.
\label{fixed_point_gauge}
\end{gather}
Using this gauge, the derivative in Eq.~\eqref{expansion_of_non_local_vev} can be replaced by a covariant derivative. In Ref.~\cite{Pascual_1984_a} the fixed-point gauge expansion of the vacuum expectation value in Eq.~\eqref{expansion_of_non_local_vev} is explicitly calculated to third order in $x$. Higher order terms in the expansion can expressed naturally in terms of higher dimensional condensates. For instance, $\mathcal{O}\left(x^2\right)$ terms in the expansion are proportional to the mixed condensate~\eqref{mixed_condensate} while $\mathcal{O}\left(x^3\right)$ terms are proportional to the dimension-six quark condensate.

So far we have only considered the leading order term in the perturbative expansion of the correlation function~\eqref{scalar_current_correlation_function}. Higher order terms that are generated by the perturbative expansion~\eqref{perturbative_expansion_of_correlation_function} can be evaluated within the OPE using an approach identical to that described above. However, this naturally leads to time ordered products that include not only the quark fields of the currents in Eq.~\eqref{scalar_current_correlation_function}, but also quark and gluon fields from the QCD action. This means that vacuum expectation values involving gluon fields will be encountered. Fixed point gauge techniques can be used to express the gluon field in terms of the gluon field strength, and hence a manifestly gauge invariant expansion of vacuum expectation values involving gluon fields can be constructed. Fixed point expansions of vacuum expectation values involving gluons are discussed in Ref.~\cite{Pascual_1984_a}. Ultimately, the 
terms in the resulting expansion will lead to contributions from the gluon condensate~\eqref{quark_and_gluon_condensate_definition} and dimension-six gluon condensate~\eqref{dimension_six_gluon_condensate}, for instance. 

In QSR calculations the OPE is usually truncated at some order and the Wilson coefficients are calculated to a certain order in the coupling $\alpha$. For instance, Chapter~\ref{chapter_3_hybrids} studies heavy quarkonium hybrids which are probed by the current
\begin{gather}
J_\mu = \frac{g}{2} \bar{Q} \lambda^a \gamma^\nu \tilde{G}^a_{\mu\nu} Q \,, \quad \tilde{G}^a_{\mu\nu} = \frac{1}{2} \epsilon_{\mu\nu\alpha\beta} G^{\alpha\beta}_a \,,
\label{hybrid_current}
\end{gather}
where $Q$ and $G^{\alpha\beta}_a$ denote a heavy quark field and the gluon field strength, respectively. In Chapter~\ref{chapter_3_hybrids} the perturbative, dimension-four $\langle \alpha G^2 \rangle$ and dimension-six $\langle g^3 G^3 \rangle$ gluon condensate contributions are included in the OPE. Because the hybrid current~\eqref{hybrid_current} contains only heavy quarks, condensates that include light quark fields contribute at higher orders in the expansion and are suppressed. The Wilson coefficients for the perturbative, dimension-four and dimension-six gluon condensate are calculated to leading order in the coupling $\alpha$. The evaluation of leading order contributions to the Wilson coefficients involves the calculation of multiple two-loop momentum integrals. These loop integrals are often quite difficult to evaluate and constitute a significant technical barrier to extending QSR calculations to higher orders. Chapter~\ref{chapter_2_loops} discusses techniques for evaluating loop integrals.

\subsection{Hadronic Spectral Function}
\label{HadronicSpectralFunction}

\begin{figure}[htb]
\includegraphics[scale=0.74]{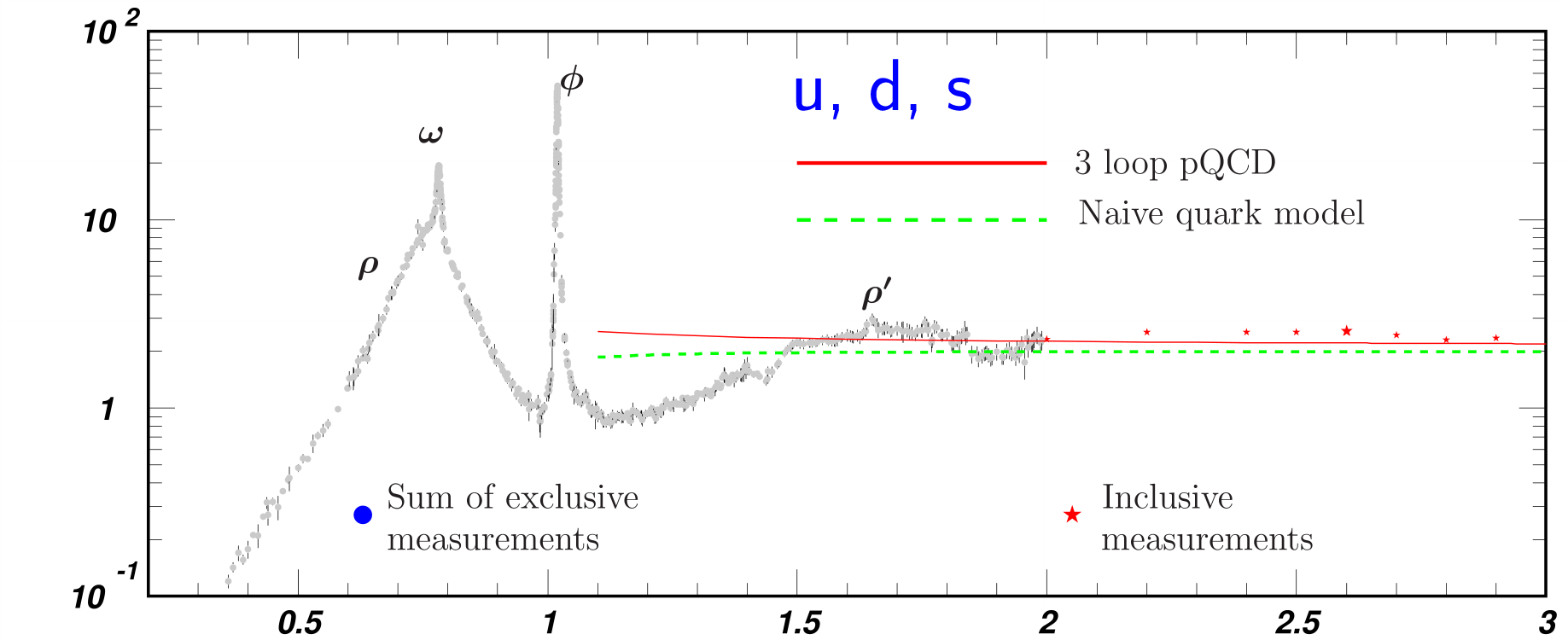}
\caption{The hadronic spectral function $R(s)$. The horizontal axis is the center of mass frame collision energy of the electron and positron in units of ${\rm GeV}$ and the vertical axis is the dimensionless number $R(s)$. The resonances labeled $\rho$, $\omega$ and $\phi$ correspond to distinct hadrons. The electron and positron annihilate through a virtual photon or $Z$ boson, both of which have the quantum numbers $J^{PC}=1^{--}$. Therefore all of these hadrons must have these quantum numbers. The horizontal location of each resonance peak is the mass of the hadron corresponding to the resonance. The region between $1.5\,{\rm GeV}$ and $3.0\,{\rm GeV}$ is the continuum which is described well by the three-loop perturbative QCD calculation. Note that in the region below $1.5\,{\rm GeV}$ the QCD prediction and resonance features agree in the sense of a global average. This is an example of the concept of quark-hadron duality which is crucial to QSR. Figure taken from Ref.~\cite{Beringer_2012_a}.}
\label{example_of_hadronic_spectral_function}
\end{figure}

As mentioned previously, the hadronic spectral function can be measured experimentally. For instance, the spectral function for hadronic states with $J^{PC}=1^{--}$ is related to the ratio of the cross sections
\begin{gather}
 R(s) = \frac{\sigma\left(e^-e^+\to\,\text{hadrons}\,\right)}{\sigma\left(e^-e^+\to\mu^-\mu^+\right)} \,.
\label{Rs}
\end{gather}
This spectral function is shown in Fig.~\ref{example_of_hadronic_spectral_function}.

Experimentally known spectral functions such as that shown in Fig.~\ref{example_of_hadronic_spectral_function} can be related to a theoretically calculated correlation function $\Pi\left(Q^2\right)$ via Eq.~\eqref{laplace_sum_rules}. In this way, QSR techniques can be used to extract QCD parameters in terms of experimentally measured quantities. 

Alternatively, a resonance model can be used to calculate hadron properties in terms of QCD parameters. This must be done in order to study exotic hadrons with QSR. For instance, a single narrow resonance can be parametrized as
\begin{gather}
\rho^{\rm res}\left(t\right) = \pi f^2 \delta\left(t-M^2\right)
\label{single_narrow_resonance_model}
\end{gather}
where $f$ and $M$ are the decay constant and mass of the hadron corresponding to the resonance. It is natural to question accuracy of this admittedly rather simple resonance model. However, it is important to remember that in QCD Laplace sum rules the resonance is multiplied by an exponential factor which tends to obscure any detailed features of the resonance. In addition, methods described in Ref.~\cite{Elias_1998_a} can be used to estimate resonance width effects. The most basic quantity of interest in any QSR analysis is the hadron mass which can be determined using this model. Inserting Eq.~\eqref{single_narrow_resonance_model} into Eq.~\eqref{laplace_sum_rules} yields
\begin{gather}
\mathcal{R}_k\left(\tau\,,s_0\right) = f^2 M^{2k} e^{-M^2\tau} \,.
\end{gather}
The hadron mass $M$ can be isolated and is given by
\begin{gather}
M\left(\tau\,,s_0\right) = \sqrt{ \frac{\mathcal{R}_{1}\left(\tau\,,s_0\right)}{\mathcal{R}_{0}\left(\tau\,,s_0\right)} }\,.
\label{mass_ratio}
\end{gather}
Using this result a hadron mass $M$ can be extracted from the theoretically calculated quantity $\mathcal{R}_k\left(\tau\,,s_0\right)$. 

Note that the hadron mass given in Eq.~\eqref{mass_ratio} is a function of the Borel parameter $\tau$ and continuum threshold $s_0$. The Borel parameter $\tau$ probes the hadronic spectral function at various energies while the continuum threshold $s_0$ is built into the model of the hadronic spectral function~\ref{resonance_and_continuum_model_of_hadronic_spectral_function} and is predicted in the QSR analysis. We must first find a region where the mass prediction~\eqref{mass_ratio} varies little with the Borel parameter. For a given value of $s_0$ we first determine a range of $\tau$ values for which the sum rule is considered reliable, called the sum rule window. In order to do this it is convenient to define the Borel mass $M_B=1/\sqrt{\tau}$. There are multiple ways in which the sum rule window can be defined. All approaches involve fixing lower and upper limits on the Borel mass. Typically contributions from condensates become significant at small values of $M_B$ whereas contributions from the 
continuum become important at large values of $M_B$. Therefore placing restrictions on condensate contributions to the sum rule can be used to place a lower bound on $M_B$, whereas restrictions on continuum contributions can be used to place an upper bound on $M_B$. The resulting range of $M_B$ values is where the sum rule is considered to be reliable, or the sum rule window. However, the width of the sum rule window varies with the value of $s_0$.

\begin{figure}[htb]
\centering
\includegraphics[scale=0.9]{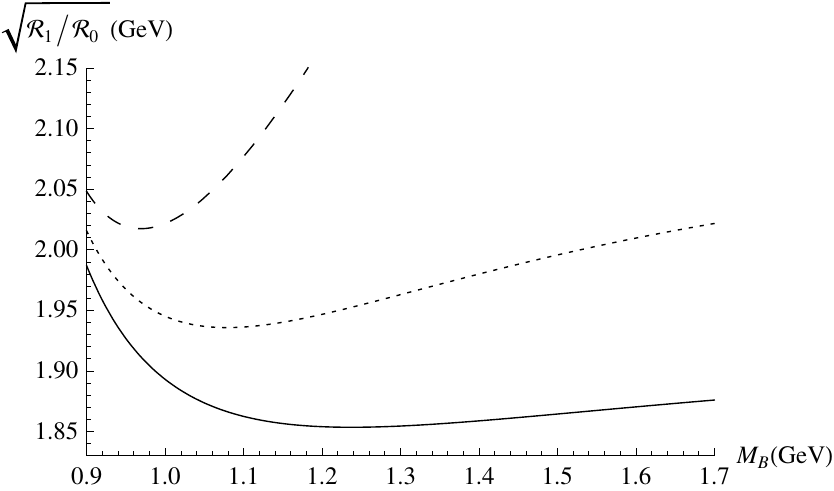}
\includegraphics[scale=0.9]{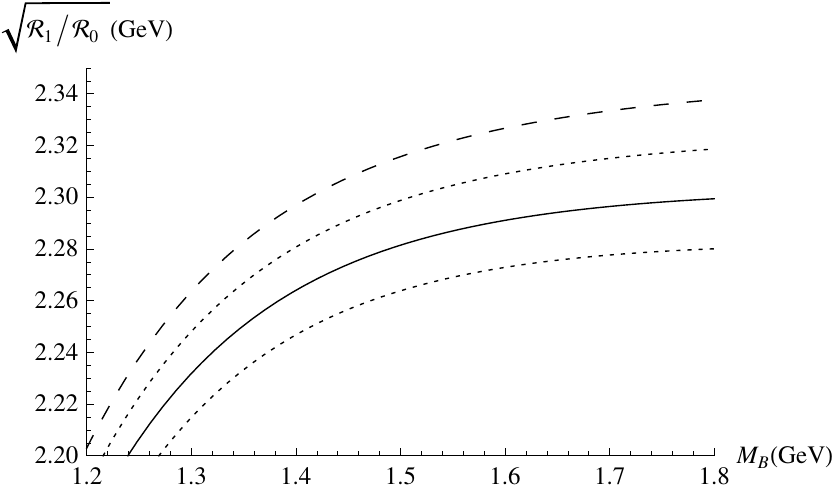}
\caption{Examples of stable (left) and unstable (right) sum rules. The stable sum rule exhibits $\tau$ stability within the sum rule window whereas the unstable sum rule does not. In the unstable case there is no region in the sum rule window where the mass prediction given by Eq.~\eqref{mass_ratio} displays weak dependence on the Borel parameter. Mass predictions made from unstable sum rules are unreliable for this reason. The left and right figures correspond to the scalar and pseudoscalar charm-light diquark sum rules from Chapter~\ref{chapter_4_Qq_diquark}, respectively.}
\label{example_of_stable_and_unstable_sum_rules}
\end{figure}

The value of $s_0$ can be constrained through the requirement of $\tau$ stability~\cite{Narison_2007_a}:
\begin{gather}
\frac{d}{d\tau} M\left(\tau\,,s_0\right) = 0 \,. 
\label{tau_stability}
\end{gather}
Examples of stable and unstable sum rules are shown in Fig.~\ref{example_of_stable_and_unstable_sum_rules}. The smallest value of $s_0$ in which the mass prediction~\eqref{mass_ratio} exhibits $\tau$ stability within the sum rule window is taken to be the minimum value $s_0^{\rm min}$. The optimal value of $s_0$ is determined using
\begin{gather}
\chi^2\left(s_0\right)=\sum_j \left( \frac{1}{M}\sqrt{\frac{{\cal R}_1\left(\tau_j,s_0\right)}{{\cal R}_0\left(\tau_j,s_0\right)}}-1 \right)^2\,, \quad  s_0 \geq s_0^{\rm min} \,.
\label{chi_squared}
\end{gather}
The optimal value of $s_0$ is that which minimizes~\eqref{chi_squared}. The sum in Eq.~\eqref{chi_squared} is calculated over the sum rule window corresponding to $s_0^{\rm min}$. Once the optimal value $s_0^{\rm opt}$ has been determined, the mass can be extracted by fitting $M\left(s_0^{\rm opt}\,,\tau\right)$ to a constant over the sum rule window.

\subsection{QCD sum rules and Heavy Quarkonium-like states}

The QSR method has been applied to a wide variety of problems in hadronic physics. In particular, QSR calculations provide crucial information on the properties of exotic hadrons. This theoretical input helps to guide the experimental search for such states. The majority of QSR studies of exotic hadrons have focused on those that are much lighter than the heavy quarkonium-like states (see Ref.~\cite{Narison_2007_a} for a detailed review). 

However, the recently discovered XYZ states have motivated QSR studies of exotic hadrons that may exist within the same mass region as heavy quarkonia. Nearly all of this work has considered four-quark states, which can be realized as molecules or tetraquarks (Ref.~\cite{Nielsen_2009_a} provides a review). A common feature of all of these studies is the use of currents that contain four quark fields. However, for reasons discussed in Chapter~\ref{chapter_4_Qq_diquark}, QSR studies that use four-quark currents cannot distinguish between the molecular and tetraquark configurations. An alternate approach is to use diquark currents, which are relevant to tetraquarks only. This approach was first applied to diquarks containing a heavy quark in Ref.~\cite{Wang_2010_a}. The research presented in Chapter~\ref{chapter_4_Qq_diquark} serves to extend this work.

Hybrid mesons could also exist in the heavy quarkonium mass region. Surprisingly, this possibility has been little explored by QSR practitioners. Refs.~\cite{Govaerts_1984_a,Govaerts_1985_a,Govaerts_1986_a} comprise the earliest QSR studies of heavy quarkonium hybrids. These studies examined a wide variety of $J^{PC}$ channels. However, the sum rules for many channels exhibited instabilities, and hence the resulting mass predictions in those channels are unreliable. Ref.~\cite{Qiao_2010_a} recently performed an updated analysis of the $1^{--}$ channel. The research presented in Chapter~\ref{chapter_3_hybrids} updates the $1^{++}$ and $0^{-+}$ heavy hybrid sum rules and extracts reliable mass predictions in both cases.

QSR analyses of heavy quarkonium-like states necessarily use currents containing heavy quarks. In practice, this means that the loop integrals that must be evaluated in order to determine the Wilson coefficients are quite complicated. This is in contrast to QSR studies of hadrons composed of light quarks, which are often performed in the chiral limit where the light quark masses are neglected. However, the heavy quark mass cannot be neglected and the resulting loop integrals lead to complicated functions in the dimensionless ratio of the external momentum and the heavy quark mass. Chapter~\ref{chapter_2_loops} discusses techniques for evaluating these integrals.

\section{Outline of Thesis}

This thesis has been prepared in the manuscript style. Chapter \ref{chapter_2_loops} develops techniques for evaluating loop integrals that are essential in subsequent chapters. Chapter \ref{chapter_3_hybrids} includes two closely related manuscripts that have been published in the Journal of Physics~G and  Physical Review~D. Chapters \ref{chapter_4_Qq_diquark}, \ref{chapter_5_diquark_renorm} and \ref{chapter_6_mixing} each consist of individual manuscripts that have been published in Physical Review~D, Journal of Physics~G and Nuclear Physics~A, respectively. The copyright agreements of the respective journals grant permission for articles to be reproduced in a thesis. Chapters \ref{chapter_3_hybrids}, \ref{chapter_4_Qq_diquark}, \ref{chapter_5_diquark_renorm} and \ref{chapter_6_mixing} each include an introduction to the research presented therein, along with a discussion of the results of the research and its relation to the thesis as a whole. Chapter \ref{chapter_7_conclusions} discusses the themes of 
the research presented in this thesis and their relation to the field of hadron spectroscopy in general. Appendices \ref{appendix_a_conventions} and \ref{appendix_b_math} discuss conventions and mathematical functions used in this thesis, respectively.

The research presented in Chapters \ref{chapter_3_hybrids}, \ref{chapter_4_Qq_diquark}, \ref{chapter_5_diquark_renorm} and \ref{chapter_6_mixing} involves three overarching themes. The first theme is the use of QSR techniques to extract mass predictions for exotic hadrons containing heavy quarks, and the comparison of these mass predictions with the XYZ states. The second theme involves the application of sophisticated loop integration techniques, which are described in Chapter~\ref{chapter_2_loops}. These techniques are essential for all of the research in this thesis. The third theme is the development of the renormalization methodology used in higher-order QSR calculations. Table~\ref{research_summary} summarizes the research presented in subsequent chapters.

\begin{table}[hbt]
\centering
\begin{tabulary}{\textwidth}{LLJJJ}
\hline
$\text{Chapter}$ & \centering Description and Key Results & Themes & &  \\
& & 1 & 2 & 3 \\
\hline
& & & &  \\
\ref{chapter_3_hybrids} & QSR study of $J^{PC}=1^{++}\,,0^{-+}$ heavy quarkonium hybrids. & \checkmark & \checkmark & \\
& & & &  \\
& $M\left(1^{++}\,,\,c\bar{c}g\right) = 5.13\pm0.25\,{\rm GeV}$, $M\left(0^{-+}\,,\,c\bar{c}g\right) = 3.82\pm0.13\,{\rm GeV}$. These results preclude the pure charmonium hybrid interpretation of the $X(3872)$ and support the charmonium hybrid interpretation of the $Y(3940)$. & & &  \\
& & & &  \\
\ref{chapter_4_Qq_diquark} & QSR study of $J^{PC}=0^{\pm}\,,1^{\pm}$ heavy-light diquarks. & \checkmark & \checkmark & \checkmark \\
& & & &  \\
& $M\left(0^{+}\,,\,cq\right) = 1.86\pm0.05\,{\rm GeV}$, $M\left(1^{+}\,,\,cq\right) = 1.87\pm0.10\,{\rm GeV}$, $M\left(0^{+}\,,\,bq\right) = M\left(1^{+}\,,\,bq\right) = 5.20\pm0.10\,{\rm GeV}$. These masses are consistent with constituent diquark models, providing QCD-based support for the tetraquark interpretation of the $X(3872)$, $Z^\pm_c\left(3895\right)$, $Y_b(10890)$, $Z^\pm_b\left(10610\right)$ and $Z^\pm_b\left(10650\right)$. The renormalization methodology needed for next-to-leading order QSR calculations is also developed. & & &  \\
& & & &  \\
\ref{chapter_5_diquark_renorm} & Calculation of two-loop scalar diquark operator renormalization factor. & & \checkmark & \checkmark \\
& & & &  \\
& The renormalization factor and anomalous dimension are calculated to $\mathcal{O}\left(\alpha^2\right)$. These results are utilized in Chapter~\ref{chapter_4_Qq_diquark}. & & &  \\
& & & &  \\
\ref{chapter_6_mixing} & QSR analysis of mixing between scalar gluonium and quark mesons. & & \checkmark & \checkmark \\
& & & &  \\
& The perturbative contribution to the non-diagonal scalar gluonium-quark meson correlation function is calculated. The renormalization methodology needed for QSR studies involving non-diagonal correlation functions is developed. & & &  \\
& & & &  \\
\hline
\end{tabulary}
\caption{Summary of thesis research.}
\label{research_summary}
\end{table}


\chapter{Loop Integrals}
\label{chapter_2_loops}
Correlation functions of quantum fields are related to experimentally observable quantities and as such they are the building blocks of all calculations in quantum field theory. When the perturbative expansion of a correlation function is extended to higher orders, integrals over the momentum variables that circulate in Feynman diagrams are encountered. These integrals can be rather difficult and many powerful techniques have been developed to evaluate them. Much of this activity is driven by steady increases in experimental accuracy: as experimental measurements become more precise, so must theoretical calculations. Refs.~\cite{Smirnov_2004_a,Steinhauser_2002_a} provide reviews of modern techniques for evaluating loop integrals. This chapter will focus on the integrals and techniques that are used in Chapters~\ref{chapter_3_hybrids}, \ref{chapter_4_Qq_diquark}, \ref{chapter_5_diquark_renorm} and \ref{chapter_6_mixing}.

\section{Properties of Loop Integrals}
\label{PropertiesOfLoopIntegrals}

Ref.~\cite{Collins_1984_a} provides a careful exposition of the properties of dimensionally regularized momentum integrals. As we shall see in Section~\ref{IntsWithAtMostOneMass}, the $i\eta$ pole prescription in the Feynman propagator permits a transition from Minkowski space to Euclidean space, thus without loss of generality we may consider all momentum integrals as existing in a $d$-dimensional Euclidean space. Dimensionally regularized integrals are completely analogous with integrals in a Euclidean space with an arbitrary integer number of dimensions. The integration operation is linear so that
\begin{gather}
\int d^d k \left[a_1 f_1(k) + a_2 f_2(k) \right] = a_1 \int d^d k f_1(k) + a_2 \int d^d k f_2(k) \,.
\end{gather}
Integration variables may be rescaled, leading to the definition
\begin{gather}
\int d^d k f(ak) = a^{-d} \int d^d k f(k) \,,
\end{gather}
which is consistent with the $d$-dimensional definition of the Jacobian. The integration operation also respects translation invariance:
\begin{gather}
\int d^d k f(k+q) = \int d^d k f(k) \,.
\end{gather}
The normalization of the integrals is fixed through the Gaussian integral
\begin{gather}
\int d^d k \, \exp{\left[-\frac{k^2}{a^2}\right]} = a^d\,\pi^{\frac{d}{2}} \,, \quad a^2 > 0 \,.
\label{d_dimensional_integral_normalization}
\end{gather}
An important property is that integrals that do not involve a mass or external momenta are identically zero, that is 
\begin{gather}
\int d^d k \,k^{2n} = 0 \,,
\label{massless_tadpole_definition} 
\end{gather}
for all values of $n$. This result is proven in Ref.~\cite{Collins_1984_a}. Integrals such as~\eqref{massless_tadpole_definition} that do not depend on any external scales are called massless tadpoles.

\section{Integrals with at most one massive propagator}
\label{IntsWithAtMostOneMass}
\begin{figure}[htb]
\centering
\includegraphics[scale=0.5]{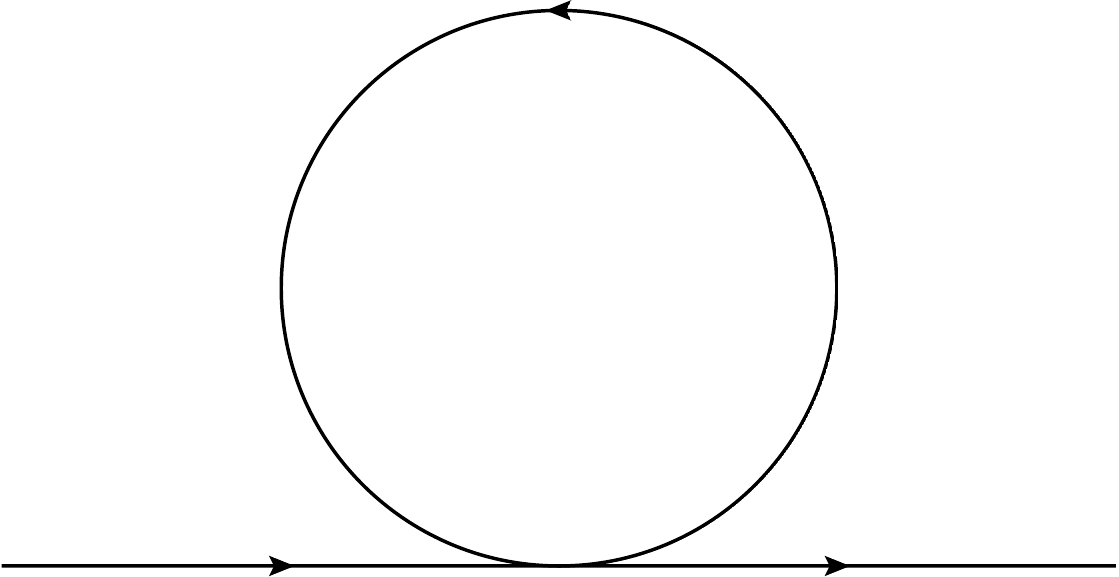}
\caption{Feynman diagram representing the $A$-type massive tadpole integral.}
\label{A_topology}
\end{figure}

The most basic loop integral is the one-loop massive tadpole, which is given by
\begin{gather}
A\left(d\,;n\,,m\right) = \lim_{\eta \to 0^+} \frac{1}{\mu^{d-4}} \int \frac{d^d k}{(2\pi)^d} \frac{1}{\left[k^2-m^2+i\eta\right]^n} \,.
\label{A_int_definition}
\end{gather}
where $\mu$ denotes the renormalization scale. A Feynman diagram representing this loop integral is shown in Fig.~\ref{A_topology}. The momentum of the particle with mass $m$ that circulates the loop is $k$, as such $k$ is called the loop momentum. There is a correspondence between the topology of a Feynman diagram and the structure of the loop integral that it represents. In particular, the flow of momenta through a diagram is closely related to the form of the corresponding integral. The Feynman diagrams in this chapter are intended to emphasize this correspondence, therefore they do not specify the spin or mass of any given particle and only serve to indicate the flow of momenta through the diagram. 

\noindent The $i\eta$ pole prescription is related to causality and is implicitly included in all propagators. As we are in Minkowski space, so the momentum $k$ has one temporal component $k_0$ and $d-1$ spatial components $k_i$. Thus the integral can be written as 
\begin{gather}
A\left(d\,;n\,,m\right) = \lim_{\eta \to 0^+} \frac{1}{\mu^{d-4}} \prod_{i=1}^{d-1} \int \frac{d k_{i}}{2\pi} \int \frac{d k_0}{2\pi} \frac{1}{\left[k_0^2-k_i^2-m^2+i\eta\right]^n} \,.
\label{A_int_components}
\end{gather}

\begin{figure}[htb]
\centering
\includegraphics[scale=0.6]{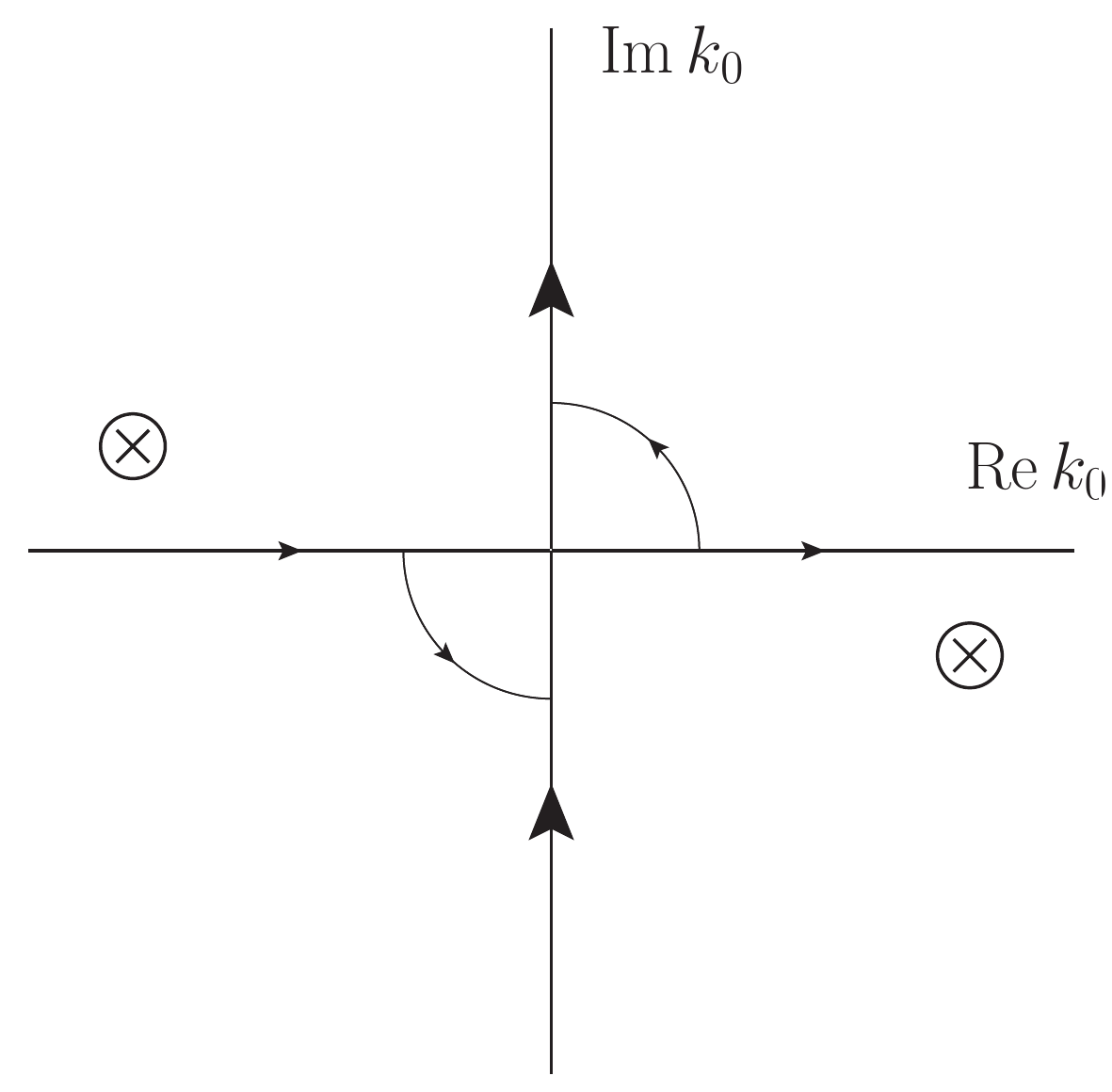}
\caption{
Integration contours for the $k_0$ integral in~\eqref{A_int_components}. Poles are indicated by $\otimes$ symbols.}
\label{Wick_Rotation}
\end{figure}

\noindent The $k_0$ integral has poles at $k_0=\pm\sqrt{k_i^2+m^2}\mp i\eta$, as shown in Fig.~\ref{Wick_Rotation}. The locations of the poles permit the integration contour to be shifted as shown in Fig.~\ref{Wick_Rotation}. Rather than integrating along the real $k_0$ axis, we perform a Wick rotation and integrate along the imaginary $k_0$ axis. This is equivalent to the following change of variables:
\begin{gather}
k_0 = i \,\ell_d \,, \quad dk_0 = i\, d\ell_d \,, \quad k_i=\ell_i \,, \quad d^dk = i \,d^d\ell \,, \quad \ell^2 = \sum_{i=1}^d \ell_i^2 \,.
\end{gather}
Formally this is equivalent to moving to a $d$-dimensional Euclidean space, therefore we can evaluate the integral in $d$-dimensional spherical coordinates. In these coordinates the volume element is given by
\begin{gather}
 d^d \ell = \ell^{d-1} d\ell \, d \Omega_d \,,
\end{gather}
where $d\ell$ denotes the integration over the radial coordinate and $d\Omega_d$ denotes the integrations over the $d-1$ angular coordinates. The $\eta \to 0^+$ limit can now be safely evaluated, and the integral~\eqref{A_int_components} becomes
\begin{gather}
A\left(d\,;n\,,m\right) = \frac{i}{(2\pi)^d}\frac{\left(-1\right)^n}{\mu^{d-4}} \int d\Omega_d \int_0^{\infty} d \ell \frac{\ell^{d-1}}{\left[\ell^2+m^2\right]^n} \,.
\label{A_int_spherical}
\end{gather}
The angular integration can be performed using the $d$-dimensional Euclidean space Gaussian integral~\eqref{d_dimensional_integral_normalization} and converting to spherical coordinates~\cite{Peskin_1995_a}. The result is
\begin{gather}
\int d\Omega_d = \frac{2\pi^\frac{d}{2}}{\Gamma\left(\frac{d}{2}\right)} \,.
\label{angular_integral}
\end{gather}
Making the change of variables $z=\frac{\ell^2}{m^2}$, the radial integral can be evaluated in terms of the Beta function~\eqref{integral_representation_of_beta_function}:
\begin{gather}
 \frac{1}{2} \left(m^2\right)^{\frac{d}{2}-n}  \int_0^{\infty} dz \, \frac{z^{\frac{d}{2}-1}}{\left[1+z\right]^n} = \frac{1}{2} \left(m^2\right)^{\frac{d}{2}-n} \frac{\Gamma\left(\frac{d}{2}\right)\Gamma\left(n-\frac{d}{2}\right)}{\Gamma\left(n\right)}\,.
\label{radial_integral}
\end{gather}
The final result for the one-loop massive tadpole integral~\eqref{A_int_definition} is 
\begin{gather}
A\left(d\,;n\,,m\right) =  \frac{i}{(4\pi)^2} \left(-m^2\right)^{2-n} \left[\frac{m^2}{4\pi\mu^2}\right]^{\frac{d}{2}-2} \frac{\Gamma\left(n-\frac{d}{2}\right)}{\Gamma\left(n\right)} \,.
\label{A_int_d_dim}
\end{gather}
Note that the mass dimension of this result is the same as that of the original integral~\eqref{A_int_definition}, namely $4-2n$. Typically, we will be interested in the behaviour of this integral near $d=4$, which can be determined by setting $d=4+2\epsilon$ and expanding around $\epsilon=0$. However, we will refrain from doing so until Section~\ref{EpsilonExpansion}. By itself, this integral is not particularly useful in QCD sum rule analyses because it has no momentum dependence. However, it is extraordinarily useful as a means for deriving more complicated integrals. 

\begin{figure}[htb]
\centering
\includegraphics[scale=0.5]{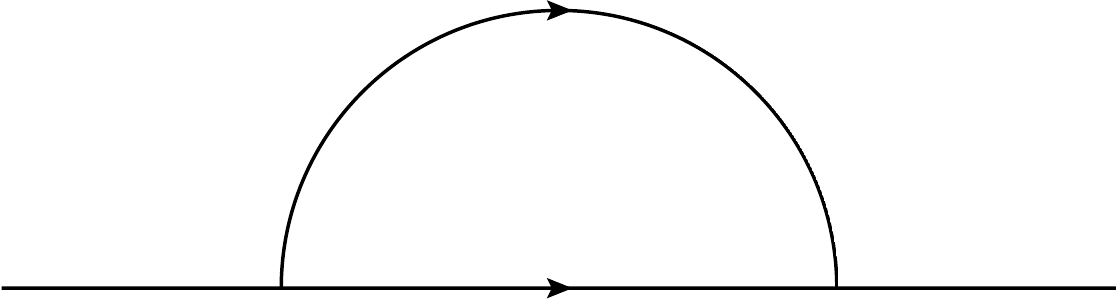}
\caption{Feynman diagram representing the $B$-type self-energy integral.}
\label{B_topology}
\end{figure}

Now consider the one-loop integral
\begin{gather}
B\left(d\,;n_1\,,0\,;n_2\,,m\right) = \frac{1}{\mu^{d-4}} \int \frac{d^dk}{(2\pi)^d} \frac{1}{k^{2n_1}\left[\right.\left(k-q\right)^2-m^2\left.\right]^{n_2}} \,,
\label{B_int_one_mass_definition}
\end{gather}
which is depicted in Fig.~\ref{B_topology}. This integral is reminiscent of a quark self-energy contribution, therefore it is called a self-energy integral. 

This integral can be evaluated with the aid of the following identity~\cite{Pascual_1984_a}:
\begin{gather}
\frac{1}{A^\alpha B^\beta} = \frac{\Gamma(\alpha+\beta)}{\Gamma(\alpha)\Gamma(\beta)} \int_0^1 dx \, \frac{x^{\alpha-1}(1-x)^{\beta-1}}{\left[Ax+B(1-x)\right]^{\alpha+\beta}} \,.
\label{feynman_parameters}
\end{gather}

\noindent The variable $x$ is called a Feynman parameter. This identity can be used to combine the two propagators in~\eqref{B_int_one_mass_definition} so that the integral can be performed using the result~\eqref{A_int_d_dim}. For instance, let $A=(k-q)^2-m^2$, $\alpha=n_2$, $B=k^2$, and $\beta=n_1$, then we can write
\begin{gather}
 Ax+B(1-x) = x(k^2+q^2-2k\cdot q-m^2) + (1-x)k^2 = (k-xq)^2 - \left[xm^2-x(1-x)q^2\right] \,.
\label{feynman_parameter_example} 
\end{gather}
Using this and making the change of variables $p=k-xq$, 
\begin{gather}
\begin{split}
B\left(d\,;n_1\,,0\,;n_2\,,m\right) = \frac{\Gamma(n_1+n_2)}{\Gamma(n_1)\Gamma(n_2)} & \int_0^1 dx \, x^{n_2-1} (1-x)^{n_1} 
\\
&\frac{1}{\mu^{d-4}} \int \frac{d^d p}{(2\pi)^d} \frac{1}{\left[\right. p^2 - \left(xm^2-x(1-x)q^2\right) \left.\right]^{n_1+n_2}} \,, 
\end{split}
\label{B_int_one_mass_feynman_parameters} 
\end{gather}
where the momentum integral can be evaluated using~\eqref{A_int_d_dim}. The result is
\begin{gather}
\begin{split}
B\left(d\,;n_1\,,0\,;n_2\,,m\right) = \frac{i}{(4\pi)^2} \left(q^2\right)^{2-n_1-n_2} & \left[-\frac{q^2}{4\pi\mu^2}\right]^{\frac{d}{2}-2}
\frac{\Gamma\left(n_1+n_2-\frac{d}{2}\right)}{\Gamma(n_1)\Gamma(n_2)} 
\\
&\int_0^1 dx \, x^{n_2-1} (1-x)^{n_1-1} \left[x(1-x)-\frac{x m^2}{q^2}\right]^{\frac{d}{2}-n_1-n_2} \,.
\end{split}
\label{B_int_one_mass_feynman_parameter_integral} 
\end{gather}
Setting $m=0$, the Feynman parameter integral can be evaluated in terms of the Beta function~\eqref{integral_representation_of_beta_function}. The result is
\begin{gather}
B\left(d\,;n_1\,,0\,;n_2\,,0\right) = \frac{i}{(4\pi)^2} \left(q^2\right)^{2-n_1-n_2}  \left[-\frac{q^2}{4\pi\mu^2}\right]^{\frac{d}{2}-2} 
\frac{\Gamma\left(\frac{d}{2}-n_1\right)\Gamma\left(\frac{d}{2}-n_2\right)\Gamma\left(n_1+n_2-\frac{d}{2}\right)}{\Gamma\left(n_1\right)\Gamma\left(n_2\right)\Gamma\left(d-n_1-n_2\right)} \,.
\label{B_int_massless_d_dim}
\end{gather}
This result agrees with an expression given for this integral in Ref.~\cite{Pascual_1984_a}. Also note that the result requires $n_1 \geq 0$ and $n_2 \geq 0$; if either of these is not satisfied the integral is a massless tadpole of the form~\eqref{massless_tadpole_definition} and is identically zero. In order to evaluate~\eqref{B_int_one_mass_feynman_parameter_integral} when the mass is non-zero, note that it can be written as
\begin{gather}
\begin{split}
B\left(d\,;n_1\,,0\,;n_2\,,m\right) = \frac{i}{(4\pi)^2} \left(q^2\right)^{2-n_1-n_2} & \left[-\frac{q^2}{4\pi\mu^2}\right]^{\frac{d}{2}-2}
\frac{\Gamma\left(n_1+n_2-\frac{d}{2}\right)}{\Gamma(n_1)\Gamma(n_2)} z^{n_1+n_2-\frac{d}{2}}
\\
& \int_0^1 dx  \, x^{\frac{d}{2}-n_1-1} (1-x)^{n_1-1} \left(1-zx\right)^{\frac{d}{2}-n_1-n_2} \,, 
\\
& z=\frac{1}{1-\frac{m^2}{q^2}} \,.
\label{B_int_one_mass_integral_rep_of_2F1} 
\end{split}
\end{gather}
The Feynman parameter integral can be evaluated in terms of the Gauss hypergeometric function~\eqref{2F1_hypergeometric_function_integral_representation}. The result is
\begin{gather}
\begin{split}
B\left(d\,;n_1\,,0\,;n_2\,,m\right) = \frac{i}{(4\pi)^2} \left(q^2\right)^{2-n_1-n_2} & \left[-\frac{q^2}{4\pi\mu^2}\right]^{\frac{d}{2}-2} z^{n_1+n_2-\frac{d}{2}} \frac{\Gamma\left(\frac{d}{2}-n_1\right)\Gamma\left(n_1+n_2-\frac{d}{2}\right)}{\Gamma(n_1)\Gamma(\frac{d}{2})} 
\\
& \phantom{}_2 F_1\left[n_1+n_2-\frac{d}{2}\,,\frac{d}{2}-n_1\,; \frac{d}{2}\,; z\right] \,.
\end{split}
\label{B_int_one_mass_q_dep}
\end{gather}
This result agrees with an expression for this integral that is given in Ref.~\cite{Pascual_1984_a}. As an additional check, we can verify that this result reproduces~\eqref{B_int_massless_d_dim} when $m=0$. Setting $z=1$ and using the identity~\eqref{2F1_at_z=1}, it is easy to verify that~\eqref{B_int_one_mass_q_dep} is consistent with the massless result~\eqref{B_int_massless_d_dim}. Note that the mass dimension of the result is carried by the term $\left(q^2\right)^{2-n_1-n_2}$ and agrees with that of the integral~\eqref{B_int_one_mass_definition}, as it must. However, for future convenience it will be helpful to recast this mass dependence in terms of the mass $m$. This can be done using the definition of the dimensionless variable $z$ given in~\eqref{B_int_one_mass_integral_rep_of_2F1}. Doing so, the final result for the one-loop self-energy integral with one massive propagator is
\begin{gather}
\begin{split}
B\left(d\,;n_1\,,0\,;n_2\,,m\right) = \frac{i}{(4\pi)^2} & \left[-\frac{m^2}{1-z}\right]^{2-n_1-n_2}  \exp{\left[\frac{d-4}{2}\left(\log\left[\frac{m^2}{4\pi\mu^2}\right]-\log{\left[1-z\right]}\right)\right]} 
\\
&\frac{\Gamma\left(\frac{d}{2}-n_1\right)\Gamma\left(n_1+n_2-\frac{d}{2}\right)}{\Gamma(n_1)\Gamma(\frac{d}{2})} 
\phantom{}_2 F_1\left[n_1+n_2-\frac{d}{2}\,,\frac{d}{2}-n_1\,; \frac{d}{2}\,; z\right] \,,
\end{split}
\label{B_int_one_mass_d_dim}
\end{gather}
where $z$ is defined as in Eq.~\eqref{B_int_one_mass_integral_rep_of_2F1}. Note that $\log{\left[1-z\right]}=-\log{\left[1+\frac{Q^2}{m^2}\right]}$, which has a branch cut on $Q^2 \in \left(-\infty\,,-m^2\right]$. It can also be shown that the hypergeometric function has the same branch cut. Therefore the result~\eqref{B_int_one_mass_d_dim} has the branch cut structure that is appropriate for a correlation function of a currents containing one massive quark with mass $m$. 

\begin{figure}[htb]
\centering
\includegraphics[scale=0.5]{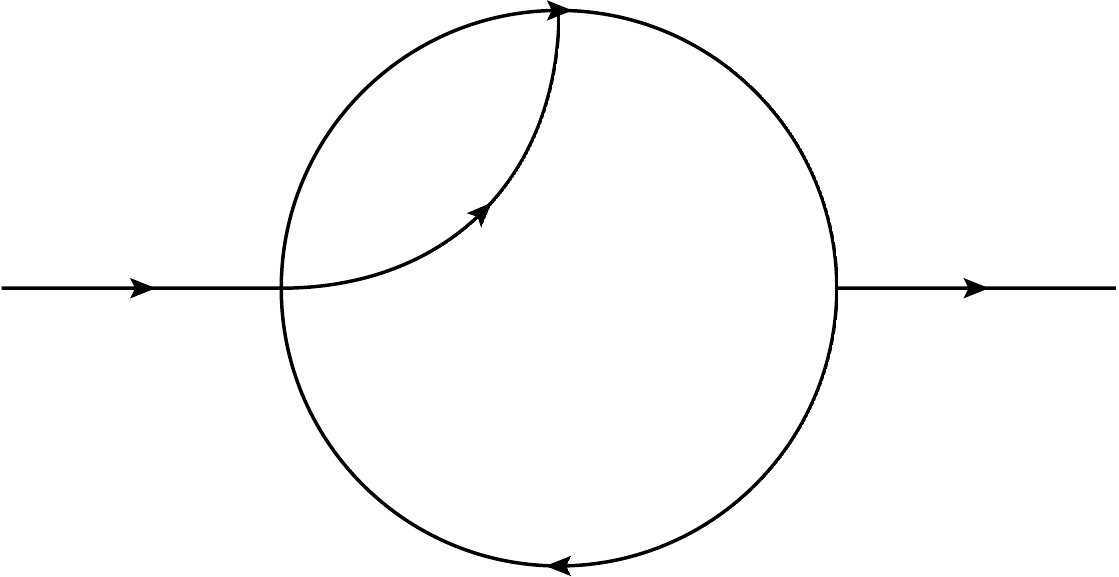}
\caption{Feynman diagram representing the $V$-type two-loop nested integral.}
\label{V_topology}
\end{figure}

The results above can be used to calculate some two-loop integrals. For instance, consider the following integral
\begin{gather}
\begin{split}
&V\left(d\,;n_1\,,0\,;n_2\,,0\,;n_3\,,0\,;n_4\,,0\right) 
\\
&= \frac{1}{\mu^{2(d-4)}} \int \frac{d^dk_1}{(2\pi)^d} \int \frac{d^dk_2}{(2\pi)^d}
\frac{1}{k_1^{2n_1}\left(k_1-q\right)^{2n_2} \left(k_2-q\right)^{2n_3} \left(k_1-k_2\right)^{2n_4}} \,,
\end{split}
\label{V_int_massless_definition}
\end{gather}
which is represented in Fig.~\ref{V_topology}. 

Note that in \eqref{V_int_massless_definition} the momentum $k_2$ only appears in the third and fourth propagators, therefore the $k_2$ integral can be performed using the result~\eqref{B_int_massless_d_dim}. Specifically, if we make the change of variables $\tilde{k}_2=k_2-q$, the $\tilde{k}_2$ integral is proportional to $(k_1-q)^{2\left(\frac{d}{2}-n_3-n_4\right)}$. Because the integrations can be performed in an iterative fashion, these are called nested loop integrals. Finally, the result for~\eqref{V_int_massless_definition} is
\begin{gather}
\begin{split}
&V\left(d\,;n_1\,,0\,;n_2\,,0\,;n_3\,,0\,;n_4\,,0\right) 
\\
&= -\frac{1}{(4\pi)^4} \left(q^2\right)^{4-n_1-n_2-n_3-n_4} 
\left[-\frac{q^2}{4\pi\mu^2}\right]^{d-4}
\frac{\Gamma\left(\frac{d}{2}-n_1\right)\Gamma\left(\frac{d}{2}-n_3\right)\Gamma\left(\frac{d}{2}-n_4\right)}{\Gamma\left(n_1\right)\Gamma\left(n_3\right)\Gamma\left(n_4\right)}
\\
&\frac{\Gamma\left(n_1+n_2+n_3+n_4-d\right)\Gamma\left(d-n_2-n_3-n_4\right)\Gamma\left(n_3+n_4-\frac{d}{2}\right)}{\Gamma\left(n_2+n_3+n_4-\frac{d}{2}\right)\Gamma\left(\frac{3d}{2}-n_1-n_2-n_3-n_4\right)\Gamma\left(d-n_3-n_4\right)} \,.
\end{split}
\label{V_int_massless_d_dim} 
\end{gather}
From the definition~\eqref{V_int_massless_definition} it is clear that this result has the proper mass dimension. Interestingly, the result is also valid for any value of $n_2$, including $n_2=0$. This technique can be extended to integrals containing one massive propagator. For example, consider the integral
\begin{gather}
\begin{split}
&V\left(d\,;n_1\,,m\,;n_2\,,0\,;n_3\,,0\,;n_4\,,0\right) 
\\
& = \frac{1}{\mu^{2(d-4)}} \int \frac{d^dk_1}{(2\pi)^d} \int \frac{d^dk_2}{(2\pi)^d}
\frac{1}{\left(k_1^2-m^2\right)^{n_1}\left(k_1-q\right)^{2n_2} \left(k_2-q\right)^{2n_3} \left(k_1-k_2\right)^{2n_4}} \,.
\end{split}
\label{V_int_one_mass_definition}
\end{gather}
This integral occurs in Chapter~\ref{chapter_4_Qq_diquark}, and also in Ref.~\cite{Jamin_2001_a}. Once again the $k_2$ integral can be integrated immediately, and the resulting $k_1$ integral can be evaluated using~\eqref{B_int_one_mass_d_dim}. Expressing the overall scale dependence in terms of the mass $m$, the result is
\begin{gather}
\begin{split}
&V\left(d\,;n_1\,,m\,;n_2 \,,0\,;n_3\,,0\,;n_4\,,0\right) 
\\
&= -\frac{1}{(4\pi)^4}  \left[-\frac{m^2}{1-z}\right]^{4-n_1-n_2-n_3-n_4}  
\exp{\left[\left(d-4\right)\left(\log\left[\frac{m^2}{4\pi\mu^2}\right]-\log{\left[1-z\right]}\right)\right]} 
\\
&\frac{ \Gamma\left(\frac{d}{2}-n_3\right) \Gamma\left(\frac{d}{2}-n_4\right)\Gamma\left(n_3+n_4-\frac{d}{2}\right)\Gamma\left(d-n_2-n_3-n_4\right)\Gamma\left(n_1+n_2+n_3+n_4-d\right)}{\Gamma\left(n_1\right) \Gamma\left(n_3\right) \Gamma\left(n_4\right)\Gamma\left(\frac{d}{2}\right)\Gamma\left(d-n_3-n_4\right)}
\\
&\phantom{}_2 F_1\left[n_1+n_2+n_3+n_4-d\,,d-n_2-n_3-n_4\,; \frac{d}{2}\,; z\right] \,, 
\end{split}
\label{V_int_one_mass_d_dim}
\end{gather}
where $z$ is as defined in Eq.~\eqref{B_int_one_mass_integral_rep_of_2F1}. Note that this result is valid for all values of $n_2$, and that the dimension of the result~\eqref{V_int_one_mass_d_dim} is the same as that of the integral~\eqref{V_int_massless_definition}. The method that we have used to calculate~\eqref{V_int_one_mass_definition} cannot be extended to integrals with more than one massive propagator. For instance, suppose that the third propagator in~\eqref{V_int_one_mass_definition} contained a mass $m$. In this case, the $k_2$ integral could be evaluated using the one-loop result~\eqref{B_int_one_mass_d_dim}. However, the resulting $k_1$ integral would involve a hypergeometric function in the argument $k_2$, and this integral cannot be evaluated in closed form. A new method is required in order to consider integrals containing external momenta and more than one massive propagator.

\section{Integrals with two massive propagators}
\label{IntsWithTwoMasses}
Consider a generalization of the integral~\eqref{B_int_one_mass_definition} where both propagators are massive
\begin{gather}
B\left(d\,;n_1\,,m\,;n_2\,,m\right) = \frac{1}{\mu^{d-4}} \int \frac{d^dk}{(2\pi)^d} \frac{1}{\left(k^2-m^2\right)^{n_1}\left[\right.\left(k-q\right)^2-m^2\left.\right]^{n_2}} \,.
\label{B_int_two_masses} 
\end{gather}
This integral can be evaluated using Feynman parameters (see, {\it e.g.} Ref.~\cite{Pascual_1984_a}). Instead, we will utilize an approach developed in Ref.~\cite{Davydychev_1990_a,Boos_1990_a}, which makes use of the Mellin-Barnes contour integral representation of the hypergeometric function $\phantom{}_1 F_0$~\eqref{1F0_Mellin_Barnes_representation} to represent a massive propagator in terms of a massless propagator. In this way, massive propagators can be represented in terms of massless propagators, at the cost of introducing a contour integration. The resulting massless integrals can be evaluated in terms of Gamma functions, and then the contour integrals involving these Gamma functions can be evaluated. The result of the contour integration is typically a generalized hypergeometric function whose argument is a dimensionless ratio of the external momentum and the mass.

In order to illustrate the Mellin-Barnes technique, we will use it to evaluate the integral~\eqref{B_int_two_masses}. Applying the Mellin-Barnes representation to the integral~\eqref{B_int_two_masses} gives
\begin{gather}
\begin{split}
B\left(d\,;n_1\,,m\,;n_2\,,m\right) &= \frac{1}{\Gamma\left(n_1\right)\Gamma\left(n_2\right)}  \int_{-i\infty}^{i\infty} \frac{ds}{2\pi i} \int_{-i\infty}^{i\infty} \frac{dt}{2\pi i} \left(-m^2\right)^{s+t} \Gamma\left(-s\right) \Gamma\left(-t\right)
\\
&  \Gamma\left(s+n_1\right) \Gamma\left(t+n_2\right)  \frac{1}{\mu^{d-4}} \int \frac{d^d k}{(2\pi)^d} \frac{1}{k^{2\left(n_1+s\right)} \left(k-q\right)^{2\left(n_2+t\right)}} \,.
\end{split}
\label{B_int_two_masses_MB_rep_1}
\end{gather}
The loop integral can be evaluated immediately using~\eqref{B_int_massless_d_dim}, which gives
\begin{gather}
\begin{split}
B\left(d\,;n_1\,,m\,;n_2\,,m\right) &= \frac{i}{(4\pi)^2}  \left[-\frac{q^2}{4\pi\mu^2}\right]^{\frac{d}{2}-2}
\frac{\left(q^2\right)^{2-n_1-n_2}}{\Gamma\left(n_1\right)\Gamma\left(n_2\right)} \int_{-i\infty}^{i\infty} \frac{ds}{2\pi i} \int_{-i\infty}^{i\infty} \frac{dt}{2\pi i} \left[-\frac{m^2}{q^2}\right]^{s+t} 
\\
&\Gamma\left(-s\right) \Gamma\left(-t\right) \frac{\Gamma\left(\frac{d}{2}-n_1-s\right) \Gamma\left(\frac{d}{2}-n_2-t\right) \Gamma\left(n_1+n_2+s+t-\frac{d}{2}\right)}{\Gamma\left(d-n_1-n_2-s-t\right)} \,.
\end{split}
\label{B_int_two_masses_MB_rep_2}
\end{gather}
Now, making the change of variables $v=s$, $w=\frac{d}{2}-n_1-n_2-s-t$, the integral becomes
\begin{gather}
\begin{split}
B\left(d\,;n_1\,,m\,;n_2\,,m\right) &= \frac{i}{(4\pi)^2}  \left[\frac{m^2}{4\pi\mu^2}\right]^{\frac{d}{2}-2}
 \frac{\left(-m^2\right)^{2-n_1-n_2}}{\Gamma\left(n_1\right)\Gamma\left(n_2\right)} \int_{-i\infty}^{i\infty} \frac{dv}{2\pi i} \int_{-i\infty}^{i\infty} \frac{dw}{2\pi i} \left(-\frac{q}{m^2}\right)^w 
\\
&\Gamma\left(n_1+w+v\right)\Gamma\left(n_1+n_2-\frac{d}{2}+w+v\right)\Gamma\left(-v\right)\Gamma\left(\frac{d}{2}-n_1-v\right)
\\
&\frac{\Gamma\left(-w\right)}{\Gamma\left(\frac{d}{2}+w\right)} \,.
\end{split}
\label{B_int_two_masses_MB_rep_3}
\end{gather}
The contour integral over the variable $v$ can now be evaluated using Barnes' Lemma~\eqref{Barnes_Lemma}. Doing so, the integral becomes
\begin{gather}
\begin{split}
B\left(d\,;n_1\,,m\,;n_2\,,m\right) = \frac{i}{(4\pi)^2}  \left[\frac{m^2}{4\pi\mu^2}\right]^{\frac{d}{2}-2} &
 \frac{\left(-m^2\right)^{2-n_1-n_2}}{\Gamma\left(n_1\right)\Gamma\left(n_2\right)}  \int_{-i\infty}^{i\infty} \frac{dw}{2\pi i} \left(-\frac{q}{m^2}\right)^w \Gamma\left(-w\right) 
\\
&\frac{\Gamma\left(n_1+w\right)\Gamma\left(n_1+n_2-\frac{d}{2}+w\right)\Gamma\left(n_2+w\right)}{\Gamma\left(n_1+n_2+2w\right)} \,.
\end{split}
\label{B_int_two_masses_MB_rep_4}
\end{gather}
The Gamma function in the denominator of~\eqref{B_int_two_masses_MB_rep_4} can be simplified using Eq.~\eqref{gamma_function_duplication_identity}, which gives
\begin{gather}
\begin{split}
B\left(d\,;n_1\,,m\,;n_2\,,m\right) = \frac{i}{(4\pi)^2}  \left[\frac{m^2}{4\pi\mu^2}\right]^{\frac{d}{2}-2} &
 \frac{\left(-m^2\right)^{2-n_1-n_2}}{\Gamma\left(n_1\right)\Gamma\left(n_2\right)}  2^{1-n_1-n_2} \pi^{\frac{1}{2}}  \int_{-i\infty}^{i\infty} \frac{dw}{2\pi i} \left(-\frac{q}{4m^2}\right)^w  
\\
&\Gamma\left(-w\right)\frac{\Gamma\left(n_1+w\right)\Gamma\left(n_1+n_2-\frac{d}{2}+w\right)\Gamma\left(n_2+w\right)}{\Gamma\left(\frac{1}{2}\left(n_1+n_2\right)+w\right)\Gamma\left(\frac{1}{2}\left(n_1+n_2+1\right)+w\right)} \,.
\end{split}
\label{B_int_two_masses_MB_rep_5}
\end{gather}
Note that the remaining contour integral can be evaluated in terms of the generalized hypergeometric function $\phantom{}_3 F_2$. The result is
\begin{gather}
\begin{split}
B\left(d\,;n_1\,,m\,;n_2\,,m\right) = \frac{i}{(4\pi)^2} \left(-m^2\right)^{2-n_1-n_2} & \left[\frac{m^2}{4\pi\mu^2}\right]^{\frac{d}{2}-2}
\frac{\Gamma\left(n_1+n_2-\frac{d}{2}\right)}{\Gamma\left(n_1+n_2\right)} 
\\
%
&\phantom{}_3 F_{2}
\left[
\begin{array}{c|}
n_1 \,,  n_2 \,,  n_1 + n_2 -\frac{d}{2} \\
\frac{1}{2}\left(n_1+n_2\right) \,, \frac{1}{2}\left(n_1+n_2+1\right)
\end{array} \, \frac{q^2}{4m^2} \right] \,.
\end{split}
\label{B_int_two_masses_d_dim}
\end{gather}
which agrees with an expression given for this integral in Ref.~\cite{Boos_1990_a}. Note that this technique can be used to calculate integrals such as~\eqref{B_int_two_masses} where the masses in each propagator are distinct. In Ref.~\cite{Boos_1990_a} this is done, and the results are given in terms of multivariable generalized hypergeometric functions functions. However, these functions are somewhat unwieldy and are not widely implemented in computer algebra systems. Fortunately, all of the integrals in Chapters~\ref{chapter_3_hybrids}, \ref{chapter_4_Qq_diquark}, \ref{chapter_5_diquark_renorm} and \ref{chapter_6_mixing} can be evaluated in terms of only two scales, the external momentum $q$ and the heavy quark mass $m$.

\begin{figure}[htb]
\centering
\includegraphics[scale=0.5]{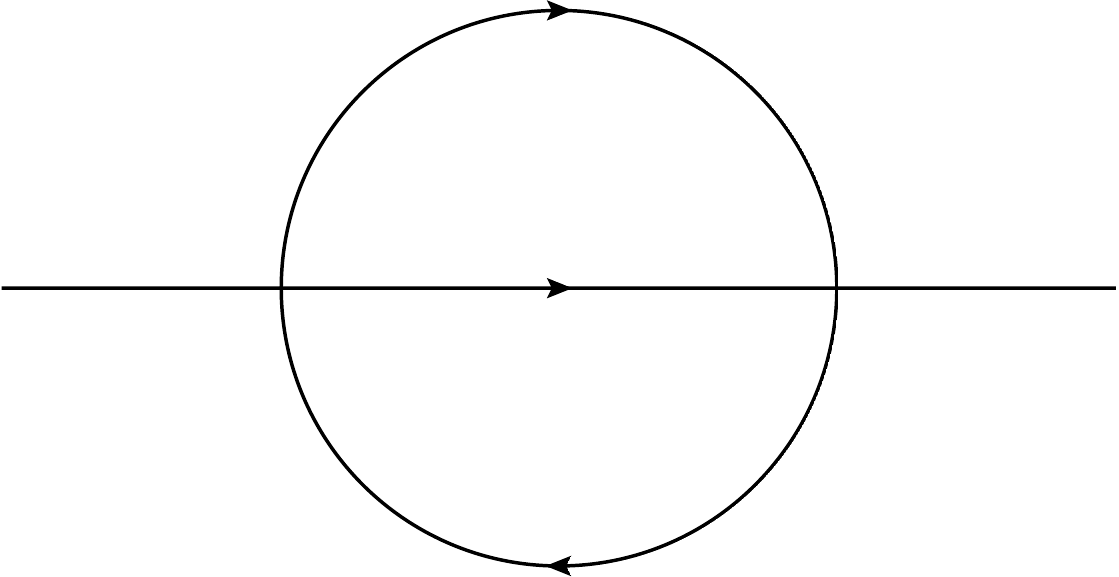}
\caption{Feynman diagram representing the $J$-type two-loop sunset integral.}
\label{J_topology}
\end{figure}

Consider now a two-loop integral with two massive propagators
\begin{gather}
\begin{split}
&J\left(d\,;n_1\,,m\,;n_2\,,m\,;n_3\,,0\right)
\\
& = \frac{1}{\mu^{2(d-4)}} \int \frac{d^dk_1}{(2\pi)^d} \int \frac{d^dk_2}{(2\pi)^d} \frac{1}{\left(k_1^2-m^2\right)^{n_1}\left[\right.\left(k_2-q\right)^2-m^2\left.\right]^{n_2} \left(k_1-k_2\right)^{2n_3} } \,.
\end{split}
\label{J_int_two_masses}
\end{gather}
This integral occurs in Chapter~\ref{chapter_3_hybrids}, and is represented by the Feynman diagram in Fig.~\ref{J_topology}. 

This integral can also be evaluated using Mellin-Barnes techniques. The result is
\begin{gather}
\begin{split}
J\left(d\,;n_1\,,m\,;n_2\,,m\,;n_3\,,0\right) &= -\frac{1}{(4\pi)^4} \left(-m^2\right)^{4-n_1-n_2-n_3} \left[\frac{m^2}{4\pi\mu^2}\right]^{d-4}
\\
&\frac{\Gamma\left(\frac{d}{2}-n_3\right)\Gamma\left(n_2+n_3-\frac{d}{2}\right)\Gamma\left(n_1+n_3-\frac{d}{2}\right)\Gamma\left(n_1+n_2+n_3-d\right)}{\Gamma\left(\frac{d}{2}\right)\Gamma\left(n_1\right)\Gamma\left(n_2\right)\Gamma\left(n_1+n_2+2n_3-d\right)}
\\
&\phantom{}_4 F_{3}
\left[
\begin{array}{c|}
n_1+n_2+n_3-d \,,  n_2+n_3-\frac{d}{2} \,,  n_1+n_3-\frac{d}{2} \,,   n_3 \\
 \frac{d}{2} \,,  n_3+\frac{1}{2}\left(n_1+n_2-d\right) \,,  n_3+\frac{1}{2}\left(n_1+n_2-d+1\right)
\end{array} \, \frac{q^2}{4m^2} \right] \,.
\end{split}
\label{J_int_two_masses_d_dim} 
\end{gather}
Note that this result agrees with results given for this integral in Ref.~\cite{Broadhurst_1993_a}. This is the most complicated integral that we will evaluate in this chapter. In the next section we will consider a technique that can be used to construct recurrence relations among loop integrals. Using this approach, all integrals encountered in Chapters~\ref{chapter_3_hybrids}, \ref{chapter_4_Qq_diquark}, \ref{chapter_5_diquark_renorm} and \ref{chapter_6_mixing} can be evaluated in terms of the integrals given so far in this chapter.

\section{Integration By Parts}
\label{IntegrationByParts}

\begin{figure}[htb]
\centering
\includegraphics[scale=0.5]{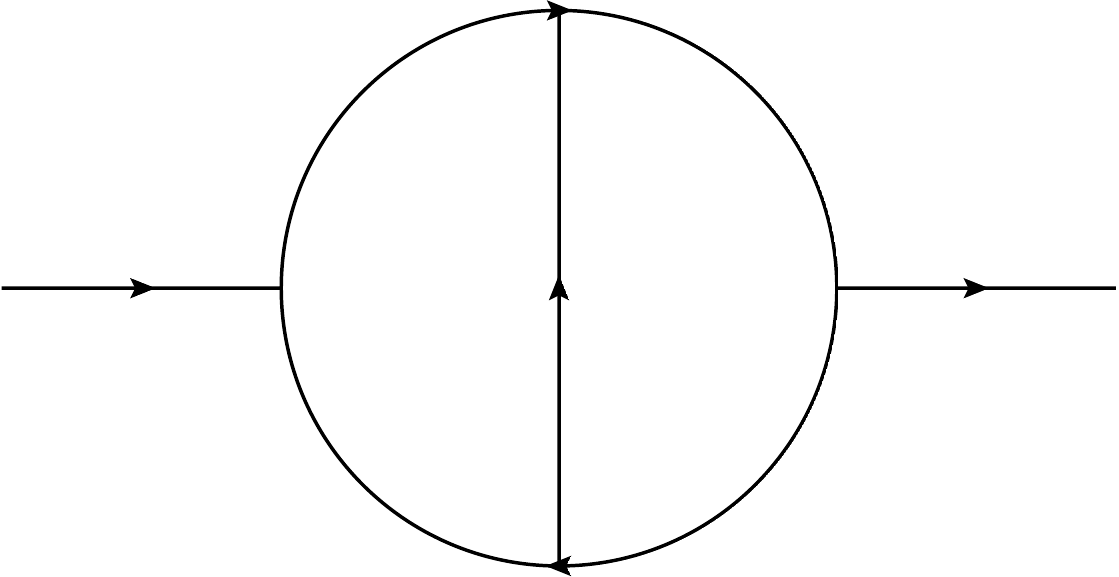}
\caption{Feynman diagram representing the $F$-type two-loop master integral.}
\label{F_topology}
\end{figure}

Consider the two-loop integral
\begin{gather}
\begin{split}
&F\left(d\,;n_1\,,m\,;n_2\,,m\,;n_3\,,0\,;n_4\,,0\,;n_5\,,0\right) 
\\
&= \frac{1}{\mu^{2(d-4)}} \int \frac{d^d k_1}{(2\pi)^d} \int \frac{d^d k_2}{(2\pi)^d} \frac{1}{\left(k_1^2-m^2\right)^{n_1} \left(k_2^2-m^2\right)^{n_2} \left(k_1-q\right)^{2n_3} \left(k_2-q\right)^{2n_4} \left(k_1-k_2\right)^{2n_5} } \,.
\end{split}
\label{F_int_massive_defn} 
\end{gather}
This integral is represented by the Feynman diagram in Fig.~\ref{F_topology}. Note that this is the most complex loop integral that can occur when calculating two-point functions at two-loop level, because at most five independent propagators can be constructed from the two loop momenta $k_1$, $k_2$, and the external momentum $q$. For this reason, the integral~\eqref{F_int_massive_defn} is occasionally referred to as the ``master'' two-loop integral for two-point functions~\cite{Broadhurst_1993_a}.

This integral cannot be evaluated using the methods discussed so far. However, it can be evaluated by appealing to one of the properties of dimensionally regularized momentum integrals. Specifically, loop integrals satisfy the relation
\begin{gather}
\frac{1}{\mu^{d-4}} \int \frac{d^d k}{(2\pi)^d} \frac{\partial}{\partial v_\mu} \left[w_\mu f\left(k\,,q\,,m\right)\right] = 0 \,, \quad v_\mu = k_\mu \,, \quad w_\mu \in \left\{k_\mu \,, \left(k \pm q\right)_\mu \right\} \,,
\label{integration_by_parts_definition}
\end{gather}
which is proven in Ref.~\cite{Collins_1984_a}. This identity is commonly referred to as integration by parts. The identity generalizes to multi-loop integrals, the restriction being that $v_\mu$ must be one of the loop momenta. The technique was developed in Ref.~\cite{Chetyrkin_1981_a} and used to calculate the renormalization group functions in scalar $\phi^4$ theory to four-loop order. In order to illustrate the technique, we will use it to calculate the integral~\eqref{F_int_massive_defn}. Using the identity~\eqref{integration_by_parts_definition}, we have
\begin{gather}
\int \frac{d^d k_1}{(2\pi)^d} \int \frac{d^d k_2}{(2\pi)^d} \frac{\partial}{\partial k_1^\mu} \left[ \frac{\left(k_1-k_2\right)^\mu}{\left(k_1^2-m^2\right)^{n_1} \left(k_2^2-m^2\right)^{n_2} \left(k_1-q\right)^{2n_3} \left(k_2-q\right)^{2n_4} \left(k_1-k_2\right)^{2n_5}}\right] = 0 \,,
\label{F_int_massive_IBP_identity} 
\end{gather}
where the derivatives can be calculated using the result
\begin{gather}
\frac{\partial}{\partial k^\mu} \frac{1}{k^{2n}} = -\frac{2nk^\mu}{k^{2(n+1)}} \,.
\end{gather}
Explicitly calculating the derivatives, we arrive at the following recurrence relation
\begin{gather}
\begin{split}
&\left(d-n_1-n_3-2n_5\right)  F\left(d\,;n_1\,,m\,;n_2\,,m\,;n_3\,,0\,;n_4\,,0\,;n_5\,,0\right)
\\
&= \left[n_1\left(\mathbf{5^-} - \mathbf{2^-}\right)\mathbf{1^+} +n_3\left(\mathbf{5^-} - \mathbf{4^-}\right)\mathbf{3^+}\right] F\left(d\,;n_1\,,m\,;n_2\,,m\,;n_3\,,0\,;n_4\,,0\,;n_5\,,0\right) \,.
\end{split}
\label{recurrence_relation_1}
\end{gather}
For convenience we have defined the operators $\mathbf{N^{+}}$ and $\mathbf{N^{-}}$ that increase and decrease the index $n_i$ by one unit when acting on the integral~\eqref{F_int_massive_defn}. Essentially, equation~\eqref{recurrence_relation_1} is a recurrence relation for integrals of the form~\eqref{F_int_massive_defn}. For instance, consider an integral of the form~\eqref{F_int_massive_defn} where each $n_i=1$. Inserting this into the recurrence relation~\eqref{recurrence_relation_1}, we find
\begin{gather}
\begin{split}
F\left(d\,;1\,,m\,;1\,,m\,;1\,,0\,;1\,,0\,;1\,,0\right) = 
\frac{1}{d-4} & \left[  F\left(d\,;2\,,m\,;1\,,m\,;1\,,0\,;1\,,0\,;0\,,0\right) \right.
\\
&- F\left(d\,;2\,,m\,;0\,,m\,;1\,,0\,;1\,,0\,;1\,,0\right) 
\\
&+F\left(d\,;1\,,m\,;1\,,m\,;2\,,0\,;1\,,0\,;0\,,0\right) 
\\
&\left.- F\left(d\,;1\,,m\,;1\,,m\,;2\,,0\,;0\,,0\,;1\,,0\right) \right] \,.
\label{recurrence_relation_example}
\end{split}
\end{gather}
This is a very important result. The integral on the left hand side of Eq.,~\eqref{recurrence_relation_example} cannot be evaluated using results given so far. However, the integrals on the right hand side can be calculated using~\eqref{B_int_massless_d_dim} and~\eqref{V_int_massless_d_dim}. The recurrence relation allows us to calculate an integral that cannot be calculated directly by expressing it as a linear combination of integrals that we can evaluate. This is the power of the integration by parts method: integrals that are incalculable by themselves can be expressed in terms of calculable integrals. Several unique recurrence relations can be developed by choosing different vectors $v^\mu$ and $w^\mu$ in~\eqref{integration_by_parts_definition}. With the aid of~\eqref{recurrence_relation_example}, all integrals that occur in the heavy-light diquark calculation in Chapter~\ref{chapter_4_Qq_diquark} can be evaluated in terms of results given in this chapter. In addition, we have used these methods to 
reproduce the result given in Ref.~\cite{Jamin_2001_a} for the next-to-leading order heavy-light pseudoscalar meson correlation function.

\section{Generalized Recurrence Relations}
\label{GeneralizedRecurrenceRelations}

Using the integration by parts technique, we can construct many recurrence relations for a certain class of loop integrals. In practice, it is often enough to use recurrence relations to calculate unknown integrals in terms of known integrals, as we did in order to calculate the integral~\eqref{F_int_massive_defn} above. However, the question arises, given all possible recurrence relations, can we determine a minimal set of basis integrals from which all others can be calculated? As we shall see, this is possible, although we must discuss some technicalities first.

So far, we have only considered scalar integrals. However, in order to determine a minimal set of basis integrals, we must first consider how to deal with integrals that include tensors composed of the loop momenta. These can always be dealt with by appealing to the Lorentz invariant nature of loop integrals. A simple example is as follows:
\begin{gather}
\frac{1}{\mu^{2(d-4)}} \int \frac{d^d k_1}{\left(2\pi\right)^d} \int \frac{d^d k_2}{\left(2\pi\right)^d} \frac{k_2^\mu}{k_2^2 \left(k_1-q\right)^2 \left(k_1-k_2\right)^2} \equiv q^\mu f\left(q^2\right) \,.
\label{example_of_irreducible_numerator}
\end{gather}
Due to Lorentz invariance, the integral must be proportional to the external momentum $q^\mu$ and a function of the Lorentz scalar $q^2$. Contracting both sides of \eqref{example_of_irreducible_numerator} with $q_\mu$, the function $f\left(q^2\right)$ can be determined and hence the integral on the left hand side of \eqref{example_of_irreducible_numerator} can be calculated. This technique can be easily generalized to accommodate more complicated tensor structures. Ref.~\cite{Passarino_1978_a} constructs an algorithm to solve this problem in general. However, we will use a far more powerful method to deal with tensor integrals such as~\eqref{example_of_irreducible_numerator}.

The method that we will utilize was developed in Ref.~\cite{Tarasov_1996_a} and relates tensor integrals in $d$-dimensions to scalar integrals in $d+2N$-dimensions. In Ref.~\cite{Tarasov_1997_a} the method was applied to two-point propagator-type integrals, that is, integrals of the type $A$, $B$, $J$, $V$ and $F$ that we have considered so far. This method can be used to find a truly minimal set of basis integrals for the family of two-loop two-point integrals that we have studied so far. The key idea of the method is that it is possible to express an arbitrary tensor integral as
\begin{gather}
\int \frac{d^d k_1}{\left(2\pi\right)^d} \int \frac{d^d k_2}{\left(2\pi\right)^d} \frac{ k_1^{\mu_1}\ldots k_1^{\mu_r} k_2^{\lambda_1}\ldots k_2^{\mu_s} }{c_1^{n_1} c_2^{n_2} c_3^{n_3} c_4^{n_4} c_5^{n_5}}
\equiv 
T^{\mu_1 \ldots \mu_r \lambda_1 \ldots \lambda_s}  \int \frac{d^d k_1}{\left(2\pi\right)^d} \int \frac{d^d k_2}{\left(2\pi\right)^d} \frac{1}{c_1^{n_1} c_2^{n_2} c_3^{n_3} c_4^{n_4} c_5^{n_5}} \,,
\\
c_1 = k_1^2 - m_1^2\,, \quad c_2 = k_2^2 - m_2^2 \,, \quad c_3 = \left(k_1-q\right)^2 - m_3^2\,, 
\\
c_4 = \left(k_2-q\right)^2 - m_4^2 \,, \quad c_5 = \left(k_1-k_2\right)^2 - m_5^2 \,.
\label{T_operator_definition}
\end{gather}
where we have omitted the renormalization scale for brevity. The tensor operator $T^{\mu_1 \ldots \mu_r \lambda_1 \ldots \lambda_s}$ is a function of the external momentum $q^\mu$, derivatives with respect to each mass $m_i$, and an operator $\mathbf{d^+}$ that increases the dimension of any given loop integral by two units, \textit{i.e.} $\mathbf{d^+}I^{(d)}=I^{(d+2)}$. We will now construct the explicit form of this operator. We will follow the derivation given in Ref.~\cite{Tarasov_1997_a}, although we will retain the loop integral normalization convention that has been used in this chapter. Introducing the auxiliary vectors $a_1$ and $a_2$, we can write
\begin{gather}
T_{\mu_1 \ldots \mu_r \lambda_1 \ldots \lambda_s} = \left. \frac{1}{i^{r+s}} \frac{\partial}{\partial a_1^{\mu_1}}\ldots\frac{\partial}{\partial a_1^{\mu_r}} \frac{\partial}{\partial a_2^{\lambda_1}}\ldots\frac{\partial}{\partial a_2^{\lambda_s}}  \exp{\left[i\left(a_1\cdot k_1 + a_2\cdot k_2\right)\right]} \right|_{a_1=a_2=0} \,.
\label{exponential_form_of_T_operator} 
\end{gather}
Now, consider the integral
\begin{gather}
G^{(d)}\left(q^2\right) =  \int \frac{d^d k_1}{\left(2\pi\right)^d} \int \frac{d^d k_2}{\left(2\pi\right)^d}
\frac{\exp{\left[i\left(a_1\cdot k_1 + a_2\cdot k_2\right)\right]}}{c_1^{n_1} c_2^{n_2} c_3^{n_3} c_4^{n_4} c_5^{n_5} } \,.
\label{G_int_definition}
\end{gather}
To evaluate this integral we will use the following identity~\cite{Smirnov_2004_a,Tarasov_1997_a}
\begin{gather}
\frac{1}{\left(k^2-m^2\right)^n} = \frac{1}{i^n\,\Gamma\left(n\right)} \int_0^\infty d\alpha \, \alpha^{n-1} \exp\left[i\alpha\left(k^2-m^2\right)\right] \,.
\label{alpha_parameters} 
\end{gather}
The integration variable $\alpha$ is called an alpha parameter and serves a similar purpose to the Feynman parameters introduced earlier. The resulting $k_1$ and $k_2$ loop integrals can be evaluated using the integral~\cite{Tarasov_1997_a}
\begin{gather}
\int d^d k \exp{\left[i\left(Ak^2+2q\cdot k\right)\right]} = i \left[ \frac{\pi}{iA} \right]^{\frac{d}{2}} \exp{\left[-\frac{iq^2}{A}\right]} \,.
\label{d_dimensional_gaussian_integral} 
\end{gather}
Doing this, the result is
\begin{gather}
G^{(d)}\left(q^2\right) = \frac{i^{2-d}}{(4\pi)^d}  \prod_{i=1}^{5} \frac{1}{i^{n_i}\,\Gamma\left(n_i\right)} \int_0^\infty \frac{d\alpha_i}{\left[D\left(\alpha\right)\right]^\frac{d}{2}} \alpha_i^{n_i-1} \exp{\left[i\left(\frac{Q\left(\alpha_i\,,a_1\,,a_2\right)}{D\left(\alpha\right)}-\sum_{j=1}^5 \alpha_j m_j^2 \right)\right]} \,.
\label{G_int_result}
\end{gather}
The functions $D\left(\alpha\right)$ and $Q\left(\alpha\,,a_1\,,a_2\right)$ are
\begin{gather}
D\left(\alpha\right) = \alpha_5 \left(\alpha_1+\alpha_2+\alpha_3+\alpha_4\right)+\left(\alpha_1+\alpha_3\right)\left(\alpha_2+\alpha_4\right) \,,
\end{gather}
\begin{gather}
\begin{split}
Q\left(\alpha_i\,,a_1\,,a_2\right) &= \left[\left(\alpha_1+\alpha_2\right)\left(\alpha_3+\alpha_4\right)\alpha_5+\alpha_1\alpha_2\left(\alpha_3+\alpha_4\right)+\alpha_3\alpha_4\left(\alpha_1+\alpha_2\right)\right]q^2
\\
&+\left(a_1\cdot q\right)Q_1 + \left(a_2\cdot q\right)Q_2 + a_1^2 Q_{11}^2 + a_2^2 Q_{22}^2 + \left(a_1 \cdot a_2\right) Q_{12} \,, 
\end{split}
\end{gather}
\begin{gather}
\begin{split}
&Q_1 = \alpha_3\alpha_5 + \alpha_4\alpha_5 + \alpha_2\alpha_3 + \alpha_3\alpha_4 \,, \quad Q_2 = \alpha_4\alpha_5 + \alpha_3\alpha_5 + \alpha_1\alpha_4 + \alpha_3\alpha_4 \,, 
\\
&Q_{11} = -\frac{1}{4}\left(\alpha_2+\alpha_4+\alpha_5\right) \,, \quad Q_{22} = -\frac{1}{4}\left(\alpha_1+\alpha_3+\alpha_5\right) \,, \quad Q_{12} = -\frac{1}{2} \alpha_5 \,.
\end{split}
\end{gather}
The explicit form of the operator $T^{\mu_1 \ldots \mu_r \lambda_1 \ldots \lambda_s}$ can be determined by taking derivatives of~\eqref{G_int_result} with respect to $a_1$, $a_2$, and then setting $a_1=a_2=0$. For instance, consider the operator corresponding to the tensor structure $k_1^\mu$:
\begin{gather}
\begin{split}
T_\mu &= \left. \frac{1}{i}\frac{\partial}{\partial a_1^\mu} G^{(d)}\left(q^2\right) \right|_{a_1=a_2=0} 
= \left.  q_\mu \left[ \alpha_3\alpha_5 + \alpha_4\alpha_5 + \alpha_2\alpha_3 + \alpha_3\alpha_4 + \ldots \right] \frac{G^{(d)}\left(q^2\right)}{D\left(\alpha\right)} \right|_{a_1=a_2=0} \,,
\label{T_operator_example_1}
\end{split}
\end{gather}
where the ellipses indicate additional terms that vanish when $a_1$ and $a_2$ are set to zero. Notice that each factor of $\alpha_i$ multiplying $G^{(d)}\left(q^2\right)$ is proportional to a derivative of $G^{(d)}\left(q^2\right)$ with respect to $m_i^2$. In addition, we can absorb the factor of $D\left(\alpha\right)$ in~\eqref{T_operator_example_1} into $G^{(d)}\left(q^2\right)$, effectively sending $d \to d+2$ in~\eqref{G_int_result}. Finally, note that when $a_1$ and $a_2$ are set to zero in $G^{(d)}\left(q^2\right)$ ({\it i.e.}\, in Eq.,~\eqref{G_int_definition}), what remains is the loop integral that the $T$ operator acts upon in Eq.,~\eqref{T_operator_definition}. Therefore, the $T$ operator corresponding to the tensor $k_1^\mu$ is given by
\begin{gather}
\begin{split}
T_\mu = -\left(4\pi\right)^2 q_\mu & \left[
\left(i\frac{\partial}{\partial m_3^2}\right)\left(i\frac{\partial}{\partial m_5^2}\right) + 
\left(i\frac{\partial}{\partial m_4^2}\right)\left(i\frac{\partial}{\partial m_5^2}\right) 
\right. 
\\
& \left. +
\left(i\frac{\partial}{\partial m_2^2}\right)\left(i\frac{\partial}{\partial m_3^2}\right) + 
\left(i\frac{\partial}{\partial m_3^2}\right)\left(i\frac{\partial}{\partial m_4^2}\right) 
\right] \mathbf{d^+} \,,
\label{T_operator_example_2}
\end{split}
\end{gather}
Generalizing this result, the $T$ operator corresponding to the tensor $k_1^{\mu_1}\ldots k_1^{\mu_r} k_2^{\lambda_1}\ldots k_2^{\lambda_s}$ is given by
\begin{gather}
\begin{split}
&T_{\mu_1 \ldots \mu_r \lambda_1 \ldots \lambda_s}  \left(q^\mu\,,\partial_k\,,\mathbf{d^+}\right)
= \frac{1}{i^{r+s}} \prod_{i=1}^r \frac{\partial}{\partial a_1^{\mu_i}} \prod_{j=1}^s \frac{\partial}{\partial a_2^{\lambda_j}} 
\\
&\left.\exp{\Biggl[i\left( Q_1\left(a_1\cdot q\right) + Q_2\left(a_2\cdot q\right) + Q_{11}a_1^2 + Q_{22}a_2^2 + Q_{12}\left(a_1\cdot a_2\right) \right)\rho\Biggr]} \right| {\begin{array}{l} a_k=0 \\ \alpha_k = i\partial_k \end{array}} \,,
\\
&\partial_k = \frac{\partial}{\partial m^2_k} \,, \quad \rho=-\left(4\pi\right)^2\mathbf{d^+} \,.
\end{split}
\label{T_operator_result}
\end{gather}

Using $T$ operators, any $d$-dimensional tensor integral which is of the same form as~\eqref{T_operator_definition} can be expressed as a linear combination of scalar integrals in $d+2N$-dimensions. Therefore, without loss of generality we can focus entirely on scalar integrals. Although this is a very helpful result, it is not the most important use of the $T$ operators. Notice that the loop integral recurrence relations derived from integration by parts identities~\eqref{integration_by_parts_definition} can only alter the indices $n_i$ of a given loop integral, and cannot change the number of dimensions $d$. However, $T$ operators effectively lead to loop integral recurrence relations in the number of dimensions. These two distinct forms of recurrence relations can be combined, creating generalized recurrence relations that shift not only the indices $n_i$ of loop integrals, but also the dimension $d$ of the loop integrals. In Ref.~\cite{Tarasov_1997_a} generalized 
recurrence relations are developed and used to determine a minimal set of basis integrals for the family of integrals that includes the $A$, $B$, $J$, $V$, and $F$ type integrals that we have considered so far. These generalized recurrence relations have been implemented in the Mathematica package Tarcer~\cite{Mertig_1998_a}. This package was utilized in the heavy quarkonium hybrid calculations in Chapter~\ref{chapter_3_hybrids}, and in the heavy-light diquark calculation in Chapter~\ref{chapter_4_Qq_diquark}. After using Tarcer, all of the Wilson coefficients in the hybrid calculations can be expressed in terms of the following set of integrals: $A\left(d\,;1\,,m\right)$\eqref{A_int_d_dim}, $B\left(d\,;1\,,m\,;1\,,m\right)$~\eqref{B_int_two_masses_d_dim}, $J\left(d\,;2\,,m\,;1\,,m\,;1\,,0\right)$ and $J\left(d\,;1\,,m\,;1\,,m\,;1\,,0\right)$~\eqref{J_int_two_masses_d_dim}. Similarly, the integrals required in the heavy-light diquark calculation are $A\left(d\,;1\,,m\right)$~\eqref{A_int_d_dim}, $B\left(d\,;
1\,,m\,;1\,,0\right)$~\eqref{B_int_one_mass_d_dim}, $J\left(d\,;2\,,m\,;1\,,0\,;1\,,0\right)$ and $J\left(d\,;1\,,m\,;1\,,0\,;1\,,0\right)$. Note that the last two of these can be calculated using~\eqref{V_int_one_mass_d_dim}. Tarcer can be applied to any two-loop calculation, with any combination of masses.

\section{The Epsilon Expansion}
\label{EpsilonExpansion}

We now have explicit $d$-dimensional results for all of the loop integrals that are encountered in Chapters~\ref{chapter_3_hybrids}, \ref{chapter_4_Qq_diquark}, \ref{chapter_5_diquark_renorm} and \ref{chapter_6_mixing}. Now we shall see that divergent integrals can be regulated by setting $d=4+2\epsilon$ and expanding around $\epsilon=0$. First we will consider the massive tadpole integral~\eqref{A_int_d_dim}. Setting $n=1$, $d=4+2\epsilon$ and using the properties of the Gamma function~\eqref{expansion_of_Gamma_z}, this integral gives
\begin{gather}
\begin{split}
A\left[4+2\epsilon\,;1\right] &= -\frac{im^2}{\left(4\pi\right)^2} \left[\frac{m^2}{4\pi\mu^2}\right]^\epsilon \Gamma\left(-1-2\epsilon\right) 
\\
&= -\frac{im^2}{\left(4\pi\right)^2} \exp{\left(\epsilon\log{\left[\frac{m^2}{4\pi\mu^2}\right]}\right)} \left(\frac{1}{\epsilon}+\gamma_E - 1 +\mathcal{O}\left(\epsilon\right)\right) 
\\
&= -\frac{im^2}{\left(4\pi\right)^2} \left[\frac{1}{\epsilon}+\gamma_E-\log\left(4\pi\right)-1+\log{\left[\frac{m^2}{\mu^2}\right]}\right] \,.
\end{split}
\label{A_int_expansion}
\end{gather}
Notice that the integral is divergent, and the divergence is parametrized as a simple pole at $\epsilon=0$. Ultimately, this divergence can be traced back to the radial integral~\eqref{A_int_spherical}. For $\ell \gg m$, the integral goes like $\ell^{d-2n}$, and hence is divergent for $d > 2n$. When this is the case, the radial integral is ill-defined. However, the expression in terms of the Gamma function~\eqref{A_int_d_dim} uniquely defines the loop integral when $d > 2n$. 

The divergence that arises in~\eqref{A_int_expansion} can be canceled through renormalization, which was discussed in Chapter~\ref{chapter_1_intro}. Recall that in the $\overline{\rm MS}$ scheme the renormalization constants are defined so that $\gamma_E$ and $\log{(4\pi)}$ terms are canceled in addition to poles at $\epsilon=0$. A convenient way of partially implementing this to rescale $\mu^2 \to \mu^2  e^{-\gamma_E}/4\pi$. Doing this, all one-loop integrals will contain a factor of
\begin{gather}
\left[\frac{M^2}{4\pi\mu^2}\right]^{\frac{d-4}{2}} \equiv \exp{\left(\frac{d-4}{2}\left[\log{\left[\frac{M^2}{\mu^2}\right]-\gamma_E}\right]\right)} \,,
\label{MS_bar_normalization}
\end{gather}
where $M^2$ is the external scale, {\it i.e.}\, the mass $m^2$ or the external momentum $-q^2$. The benefit of this replacement is that it automatically cancels all factors of $\gamma_E$ and $\log{(4\pi)}$ that would otherwise emerge when performing the epsilon expansion. This replacement easily generalizes to multi-loop integrals: an $n$-loop integral would have $n$ factors of~\eqref{MS_bar_normalization}. All calculations in Chapters \ref{chapter_3_hybrids}, \ref{chapter_4_Qq_diquark}, \ref{chapter_5_diquark_renorm} and \ref{chapter_6_mixing} use the $\overline{\rm MS}$ renormalization scheme, therefore all loop integrals encountered there implicitly include~\eqref{MS_bar_normalization}. 

Now let us consider a more complicated example. For example, consider the integral
\begin{gather}
\begin{split}
&F\left(d\,;1\,,0\,;1\,,0\,;1\,,0\,;1\,,0\,;1\,,0\right)
\\
&=\frac{1}{\mu^{2(d-4)}} \int \frac{d^d k_1}{(2\pi)^d} \int \frac{d^d k_2}{(2\pi)^d} \frac{1}{k_1^2 \, k_2^2 \left(k_1-q\right)^2 \left(k_2-q\right)^2 \left(k_1-k_2\right)^2} \,,
\end{split}
\label{F_int_massless_definition}
\end{gather}
which corresponds to the integral~\eqref{F_int_massive_defn} with $m=0$. This integral occurs in Chapter~\ref{chapter_6_mixing}. Setting the mass to zero does not change the recurrence relation~\eqref{recurrence_relation_1}, and therefore the integral above can be written as
\begin{gather}
\begin{split}
F\left(d\,;1\,,0\,;1\,,0\,;1\,,0\,;1\,,0\,;1\,,0\right) = 
\frac{1}{d-4} &\left[F\left(d\,;2\,,0\,;1\,,0\,;1\,,0\,;1\,,0\,;0\,,0\right) \right.
\\
&-F\left(d\,;2\,,0\,;0\,,0\,;1\,,0\,;1\,,0\,;1\,,0\right) 
\\
&+F\left(d\,;1\,,0\,;1\,,0\,;2\,,0\,;1\,,0\,;0\,,0\right) 
\\
&\left.- F\left(d\,;1\,,0\,;1\,,0\,;2\,,0\,;0\,,0\,;1\,,0\right) \right] \,.
\label{recurrence_relation_example_massless}
\end{split}
\end{gather}
The integrals on the right hand side can be calculated using \eqref{B_int_massless_d_dim} and \eqref{V_int_massless_d_dim}. Setting $d=4+2\epsilon$ and expanding using the properties of the Gamma function to perform the expansion, we find
\begin{gather}
F\left(d\,;1\,,0\,;1\,,0\,;1\,,0\,;1\,,0\,;1\,,0\right) = -\frac{1}{(4\pi)^4} \frac{1}{q^2} \left[-\frac{q^2}{4\pi\mu^2}\right]^\epsilon \zeta\left(3\right) \,,
\label{F_int_massless_result}
\end{gather}
where $\zeta$ denotes the Riemann Zeta function. This integral is calculated in Ref.~\cite{Pascual_1984_a} using position space methods~\cite{Chetyrkin_1980_a}, and is in complete agreement with~\eqref{F_int_massless_result}. Note that loop integrals that do not involve massive propagators can always be expressed in terms of Gamma functions which can be expanded easily using any computer algebra system.

As we have seen, integrals that involve an external momentum and massive propagators tend to lead to hypergeometric functions whose indices are $d$-dependent. Therefore, after setting $d=4+2\epsilon$ we are required to expand around $\epsilon=0$ in the indices of a hypergeometric function, which can often be a non-trivial task. For example, consider the integral 
\begin{gather}
\begin{split}
&V\left(d\,;1\,,m\,;1\,,0\,;1\,,0\,;1\,,0\right) 
\\
&= \frac{1}{\mu^{2(d-4)}} \int \frac{d^d k_1}{(2\pi)^d} \int \frac{d^d k_2}{(2\pi)^d} \frac{1}{\left(k_1^2-m^2\right) \left(k_1-q\right)^2 \left(k_2-q\right)^2 \left(k_1-k_2\right)^2} 
\end{split}
\label{V_int_one_mass_expansion_1} 
\end{gather}
Using the result~\eqref{V_int_one_mass_d_dim}, setting $d=4+2\epsilon$, and working in the $\overline{\rm MS}$ scheme, we find
\begin{gather}
\begin{split}
V\left(d\,;1\,,m\,;1\,,0\,;1\,,0\,;1\,,0\right) &= -\frac{1}{(4\pi)^4} \exp{\left[2\epsilon\left(\log{\left[\frac{m^2}{\mu^2}\right]-\gamma_E-\log{\left[1-z\right]}}\right)\right]} 
\\
&\frac{\Gamma^2\left(1+\epsilon\right)\Gamma\left(-\epsilon\right)\Gamma\left(1+2\epsilon\right)\Gamma\left(-2\epsilon\right)}{\Gamma\left(2+2\epsilon\right)\Gamma\left(2+\epsilon\right)}
\\
&\phantom{}_2 F_{1} \left[ -2\epsilon \,,  1+2\epsilon \,; 2+\epsilon \,; z \right] \,, 
\quad z=\frac{1}{1-\frac{m^2}{q^2}} \,.
\end{split}
\label{V_int_one_mass_expansion_2} 
\end{gather}
In order to expand the hypergeometric function, it is useful to write it in series form
\begin{gather}
\begin{split}
V\left(d\,;1\,,m\,;1\,,0\,;1\,,0\,;1\,,0\right) &= -\frac{1}{(4\pi)^4} \exp{\left[2\epsilon\left(\log{\left[\frac{m^2}{\mu^2}\right]-\gamma_E-\log{\left[1-z\right]}}\right)\right]} 
\\
&\frac{\Gamma^2\left(1+\epsilon\right)\Gamma\left(-\epsilon\right)}{\Gamma\left(2+2\epsilon\right)}
\sum_{n=0}^\infty \frac{\Gamma\left(-2\epsilon+n\right)\Gamma\left(1+2\epsilon+n\right)}{\Gamma\left(2+\epsilon+n\right)} \frac{z^n}{n!} \,.
\end{split}
\label{V_int_one_mass_expansion_3} 
\end{gather}
Now we must expand the sum around $\epsilon=0$, and because of the overall $\Gamma\left(-\epsilon\right)$ term we must expand the sum to $\mathcal{O}\left(\epsilon\right)$ in order to expand the entire integral to $\mathcal{O}\left(\epsilon^0\right)$. First, note that for $n\geq1$ we may safely set $\epsilon=0$ in the sum. Extracting the $n=0$ term, we have
\begin{gather}
\begin{split}
\sum_{n=0}^\infty f\left(\epsilon\,,n\right) \frac{z^n}{n!} 
&= f\left(\epsilon\,,0\right) + \sum_{n=1}^\infty \left. \left[ f\left(0\,,n\right) + \epsilon \frac{d}{d\epsilon} f\left(\epsilon\,,n\right)  \right|_{\epsilon=0} \, \right]  \frac{z^n}{n!} \,, 
\\
&f\left(\epsilon\,,n\right) = \frac{\Gamma\left(-2\epsilon+n\right)\Gamma\left(1+2\epsilon+n\right)}{\Gamma\left(2+\epsilon+n\right)} \,.
\end{split}
\label{V_int_one_mass_expansion_4} 
\end{gather}
Using the properties of the Gamma function, this can be written as
\begin{gather}
\begin{split}
\sum_{n=0}^\infty f\left(\epsilon\,,n\right) \frac{z^n}{n!} 
= \frac{\Gamma\left(-2\epsilon\right)\Gamma\left(1+2\epsilon\right)}{\Gamma\left(2+\epsilon\right)} + \sum_{n=1}^{\infty} \frac{z^n}{n\left(n+1\right)} + \epsilon \sum_{n=1}^{\infty}  \left[\frac{1}{n^2\left(1+n\right)^2} - \frac{\psi\left(n\right)}{n\left(1+n\right)}\right] z^n \,,
\end{split}
\label{V_int_one_mass_expansion_5} 
\end{gather}
where $\psi\left(n\right)$ is the Polygamma function~\eqref{polygamma_function}. Evaluating the sums and performing the epsilon expansion, the result for this integral is
\begin{gather}
\begin{split}
V\left(d\,;1\,,m\,;1\,,0\,;1\,,0\,;1\,,0\right) = \frac{1}{512\pi^4} & \Biggl[ -\frac{1}{\epsilon^2} + \frac{1}{\epsilon} 
\left( 5+\frac{2\log{\left[1-z\right]}}{z} - 2 \log{\left[\frac{m^2}{\mu^2}\right]}  \right) 
 \Biggr.
\\
&-\frac{1}{2}\left(38+\pi^2-20\log{\left[\frac{m^2}{\mu^2}\right]}+4\log^2{\left[\frac{m^2}{\mu^2}\right]}\right)
\\
& + \frac{2\log{\left[1-z\right]}}{z} \left(2 \log{\left[\frac{m^2}{\mu^2}\right]}-5\right) 
\\
&\Biggl.+2\left(1+\frac{1}{z}\right) {\rm Li}_2\left(z\right)+\left(1-\frac{3}{z}\right)\log^2{\left[1-z\right]}  \Biggr] \,,
\end{split}
\label{V_int_one_mass_expansion_result} 
\end{gather}
where ${\rm Li}_2\left(z\right)$ denotes the dilogarithm function~\eqref{definition_of_dilogarithm_function}. Polylogarithm functions often appear when hypergeometric functions such as \eqref{V_int_one_mass_expansion_1} are expanded. Note the divergent terms proportional to $\epsilon^{-2}$ and $\epsilon^{-1}$. In a QCD sum rule calculation most of these terms would correspond to dispersion relation subtraction constants that would be eliminated when the Borel transform is applied. However, the divergent term proportional to $\log{\left[1-z\right]}$ is a non-local divergence that will not be eliminated by the Borel transform. Such a divergence must be dealt with through renormalization. 

Using the same method as was used to expand the integral~\eqref{V_int_one_mass_expansion_1}, the epsilon expansion can be performed for all of the integrals that occur in Chapter~\ref{chapter_4_Qq_diquark}, and for all those in Ref.~\cite{Jamin_2001_a}. Ref.~\cite{Kalmykov_2006_a} provides a result for the epsilon expansion of the $\phantom{}_2 F_1$ hypergeometric function up to fifth order in epsilon. In conjunction with hypergeometric function identities, this result has be used to verify the results in Chapter~\ref{chapter_4_Qq_diquark} and in Ref.~\cite{Jamin_2001_a}. In addition, the Mathematica package HypExp~\cite{Huber_2005_a,Huber_2007_a} can perform epsilon expansions of many different hypergeometric functions. This package has also been used to verify the results in Chapter~\ref{chapter_4_Qq_diquark} and Ref.~\cite{Jamin_2001_a}. 

Finally, we will consider a typical integral occurring in the hybrid calculations in Chapter~\ref{chapter_3_hybrids},
\begin{gather}
J\left(d\,;1\,,m\,;1\,,m\,;1\,,0\right) = \frac{1}{\mu^{2(d-4)}} \int \frac{d^d k_1}{(2\pi)^d} \int \frac{d^d k_2}{(2\pi)^d} \frac{1}{\left(k_1^2-m^2\right) \left[\right.\left(k_2-q\right)^2-m^2\left.\right] \left(k_1-k_2\right)^2} \,.
\label{J_int_two_masses_expansion_1} 
\end{gather}
Using the result~\eqref{J_int_two_masses_d_dim}, setting $d=4+2\epsilon$ and working in the $\overline{\rm MS}$ renormalization scheme, 
\begin{gather}
\begin{split}
J\left(d\,;1\,,m\,;1\,,m\,;1\,,0\right) = \frac{m^2}{(4\pi)^4} \exp{\left[2\epsilon\left(\log{\left[\frac{m^2}{\mu^2}\right]-\gamma_E}\right)\right]}
&\frac{\Gamma^2\left(-\epsilon\right)\Gamma\left(1+\epsilon\right)\Gamma\left(-1-2\epsilon\right)}{\Gamma\left(2+\epsilon\right)\Gamma\left(-2\epsilon\right)}
\\
&\phantom{}_4 F_{3}
\left[
\begin{array}{c|}
-1-2\epsilon \,, -\epsilon \,, -\epsilon \,, 1 \\
 2+\epsilon \,, -\epsilon \,, \frac{1}{2}-\epsilon 
\end{array} \; w \; \right] \,, 
\\
&w=\frac{q^2}{4m^2} \,.
\end{split}
\label{J_int_two_masses_expansion_2} 
\end{gather}
Because the $\phantom{}_4 F_3$ hypergeometric function has one common upper and lower index, it reduces to a $\phantom{}_3 F_2$ hypergeometric function.  Using the series representation of the hypergeometric function, the result can be written as
\begin{gather}
\begin{split}
J\left(d\,;1\,,m\,;1\,,m\,;1\,,0\right) = \frac{m^2}{(4\pi)^4} & \exp{\left[2\epsilon\left(\log{\left[\frac{m^2}{\mu^2}\right]-\gamma_E}\right)\right]}
\frac{\Gamma\left(-\epsilon\right)\Gamma\left(1+\epsilon\right)\Gamma\left(\frac{1}{2}-\epsilon\right)}{\Gamma\left(-2\epsilon\right)}
\\
&\sum_{n=0}^\infty \frac{\Gamma\left(-1-2\epsilon+n\right)\Gamma\left(-\epsilon+n\right)\Gamma\left(1+n\right)}{\Gamma\left(2+\epsilon+n\right)\Gamma\left(\frac{1}{2}-\epsilon+n\right)} \frac{w^n}{n!} \,.
\end{split}
\label{J_int_two_masses_expansion_3} 
\end{gather}
Noting that the factor multiplying the sum is $\mathcal{O}\left(\epsilon^0\right)$, we only need to expand the sum to this order. Also, note that the for $n\geq2$ we may safely set $\epsilon=0$ in the sum, so we can extract the $n=0$ and $n=1$ terms. Doing so, we have
\begin{gather}
\begin{split}
J\left(d\,;1\,,m\,;1\,,m\,;1\,,0\right) &= \frac{m^2}{(4\pi)^4}  \exp{\left[2\epsilon\left(\log{\left[\frac{m^2}{\mu^2}\right]-\gamma_E}\right)\right]}
\Gamma\left(-\epsilon\right)\Gamma\left(\frac{1}{2}-\epsilon\right) 
\\
&\Biggl[ \frac{\Gamma\left(-1-2\epsilon\right)\Gamma\left(-\epsilon\right)}{\Gamma\left(2+\epsilon\right)\Gamma\left(\frac{1}{2}-\epsilon\right)} 
+\frac{\Gamma\left(-2\epsilon\right)\Gamma\left(1-\epsilon\right)}{\Gamma\left(3+\epsilon\right)\Gamma\left(\frac{3}{2}-\epsilon\right)}w \Biggr.
\\
&\Biggl. +\sum_{n=2}^\infty \frac{\Gamma\left(-1-2\epsilon+n\right)\Gamma\left(-\epsilon+n\right)\Gamma\left(1+n\right)}{\Gamma\left(2+\epsilon+n\right)\Gamma\left(\frac{1}{2}-\epsilon+n\right)} \frac{w^n}{n!}
\Biggr] \,.
\end{split}
\label{J_int_two_masses_expansion_4} 
\end{gather}
Letting $k=n-2$ and using the properties of the Gamma function, the sum can be expressed as a hypergeometric function
\begin{gather}
\begin{split}
w^2 \sum_{k=0}^\infty \frac{\Gamma\left(1+k\right)\Gamma\left(2+k\right)\Gamma\left(3+k\right)}{\Gamma\left(4+k\right)\Gamma\left(\frac{5}{2}+k\right)} \frac{w^k}{\Gamma\left(3+k\right)}
= 
\frac{2w^2}{9\sqrt{\pi}} \, \phantom{}_3 F_2 \left[1\,,1\,,2\,; 4\,, \frac{5}{2} \,; w\right] \,.
\end{split}
\label{J_int_two_masses_expansion_5} 
\end{gather}
Finally, expanding the remaining functions we find
\begin{gather}
\begin{split}
J\left(d\,;1\,,m\,;1\,,m\,;1\,,0\right) = \frac{m^2}{256\pi^4} &\Biggl[
-\frac{1}{6} \left( 42 + \pi^2 +3w + 12 \log{\left[\frac{m^2}{\mu^2}\right]} \left[w-3+\log{\left[\frac{m^2}{\mu^2}\right]}\right]  \right)
\Biggr.
\\
& 
-\frac{1}{\epsilon^2} + \frac{1}{\epsilon}\left(3-w-2\log{\left[\frac{m^2}{\mu^2}\right]}\right)
\\
&\Biggl.
+\frac{4w^2}{9} \phantom{}_3 F_2 \left[1\,,1\,,2\,; 4\,, \frac{5}{2} \,; w\right] 
\Biggr] \,.
\end{split}
\label{J_int_two_masses_expansion_result} 
\end{gather}
Note that the hypergeometric function can be expressed in terms of inverse sine functions. However, there are several reasons for leaving the result in terms of a hypergeometric function. First, all hypergeometric functions with argument $x$ have a branch cut on $x\in\left[1\,,\infty\right)$, therefore it is clear that the result~\eqref{J_int_two_masses_expansion_result} has appropriate branch cut structure, namely a branch cut on $q^2\in \left[4m^2\,,\infty\right) $. However, when the result is expressed in terms of inverse sine functions, this branch cut structure is obscured. Second, the result is very compact. Apart from the hypergeometric function, all terms in~\eqref{J_int_two_masses_expansion_result} are dispersion relation subtraction constants.

\section{Analytic Continuation}
\label{AnalyticContinuation}

\begin{figure}[htb]
\centering
\includegraphics[scale=0.5]{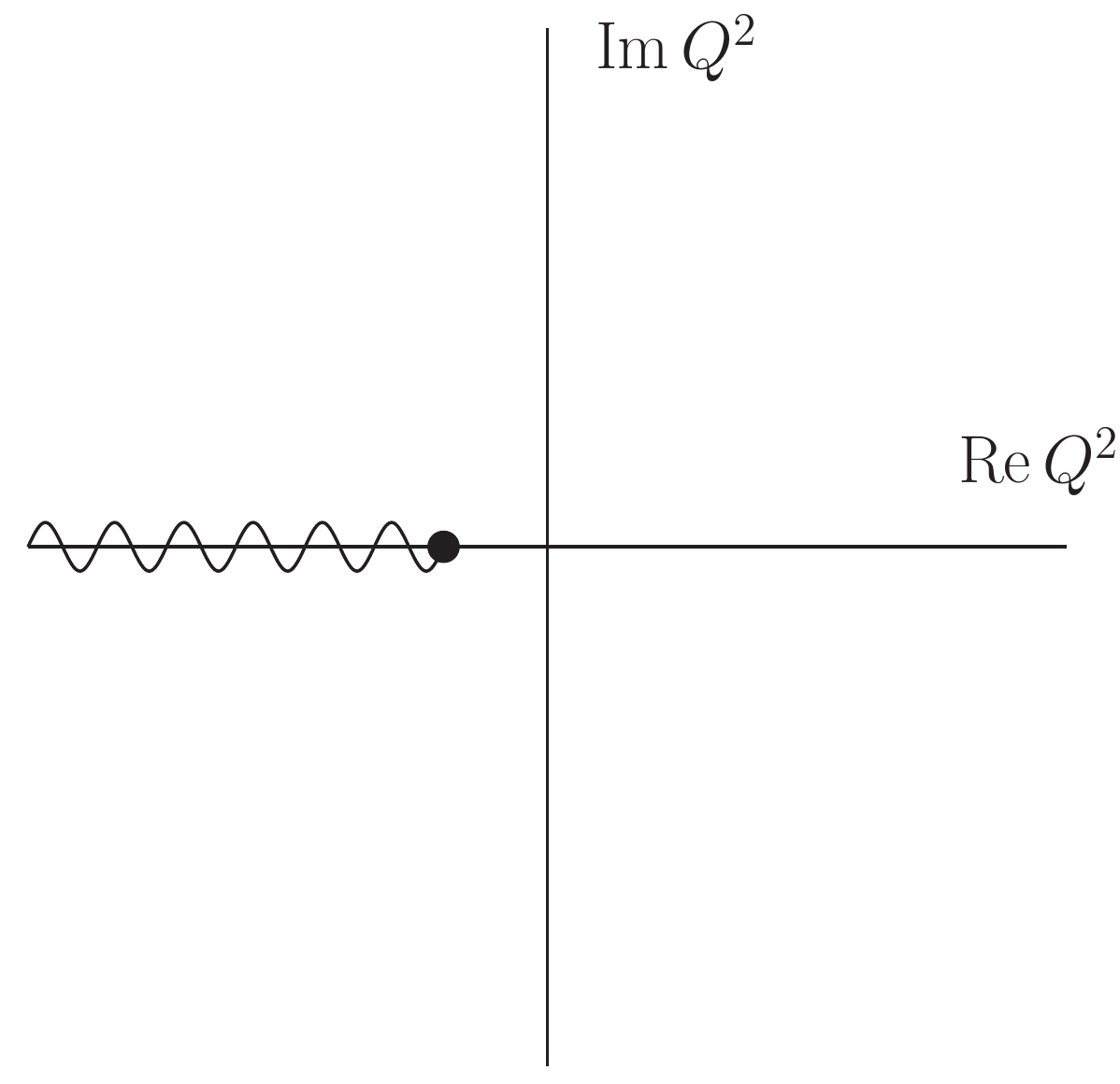}
\caption{A branch cut singularity in the complex Euclidean momentum plane. The branch cut is on the interval $Q^2 \in \left(-\infty \,, -t_0 \right]$, where $t_0$ is the hadronic threshold.}
\label{branch_cut_structure}
\end{figure}

The methods discussed so far in this chapter are sufficient to calculate the correlation functions that are studied in Chapters \ref{chapter_3_hybrids}, \ref{chapter_4_Qq_diquark} and \ref{chapter_6_mixing}. However, it is the singularities of the Wilson coefficients in the complex Euclidean momentum plane that are of interest in QSR analyses. Typically, these singularities appear as isolated poles or as branch cuts. Fig.~\ref{branch_cut_structure} shows a typical branch cut singularity. 

The branch point corresponds to the hadronic threshold, which is related to the total mass of the hadronic constituents. For instance, in the heavy quarkonium hybrid calculations in Chapter~\ref{chapter_3_hybrids}, the currents used contain two identical heavy quarks so that $t_0=4m^2$. In calculations that involve light quarks we work in the chiral limit, ignoring the light quark mass. Accordingly, $t_0=m^2$ in the heavy-light diquark calculation in Chapter~\ref{chapter_4_Qq_diquark}, while in the glueball quark meson mixing calculation in Chapter~\ref{chapter_6_mixing}, $t_0=0$. Recall that in QSR analyses the discontinuity of the correlation function across the branch cut is required. As discussed in Chapter~\ref{chapter_1_intro}, due to the analytic properties of the correlation function, the discontinuity is related to the imaginary part. Therefore, once the correlation function has been calculated, the imaginary part must be extracted. This can be achieved via analytic continuation. 

The complexity of the functions that arise in QCD sum rule calculations depends on the number of external scales involved in the calculation. In the chiral limit, typically only logarithms involving dimensionless ratios of the Euclidean external momentum $Q^2$ and renormalization scale $\mu$ occur. In order to deal with these, we define the complex logarithm as follows:
\begin{gather}
\log{\left(w\right)} \equiv \log{\left| w \right|} + i {\rm Arg}\left(w\right) \,, \quad {\rm Arg}\left(w\right) \in \left[-\pi \,, \pi \right) \,.
\label{complex_logarithm_definition}
\end{gather}
Using this, we can define
\begin{gather}
\left. \log{\left[\frac{Q^2}{\mu^2}\right]} \right|_{Q^2 \, = \, te^{-i\pi} \,,\, t>0}  \equiv \log{\left[\frac{t}{\mu^2}\right]} - i\pi,
\label{logarithm_definition}
\end{gather}
where we have considered a point below the branch cut, as described in Chapter~\ref{chapter_1_intro}. From \eqref{logarithm_definition} it can be shown that
\begin{gather}
 {\rm Im} \log{\left[\frac{Q^2}{\mu^2}\right]} = -\pi \,, \quad {\rm Im} \log^2{\left[\frac{Q^2}{\mu^2}\right]} = -2\pi\log{\left[\frac{t}{\mu^2}\right]}\,,
\end{gather}
for example. In the heavy-light diquark calculation in Chapter~\ref{chapter_4_Qq_diquark} , more complicated functions involving dimensionless ratios of the external momentum $Q^2$ and the quark mass $m$ occur. For instance, in the integral~\eqref{V_int_one_mass_expansion_result} the logarithms can be written as
\begin{gather}
\log{\left[1-z\right]} = - \log{\left[1+\frac{Q^2}{m^2}\right]} \,, \quad z=\frac{1}{1+\frac{m^2}{Q^2}} \,.
\end{gather}
which has the appropriate branch cut structure. Using~\eqref{complex_logarithm_definition}, we can define
\begin{gather}
\left. \log{\left[1+\frac{Q^2}{m^2}\right]}  \right|_{Q^2 \, = \, te^{-i\pi} \,,\, t>m^2}  \equiv \log{\left[\frac{t}{m^2}-1\right]} - i\pi \,,
\end{gather}
so that, for instance, 
\begin{gather}
 {\rm Im} \log{\left[1+\frac{Q^2}{m^2}\right]} = -\pi \,, \quad {\rm Im} \log^2{\left[1+\frac{Q^2}{m^2}\right]} = -2\pi \log{\left[\frac{t}{m^2}-1\right]} \,.
\end{gather}
Note the occurrence of the dilogarithm function in the integral~\eqref{V_int_one_mass_expansion_result}. Using the definition of the dilogarithm function~\eqref{definition_of_dilogarithm_function} and the properties of the logarithm above, the imaginary part can be shown to be
\begin{gather}
\left. {\rm Im} \, {\rm Li}_2 \left[\frac{1}{1+\frac{m^2}{Q^2}}\right] \right|_{Q^2 = t e^{-i\pi} \,, \, t>m^2} = i\pi \log{\left[1-\frac{m^2}{t}\right]} \,.
\end{gather}
Finally, we will consider the hypergeometric functions occurring in the heavy quarkonium hybrid calculations in Chapter~\ref{chapter_3_hybrids}. As mentioned previously, the hypergeometric function that occurs in \eqref{J_int_two_masses_expansion_result} can be expressed in terms of inverse sine functions. Once this has been done, the imaginary part can be extracted using the identity
\begin{gather}
\sin^{-1}{\left(w\right)} = -i \log{\left[iw + \sqrt{1-w^2}\right]} \,, 
\end{gather}
where the logarithm can be dealt with as above. Using these techniques, closed form expressions for the imaginary parts of the correlation functions in Chapters \ref{chapter_3_hybrids}, \ref{chapter_4_Qq_diquark} and \ref{chapter_6_mixing} can be determined. 

\chapter{Heavy Quarkonium Hybrid mass predictions}
\label{chapter_3_hybrids}
\section{Introduction}
The research presented in this chapter involves two closely related publications:
\begin{itemize}
 \item D.~Harnett, R.T.~Kleiv, T.G.~Steele, and Hong-Ying~Jin, Axial Vector $J^{PC}=1^{++}$ Charmonium and Bottomonium Hybrid Mass Predictions with QCD Sum-Rules, J.~Phys. G39~(2012)~125003.
\item R.~Berg, D.~Harnett, R.T.~Kleiv, and T.G.~Steele, Mass Predictions for Pseudoscalar $J^{PC}=0^{-+}$ Charmonium and Bottomonium Hybrids in QCD Sum-Rules, Phys.~Rev. D86~(2012)~034002.
\end{itemize}

The publications above (Refs.~\cite{Harnett_2012_a,Berg_2012_a}) extract mass predictions for heavy quarkonium hybrids with the quantum numbers $J^{PC}=1^{++}$ and $0^{-+}$, respectively. Heavy quarkonium hybrids are widely suspected to exist in the same mass region as charmonia and bottomonia, so it is entirely possible that some of the heavy quarkonium-like states that have been discovered so far could be heavy quarkonium hybrids. For instance, the $Y(4260)$ is considered to be a strong candidate for a charmonium hybrid~\cite{Zhu_2005_a}.

Surprisingly, this possibility has been little explored by QSR practitioners. The original QSR studies of heavy quarkonium hybrids were performed by Govaerts {\it et al.} in Refs.~\cite{Govaerts_1984_a,Govaerts_1985_a,Govaerts_1986_a}. Many different heavy quarkonium hybrid $J^{PC}$ channels were examined, however, only the perturbative and dimension-four gluon condensate $\langle \alpha G^2 \rangle$ were included in the OPE of the correlation functions. Consequently many of the sum rules that were derived were unstable, meaning that the resulting heavy quarkonium hybrid mass predictions are unreliable. However, the authors of Ref.~\cite{Qiao_2010_a} recently updated the sum rule for the vector $\left(1^{--}\right)$ channel, which was unstable in Refs.~\cite{Govaerts_1984_a,Govaerts_1985_a,Govaerts_1986_a}. It was found that inclusion of the dimension-six gluon condensate $\langle g^3 G^3 \rangle$ stabilizes the sum rule in this channel, permitting reliable mass predictions to be made. The publications in 
this chapter update the $J^{PC}=1^{++}$ and $0^{-+}$ heavy quarkonium hybrid sum rules to include the effects of the dimension-six gluon condensate.

\section{Results}
The axial vector $\left(1^{++}\right)$ and pseudoscalar $\left(0^{-+}\right)$ heavy quarkonium hybrids can be studied within QSR using the following current and correlation function~\cite{Govaerts_1984_a,Govaerts_1985_a,Govaerts_1986_a}:
\begin{gather}
\Pi_{\mu\nu}\left(q\right) = i \int d^4 x \, e^{iq\cdot x} \langle \Omega | T\left[\right. J_\mu\left(x\right) J_\nu\left(0\right) \left.\right] | \Omega \rangle \,,
\\
J_\mu = \frac{g}{2} \bar{Q} \lambda^a \gamma^\nu \tilde{G}^a_{\mu\nu} Q \,, \quad \tilde{G}^a_{\mu\nu} = \frac{1}{2} \epsilon_{\mu\nu\alpha\beta} G^{\alpha\beta}_a \,,
\end{gather}
where $q$ is the external momentum, $Q$ denotes a heavy (charm or bottom) quark field and $G^a_{\mu\nu}$ denotes the gluon field strength tensor. Because the correlation function is Lorentz invariant, it can be decomposed into 
\begin{gather}
\Pi_{\mu\nu}\left(q\right) = \left[\frac{q_\mu q_\nu}{q^2}-g_{\mu\nu}\right] \Pi_V\left(q^2\right) + \frac{q_\mu q_\nu}{q^2} \Pi_S\left(q^2\right) \,,
\label{hybrid_correlator_decomposition}
\end{gather}
where $\Pi_V\left(q^2\right)$ and $\Pi_S\left(q^2\right)$ couple to axial vector and pseudoscalar heavy quarkonium hybrids, respectively~\cite{Govaerts_1986_a}. These functions can be isolated by contracting Eq.~\eqref{hybrid_correlator_decomposition} with appropriate combinations of the metric and the external momentum. Note that the contractions must be performed in $d$-dimensions when dimensional regularization is used. 

The axial vector channel sum rule analysis resulted in mass predictions of $5.13\pm0.25\,{\rm GeV}$ and $11.32\pm0.32\,{\rm GeV}$ for the charmonium and bottomonium hybrids, respectively. Interestingly, the dimension-six gluon condensate had little effect on the sum rules in this channel. Although the axial vector channel sum rule was stable in the original analysis~\cite{Govaerts_1984_a,Govaerts_1985_a,Govaerts_1986_a}, it was important to examine the effects of the dimension-six condensate on this channel. Furthermore, because the quantum numbers of the $X(3872)$ are now firmly established to be $1^{++}$~\cite{Aaij_2013_a}, clear mass predictions for all axial vector exotic hadrons are needed. As discussed in Chapter~\ref{chapter_1_intro}, the $X(3872)$ is most often interpreted as a four-quark state. There has also been an attempt to describe it as a charmonium hybrid~\cite{Li_2004_a}. However, the hybrid interpretation has been largely set aside due to the fact that several different theoretical 
approaches predict an axial vector charmonium hybrid mass that is much greater than that of the $X(3872)$~\cite{Barnes_1995_a,Perantonis_1990_a,Liu_2011_a,Liu_2012_a}. With the mass predictions that we have extracted, QSR is now in agreement with these other theoretical approaches. Therefore this work has helped to rule out the pure charmonium hybrid interpretation of the $X(3872)$. However, it should be noted that the results of this work cannot exclude the possibility that the $X(3872)$ could be a mixture of various hadronic structures, perhaps with a hybrid component. The latter possibility is explored in Ref.~\cite{Chen_2013_a}.

The pseudoscalar channel sum rule analysis lead to mass predictions of $3.82\pm0.13\,{\rm GeV}$ and $10.64\pm0.19\,{\rm GeV}$ for the charmonium and bottomonium hybrids, respectively. Both of these mass predictions are significantly lower than the mass predictions of the original studies of Govaerts {\it et al.} which were derived from unstable sum rules. Similar to the recent work in the vector channel, inclusion of the dimension-six gluon condensate was found to stabilize the pseudoscalar channel sum rules. Including the theoretical uncertainty, the pseudoscalar charmonium hybrid mass prediction is comparable to the mass of the $Y(3940)$~\cite{Abe_2004_a,Aubert_2007_a}. This particle has been identified as a charmonium hybrid candidate~\cite{Abe_2004_a}, and our mass prediction supports this claim. However, to date the quantum numbers of this state have not yet been firmly established. More experimental work is needed to determine the true nature of the $Y(3940)$. 

The research in this chapter contributes to several of the themes of this thesis. First, the heavy quarkonium hybrid mass predictions presented here will help to unravel the true nature of the enigmatic heavy quarkonium-like states. It is also interesting to note that the vector~\cite{Qiao_2010_a}, axial vector~\cite{Harnett_2012_a} and pseudoscalar~\cite{Berg_2012_a} charmonium hybrid mass predictions derived from QSR are in qualitative agreement with the charmonium hybrid multiplet structure predicted using lattice QCD~\cite{Liu_2012_a}. Second, the calculations of the axial vector and pseudoscalar heavy quarkonium hybrid correlation functions profitably apply the loop integration techniques discussed in Chapter~\ref{chapter_2_loops}. In order to properly formulate the contributions of the dimension-six gluon condensate to the sum rules the entire correlation function must be calculated explicitly. Specifically, $\Pi_{\rm GGG}\left(Q^2\right)$ is singular at the hadronic threshold $t_0=4m^2$. This 
singularity also appears in the imaginary part ${\rm Im}\Pi_{\rm GGG}\left(Q^2\right)$, and hence when the sum rules are formulated the integration in Eq.~\eqref{final_sum_rule_contribution_of_branch} is singular at the lower limit. This difficulty can be overcome by noting that Laplace sum rules~\eqref{laplace_sum_rules} involve the inverse Laplace transform of the entire correlation function. The inverse Laplace transform can be calculated via a limiting procedure so that the sum rules are well defined at the hadronic threshold. However, the imaginary part of the correlation function alone is insufficient to do this. Therefore the the entire correlation function must be calculated explicitly, and the loop integration techniques discussed in Chapter~\ref{chapter_2_loops} are indispensable for this.

\section{Published Articles}

The $J^{PC}=1^{++}$ heavy quarkonium hybrid paper was published in the Journal of Physics G in 2012, while the $J^{PC}=0^{-+}$ heavy quarkonium hybrid paper was published in Physical Review D in 2012. Links to the published journal articles and preprints are included below for each paper.
\begin{itemize}
 \item D.~Harnett, R.T.~Kleiv, T.G.~Steele, and Hong-Ying~Jin, Axial Vector $J^{PC}=1^{++}$ Charmonium and Bottomonium Hybrid Mass Predictions with QCD Sum-Rules, \href{http://iopscience.iop.org/0954-3899/39/12/125003/}{J.~Phys. G39~(2012)~125003}, \href{http://arxiv.org/abs/1206.6776}{arXiv:1206.6776 [hep-ph]}.
 \item R.~Berg, D.~Harnett, R.T.~Kleiv, and T.G.~Steele, Mass Predictions for Pseudoscalar $J^{PC}=0^{-+}$ Charmonium and Bottomonium Hybrids in QCD Sum-Rules, \href{http://journals.aps.org/prd/abstract/10.1103/PhysRevD.86.034002}{Phys.~Rev. D86~(2012)~034002}, \href{http://arxiv.org/abs/1204.0049}{arXiv:1204.0049 [hep-ph]}.
\end{itemize}


\chapter{Heavy-Light Diquark mass predictions}
\label{chapter_4_Qq_diquark}


\section{Introduction}
The research in this chapter is based upon the following publication:

\begin{itemize}
 \item R.T.~Kleiv, T.G.~Steele, Ailin Zhang, and Ian Blokland, Heavy-light diquark masses from QCD sum rules and constituent diquark models of tetraquarks, Phys. Rev. D87 (2013) 125018.
\end{itemize}

The manuscript above (Ref.~\cite{Kleiv_2013_a}) uses QSR to determine the masses of diquarks with $J^P=0^\pm\,,1^\pm$ that are composed of one heavy (charm or bottom) quark and one light quark. As described in Chapter~\ref{chapter_1_intro}, many of the heavy quarkonium-like states have been interpreted as four-quark states. These can be realized as weakly bound molecular states, or as tightly bound tetraquarks composed of diquark clusters. Heavy quarkonium-like four-quark states have been widely studied using QSR (see Ref.~\cite{Nielsen_2009_a} for a review). A universal feature of these approaches has been the use of currents containing four quark fields, which are in either the molecule or tetraquark configurations
\begin{gather}
 J_{\rm molecule} = \left(\bar{Q}\Gamma Q\right) \left(\bar{q}\Gamma q\right) \,, \quad J_{\rm tetraquark} = \left[\bar{Q}\tilde{\Gamma} \bar{q}\right] \left[Q\tilde{\Gamma} q\right] \,, 
\label{four_quark_currents}
\end{gather}
where $Q$ and $q$ denote heavy and light quark fields, respectively. The Dirac matrices $\Gamma$ and $\tilde{\Gamma}$ are related to the quantum numbers of the hadrons probed by each current. The composite operator in the round brackets in Eq.~\eqref{four_quark_currents} is a current that couples to heavy-light mesons, while that in the square brackets is a current that couples to heavy-light diquarks. However, the two currents in Eq.~\eqref{four_quark_currents} are not truly independent because they can be transformed into one another through Fierz transformations. In Ref.~\cite{Zhang_2006_a} it was pointed out that this ambiguity obscures the nature of the hadronic states that are probed by these currents. For this reason QSR studies that utilize currents containing four quark fields cannot distinguish between the molecular and tetraquark scenarios. 

An alternative approach to studying four-quark states within QSR is to use diquark currents. Using QSR the diquark mass can be calculated and can be thought of as a constituent diquark mass. This in turn can be used in constituent diquark models of tetraquarks. This approach was first used in Ref.~\cite{Zhang_2006_a} to study tetraquarks composed of light quarks. There are several benefits to this approach for studying four-quark states in QSR. First, it avoids the Fierz transformation ambiguities associated with four-quark currents. This is perhaps the only way that pure tetraquark states can be studied using QSR. Second, the composite operators in Eq.~\eqref{four_quark_currents} mix under renormalization~\cite{Narison_1983_a,Jamin_1985_a}. For this reason it is challenging to extend QSR studies using four-quark currents to higher orders. However, as discussed in Chapter~\ref{chapter_1_intro}, the diquark current does not mix with other operators under renormalization. The renormalization factor of the 
scalar $(J^P=0^+)$ diquark operator is determined to two-loop order in Ref.~\cite{Kleiv_2010_a} and is the subject of Chapter~\ref{chapter_5_diquark_renorm}. For these reasons QSR studies using diquark currents can be extended to higher order in the perturbative expansion much more easily than those that use four-quark currents.

In Ref.~\cite{Maiani_2004_a} the $X(3872)$ is interpreted as a tetraquark using a constituent diquark model where the scalar $\left(0^+\right)$ and axial vector $\left(1^+\right)$ charm-light diquark masses are assumed to be degenerate due to heavy quark symmetry. The constituent charm-light diquark mass is determined to be $1.93\,{\rm GeV}$ from a fit to the $X(3872)$. The model also predicts the existence of electrically charged tetraquarks that are members of the same nonet as the $X(3872)$. The recently discovered $Z_c^{\pm}\left(3895\right)$ appears to be compatible with this prediction~\cite{Faccini_2013_a}. A similar analysis was performed in Ref.~\cite{Ali_2011_a} where a bottom-light constituent diquark mass of $5.20\,{\rm GeV}$ was extracted from a fit to the tetraquark candidate $Y_b\left(10890\right)$~\cite{Abe_2007_a}. The results of the analysis support the tetraquark interpretation of the charged bottomonium-like states $Z_b^{\pm}(10610)$ and $Z_b^{\pm}(10650)$ which were discovered by the 
Belle collaboration~\cite{Bondar_2011_a}. Essential features of these constituent diquark models are that the masses of the scalar $\left(0^+\right)$ and axial vector $\left(1^+\right)$ diquarks are assumed to be identical, and that the constituent diquark masses are extracted from fits to tetraquark candidates among the XYZ states. 

The main goal of the research in this chapter was to calculate the constituent heavy-light diquark mass using QSR, so as to provide a QCD-based test of the constituent diquark models used in Refs.~\cite{Maiani_2004_a,Ali_2011_a}. Constituent masses of diquarks composed of light quarks only were determined in  Refs.~\cite{Dosch_1988_a,Jamin_1989_a,Zhang_2006_a,Wang_2011_a}. In Ref.~\cite{Wang_2010_a} heavy-light diquarks with $J^P=0^+\,,1^+$ were studied using QSR, however, only leading-order perturbative contributions to the OPE were considered. The research presented in this chapter builds upon the work of Ref.~\cite{Wang_2010_a} by including next-to-leading order perturbative contributions and diquarks with $J^P=0^-\,,1^-$.

\section{Results}

The correlation function and currents used to study heavy-light diquarks with $J^P=0^\pm\,,1^\pm$ are given by
\begin{gather}
\Pi\left(Q^2\right) = i \int d^4 x \, e^{i q \cdot x} \langle \Omega | T\left[ \right. J_{\alpha}\left(x\right) S_{\alpha\omega} \left[x\,,0\right] J^\dag_{\omega}\left(0\right) \left. \right] |\Omega\rangle \,,
\label{correlation_fcns}
\end{gather}
where $\alpha$, $\omega$ are color indices. The heavy-light diquark currents are
\begin{gather}
J_\alpha = \epsilon_{\alpha\beta\gamma} Q^{T}_{\beta} C\mathcal{O} q_{\gamma} \,,
\label{diquark_currents}
\end{gather}
where the Lorentz structures $\mathcal{O}=\gamma_5\,,I\,,\gamma_\mu\,,\gamma_\mu\gamma_5$ respectively probe scalar $\left(J^P=0^+\right)$, pseudoscalar $\left(0^-\right)$, axial vector $\left(1^+\right)$, and vector $\left(1^-\right)$ heavy-light diquarks~\cite{Dosch_1988_a,Jamin_1989_a}. In Eq.~\eqref{diquark_currents} $C$~is the charge conjugation operator~\eqref{definition_of_charge_conjugation_operator}, $T$ denotes the transpose, $Q$ is a heavy (charm or bottom) quark field, and $q$ is a light quark field. The axial vector and vector correlation functions are given by
\begin{gather}
\Pi^{\rm \left(A,V\right)} \left(q\right) = \frac{1}{d-1}\left(\frac{q^\mu q^\nu}{q^2} - g^{\mu\nu} \right) \Pi^{\rm \left(A,V\right)}_{\mu\nu} \left(q\right) \,,
\label{vector_projection}
\end{gather}
where the number of spacetime dimensions $d$ is kept arbitrary because dimensional regularization is used. The correlation function in Eq.~\eqref{correlation_fcns} includes a path-ordered exponential, also known as a Schwinger string, defined as
\begin{gather}
\begin{split}
S_{\alpha\omega} \left[x\,,0\right] = P \exp\left[{ig\frac{\lambda^a_{\alpha\omega}}{2}\int_0^x dz^\mu\ A^a_\mu\left(z\right)}\right]
\,,
\label{general_schwinger_string}
\end{split}
\end{gather}
where $P$ denotes path-ordering and $g$ is the strong coupling. In Ref.~\cite{Dosch_1988_a} the correlation function~\eqref{correlation_fcns} was calculated for diquarks composed of light quarks, and it was demonstrated that the correlation function is gauge invariant to next-to-leading order in the strong coupling. Because physical observables are gauge invariant, only gauge invariant correlation functions can be used in QSR analyses. Therefore, in order to extract physically meaningful heavy-light diquark masses, it is crucial to verify the gauge invariance of the heavy-light diquark correlation function. In this chapter we perform an explicit calculation that confirms that Eq.~\eqref{correlation_fcns} is gauge invariant to next-to-leading order in the strong coupling. Using a straight line geometry, the Schwinger string is given by
\begin{gather}
\begin{split}
S_{\alpha\omega} \left[x\,,0\right] &= \delta_{\alpha\omega} + ig\frac{\lambda^a_{\alpha\omega}}{2}\int_0^1 d\xi A^a_\mu\left(\xi x\right) x^\mu + \mathcal{O}\left(g^2\right) 
\,.
\label{schwinger_string} 
\end{split}
\end{gather}
The first term in Eq.~\eqref{schwinger_string} simply generates a trace over the colour indices in the correlation function~\eqref{correlation_fcns}. However, the second term in Eq.~\eqref{schwinger_string} leads to a non-trivial contribution. Note that this term is not calculated in the leading-order analysis performed in Ref.~\cite{Wang_2010_a}. In order to verify that the heavy-light correlation function is gauge invariant, all calculations must be performed in a general covariant gauge. That is, the gauge parameter $a$ in the gluon propagator~\eqref{gluon_propagator} must be retained in all calculations. 

The next-to-leading order perturbative contributions to the heavy-light diquark correlation function also introduce gauge dependent terms. In this chapter it is shown that the gauge dependent contributions of the Schwinger string~\eqref{schwinger_string} exactly cancel the gauge dependence in the next-to-leading order perturbative contribution. Therefore the heavy-light diquark correlation function~\eqref{correlation_fcns} is gauge invariant to next-to-leading order in the strong coupling and can be utilized in QSR to determine the heavy-light diquark mass. 

Once the gauge invariance of the heavy-light diquark correlation has been established the bare correlation function must be renormalized. This can be achieved by renormalizing the heavy quark mass and the diquark current, whose renormalization factor is calculated in Chapter~\ref{chapter_5_diquark_renorm}. The renormalization can be implemented using the methods discussed in Chapter~\ref{chapter_1_intro}. However, in order to perform the renormalization in a self-consistent fashion, renormalization-induced contributions must be included. In practical terms this means that the explicit $\mathcal{O}\left(\epsilon\right)$ terms in the leading order perturbative contribution must be calculated. These terms can be calculated using the loop integration methods discussed in Chapter~\ref{chapter_2_loops}. After the correlation function has been renormalized, the imaginary part is needed for the QSR analysis. A closed form expression for the imaginary part can be determined using methods discussed in Chapter~\ref{chapter_2_loops}.

Mass predictions were successfully extracted for all positive parity diquarks. However, mass predictions could not be extracted for any negative parity diquarks due to instabilities in those sum rules. The scalar and axial vector charm-light diquark masses were found to be $1.86\pm0.05\,{\rm GeV}$ and $1.87\pm0.10\,{\rm GeV}$, respectively. These mass predictions are degenerate within uncertainty as expected by heavy quark symmetry and in excellent agreement with the constituent charm-light diquark mass of $1.93\,{\rm GeV}$ predicted by Maiani {\it et al.}~\cite{Maiani_2004_a}. Similarly, the scalar and axial vector bottom-light diquark masses were both found to be $5.08\pm0.04\,{\rm GeV}$, which is in reasonable agreement with the mass of $5.20\,{\rm GeV}$ determined by Ali {\it et al.}~\cite{Ali_2011_a}. Therefore, these heavy-light diquark mass predictions support interpreting the $X(3872)$ and the $Y_b\left(10890\right)$ as tetraquarks. This QCD-based test supports the constituent diquark model of 
tetraquarks, and provides indirect support for the tetraquark interpretation of the charged heavy quarkonium-like states $Z_c^{\pm}\left(3895\right)$, $Z_b^{\pm}(10610)$ and $Z_b^{\pm}(10650)$. 

The research presented in this chapter will contribute to the ongoing effort to understand the $X(3872)$, $Y_b\left(10890\right)$ and the electrically charged heavy quarkonium-like states. There are several technical challenges that are involved in calculating the next-to-leading order perturbative contributions to the heavy-light diquark correlation function. Although only the imaginary part of the correlation function is required for the QSR analysis, the entire correlation function must be calculated in order to properly deal with the gauge invariance and renormalization issues that arise in this calculation. The loop integration techniques discussed in Chapter~\ref{chapter_2_loops} are essential for this. In order to verify that the heavy-light diquark correlation function is gauge invariant, and hence is suitable for use in a QSR analysis, the entire correlation function must be calculated in a general covariant gauge. In addition, the entire correlation function is needed in order to renormalize the 
next-to-leading order perturbative contributions. The research in this chapter develops a renormalization methodology that can be applied to next-to-leading order QSR calculations. Key features of this methodology are the renormalization of the diquark current, which is discussed in Chapter~\ref{chapter_5_diquark_renorm}, and the generation of renormalization-induced contributions to the correlation function. 

\section{Published Article}

The Heavy-light diquark article was published in Physical Review D in 2013. Links to the published journal article and preprint are included below.
\begin{itemize}
 \item R.T.~Kleiv, T.G.~Steele, Ailin Zhang, and Ian Blokland, Heavy-light diquark masses from QCD sum rules and constituent diquark models of tetraquarks, \href{http://journals.aps.org/prd/abstract/10.1103/PhysRevD.87.125018}{Phys. Rev. D87 (2013) 125018}, \href{http://arxiv.org/abs/1304.7816}{arXiv:1304.7816 [hep-ph]}.
\end{itemize}


\chapter{Scalar Diquark Operator Renormalization}
\label{chapter_5_diquark_renorm}
\section{Introduction}
The research in this chapter is based upon the publication:

\begin{itemize}
 \item R.T.~Kleiv and T.G.~Steele, Two-loop QCD renormalization and anomalous dimension of the scalar diquark operator, J. Phys. G38 (2011) 025001.
\end{itemize}

In QSR calculations hadronic states are probed by currents which are composite local operators constructed from quark and gluon fields. As discussed in Chapter~\ref{chapter_1_intro}, composite operators can mix under renormalization with operators of lower dimension and the same quantum numbers. This presents a significant challenge to extending QSR studies to higher orders. However, some composite operators are protected from this mixing by the fact that there are no lower dimensional operators with which they could mix. Such operators must renormalize multiplicatively, and the renormalization factor can be determined using the methods described in Chapter~\ref{chapter_1_intro}. Therefore it is much easier to perform higher order QSR analyses using operators that do not mix under renormalization. 

An example of a composite operator that does not mix under renormalization is the scalar diquark operator, which is given by
\begin{gather}
J^d_\alpha = \epsilon_{\alpha\beta\gamma} Q_\beta^T C\gamma_5 q_\gamma
\label{scalar_diquark_operator}
\end{gather}
where the notation used here is identical to that of Eq.~\eqref{diquark_currents}. The current couples to diquarks with $J^P=0^+$. However, because diquarks have a net colour there are no lower dimensional operators that could mix with the scalar diquark current~\eqref{scalar_diquark_operator}. Therefore, the scalar diquark operator must renormalize multiplicatively. The publication above (Ref.~\cite{Kleiv_2010_a}) determines the renormalization factor of the scalar $\left(J^P=0^+\right)$ diquark operator to second (two-loop) order in the strong coupling $\alpha$. This builds upon the work of Ref.~\cite{Dosch_1988_a}, which gives the scalar diquark renormalization factor to first order.

\section{Results}

The renormalization factor of the scalar diquark operator can be determined by considering the correlation function
\begin{gather}
\Gamma^d = \langle \Omega | \, T\left[ Q\left(x\right) J^d\left(0\right) q\left(y\right) \right] | \Omega \rangle \,,
\label{scalar_diquark_operator_renorm_correlation_function}
\end{gather}
where $J^d$ is the scalar diquark operator~\eqref{scalar_diquark_operator} and colour indices have been omitted for brevity. The correlation function can be calculated using the perturbative expansion~\eqref{perturbative_expansion_of_correlation_function} in momentum space with the external quark propagators amputated. Because we are calculating a renormalization factor which is momentum independent, the diquark operator inserted into Eq.~\eqref{scalar_diquark_operator_renorm_correlation_function} can be taken to have zero momentum without loss of generality. We will use the $\overline{\rm MS}$ renormalization scheme which is mass independent so we can work in the chiral limit, ignoring the quark masses in Eq.~\eqref{scalar_diquark_operator_renorm_correlation_function}. The bare and renormalized correlation functions are related by
\begin{gather}
\Gamma^d_R\left(q\,;m_R\,,a_R\,,\alpha_R\right) = \lim_{\epsilon\to0}\left[Z_{\rm d} \, Z^{-1}_{\rm 2F}  \Gamma^d_B\left(q\,;m_B\,,a_B\,,\alpha_B\right) \right] \,.
\label{scalar_diquark_operator_renorm_bare_and_renormalized_correlation_functions}
\end{gather}
As discussed in Chapter~\ref{chapter_1_intro}, the scalar diquark renormalization factor $Z_{\rm d}$ is the additional renormalization factor that is required in order to evaluate the limit in Eq.~\eqref{scalar_diquark_operator_renorm_bare_and_renormalized_correlation_functions}. This relationship can be used to calculate the scalar diquark operator renormalization factor $Z_{\rm d}$ to any order in the coupling $\alpha$.

In order to calculate scalar diquark operator renormalization factor, it is helpful to exploit the similarity between the scalar diquark and scalar meson operators. The scalar meson operator renormalizes as
\begin{gather}
J^s = \bar{Q}q \,, \quad \left[J^s\right]_R = Z_m \left[J^s\right]_B \,,
\label{scalar_meson_operator}
\end{gather}
where $Z_m$ corresponds to the quark mass renormalization factor in the $\overline{\rm MS}$ scheme and this expression is valid to all orders in the coupling $\alpha$. The renormalization factor $Z_{\rm m}$ is given to $\mathcal{O}\left(\alpha^2\right)$ in Ref.~\cite{Pascual_1984_a}. Equivalently, the scalar meson operator renormalization factor can be calculated directly using the relation
\begin{gather}
\Gamma^s_R\left(q\,;m_R\,,a_R\,,\alpha_R\right) = \lim_{\epsilon\to0}\left[Z_{\rm m} \, Z^{-1}_{\rm 2F}  \Gamma^s_B\left(q\,;m_B\,,a_B\,,\alpha_B\right) \right] \,,
\label{scalar_meson_operator_renorm_bare_and_renormalized_correlation_functions}
\end{gather}
where $\Gamma^s$ is a correlation function similar to that in Eq.~\eqref{scalar_diquark_operator_renorm_correlation_function}, except with a zero momentum insertion of $J^s$ rather than $J^d$. Because the scalar diquark~\eqref{scalar_diquark_operator} and scalar meson~\eqref{scalar_meson_operator} operators are very similar in structure, the correlation functions given in Eqs.~\eqref{scalar_diquark_operator_renorm_bare_and_renormalized_correlation_functions} and \eqref{scalar_meson_operator_renorm_bare_and_renormalized_correlation_functions} are closely related. In fact, to any order in perturbation theory, each diagram contributing to the scalar diquark correlation function is proportional to a corresponding diagram contributing to the scalar meson operator. This relationship and the known two loop expression for the scalar meson operator renormalization factor provide a useful benchmark for the direct calculation of the scalar diquark renormalization factor via Eq.~\eqref{scalar_diquark_operator_renorm_bare_and_renormalized_correlation_functions}. At one-loop order, there is only one diagram that contributes to each correlation function, so the one-loop renormalization factors are proportional. However, at two-loop order there are eleven Feynman diagrams that contribute to each correlation function. Thus the simple proportionality between the scalar diquark and scalar meson operator renormalization factors does not persist at two-loop level. The complete expression for the two-loop scalar diquark renormalization factor in the $\overline{\rm MS}$ scheme is determined to be
\begin{gather}
\begin{split}
Z_{\rm d} = 1 + \frac{\alpha}{\pi}\left[\frac{3-a}{6\epsilon}\right] + \left(\frac{\alpha}{\pi}\right)^2 & \left[\frac{1}{\epsilon}\left(\frac{1545-40n_f}{2880}-\frac{a}{8}-\frac{a^2}{64}\right) \right. 
\\
&\left.+ \frac{1}{\epsilon^2}\left(\frac{234-12n_f}{288}-\frac{17a}{96}-\frac{5a^2}{288}\right) \right]  \,,
\end{split}
\label{ZD}
\end{gather}
where $a$ is the covariant gauge parameter, $n_f$ is the number of quark flavours and dimensional regularization with $d=4+2\epsilon$ has been used.

The two-loop scalar diquark operator renormalization factor given in Eq.~\eqref{ZD} can be used to extend existing QSR studies of diquarks to higher order. It is possible that these higher order corrections could have a significant effect on QSR mass predictions for diquarks. Note that the renormalization factor was calculated in the $\overline{\rm MS}$ renormalization scheme, where all quark flavours renormalize in the same way. Therefore the renormalization factor determined in Ref.~\cite{Kleiv_2010_a} applies to all scalar diquark operators, regardless of the flavour of the quarks composing the diquark operator. In Chapter~\ref{chapter_4_Qq_diquark}, mass predictions were determined for heavy-light diquarks with $J^P=0^\pm\,,1^\pm$. The unknown renormalization factors for the pseudoscalar $\left(0^-\right)$, axial vector $\left(1^+\right)$ and vector $\left(1^-\right)$ diquark operators were determined by utilizing the one-loop relationship between diquark and quark meson operators established in this 
chapter. 

Although this research presented in this chapter is not directly relevant to the heavy quarkonium-like states, it has been applied in the QSR study of heavy-light diquarks in Chapter~\ref{chapter_4_Qq_diquark}. The renormalization of the diquark current is an essential aspect of the renormalization methodology used in Chapter~\ref{chapter_4_Qq_diquark}. In order to calculate the two-loop scalar diquark operator renormalization factor, a large number of loop integrals must be calculated. Because the $\overline{\rm MS}$ renormalization scheme is mass independent, these integrals can be evaluated in the chiral limit. The loop integration methods discussed in Chapter~\ref{chapter_2_loops} are needed in order to evaluate these integrals. Finally, the result presented here for the two-loop scalar diquark renormalization factor could permit higher order QSR studies of scalar diquarks.

\section{Published Article}

The two-loop scalar diquark renormalization paper was published in the Journal of Physics G in 2011. Note that a corrigendum correcting a minor error was published in the same journal in 2012. Links to the preprint, published article and corrigendum are included below.
\begin{itemize}
 \item R.T.~Kleiv and T.G.~Steele, Two-loop QCD renormalization and anomalous dimension of the scalar diquark operator, \href{http://iopscience.iop.org/0954-3899/38/2/025001/}{J. Phys. G38 (2011) 025001}; Corrigendum ibid \href{http://iopscience.iop.org/0954-3899/39/3/039501}{ J. Phys. G39 (2012) 039501}, \href{http://arxiv.org/abs/1010.2971}{arXiv:1010.2971 [hep-ph]}.
\end{itemize}


\chapter{Mixing of Scalar Gluonium and Quark Mesons}
\label{chapter_6_mixing}


\section{Introduction}

The research in this chapter is based upon the publication:

\begin{itemize}
 \item D.~Harnett, R.T.~Kleiv, K.~Moats and T.G.~Steele, Near-maximal mixing of scalar gluonium and quark mesons: a Gaussian sum-rule analysis, Nucl. Phys. A850 (2011) 110.
\end{itemize}

The publication above (Ref.~\cite{Harnett_2008_a}) explores mixing between scalar $\left(J^{PC}=0^{++}\right)$ glueballs and quark mesons. As described in Chapter~\ref{chapter_1_intro}, glueballs (or gluonia) are hadrons that are composed entirely of gluons. The scalar glueball is predicted to be the lightest glueball, with a mass in the range of approximately $1.0-1.7\,{\rm GeV}$. The heavy quarkonium-like states have masses in the range $3.8-4.7\,{\rm GeV}$, therefore the research in this chapter is not directly relevant to the heavy quarkonium-like states. Rather, the research in this chapter is related to the problem of the light scalar mesons. Below $2.0\,{\rm GeV}$, there are too many hadrons with $J^{PC}=0^{++}$ to be explained in terms of conventional mesons. It is widely suspected that some of these supernumerary states could be exotic hadrons, with the scalar glueball among them. The research in this chapter considers the possibility that some of the light scalars could be mixtures of a glueball 
and a conventional quark meson. Refs.~\cite{Mathieu_2008_a,Ochs_2013_a} review the current experimental and theoretical status of glueballs.

\section{Results}

The emphasis of this chapter is on the field-theoretic aspects of the publication above. As mentioned in Chapter~\ref{chapter_1_intro}, multiple currents may couple to a single hadronic state. For instance, consider a state $|h\rangle$ that couples to both scalar meson and glueball currents:
\begin{gather}
\langle \Omega | J_q | h \rangle \neq 0 \,, \quad \langle \Omega | J_g | h \rangle \neq 0 \,.
\end{gather}
Hadrons that couple to multiple currents can be studied within QSR using non-diagonal correlation functions. In this case the non-diagonal correlation function contains scalar glueball and quark meson currents
\begin{gather}
\Pi_{gq}\left(Q^2\right) = i \int d^4 x \, e^{iq\cdot x} \, \langle \Omega | T\left[\right. J_g\left(x\right) J_q\left(0\right) \left.\right] | \Omega \rangle \,, \quad Q^2 = -q^2 \,,
\\
J_q = m_q \left(\bar{u}u + \bar{d}d\right) \,, \quad J_g = \alpha G^2 \,, \quad G^2=G^a_{\mu\nu} G_a^{\mu\nu} \,.
\label{glue_meson_non_diagonal_correlator}
\end{gather}
This correlation function can be calculated using the perturbative expansion~\eqref{perturbative_expansion_of_correlation_function} and the OPE~\eqref{momentum_space_ope} as usual. However, the leading order contribution to the perturbative Wilson coefficient contains a non-local divergence. Because this divergence arises at leading order, it cannot be canceled through a multiplicative renormalization. 

This problem can be solved by considering the renormalization of the composite operator representing the scalar glueball current, which mixes with the scalar meson current under renormalization. The renormalized scalar glueball operator is given by
\begin{gather}
G^2_R = \left[1+\frac{1}{\epsilon}\frac{\alpha}{\pi}\left(\frac{11}{4}-\frac{n_f}{6}\right)\right] G_B^2 - \frac{4}{\epsilon}\frac{\alpha}{\pi} \left[m_u \bar{u}u + m_d \bar{d}d\right]_B \,, 
\label{renormalized_scalar_glueball_operator}
\end{gather}
where $n_f$ is the number of quark flavours and the subscripts $R$ and $B$ denote renormalized and bare quantities, respectively.~\cite{Pascual_1984_a,Narison_2007_a}. The renormalized scalar glueball operator must be used in order to renormalize the non-diagonal correlation function~\eqref{glue_meson_non_diagonal_correlator}. Note the appearance of the second term in Eq.~\eqref{renormalized_scalar_glueball_operator} which is divergent and proportional to the scalar meson current. This arises due to operator mixing and must be included in the QSR analysis. This term amounts to a renormalization-induced contribution to the non-diagonal correlation function, and serves to precisely cancel the non-local divergence that appears in the bare non-diagonal correlation function. The renormalized non-diagonal correlation function is free of divergences, as it must be. This represents the perturbative contribution to the OPE of the non-diagonal correlation function~\eqref{glue_meson_non_diagonal_correlator}, and hence represents purely perturbative contributions to the mixing between scalar mesons and gluonia. 

The research presented in this chapter emphasizes the renormalization methodology used in QSR analyses. In particular, the composite local operators used to represent currents that probe hadronic states can mix under renormalization. Divergent terms that appear at leading order in the expansion of Wilson coefficients cannot be renormalized multiplicatively and hence must be due to operator mixing. Conversely, when divergent terms appear in higher order terms in the Wilson coefficients, such as in Chapter~\eqref{chapter_4_Qq_diquark} they can be removed through a multiplicative renormalization. In both cases renormalization-induced contributions are generated and must be included. The loop integration techniques developed in Chapter~\ref{chapter_2_loops} are needed in order to perform these calculations. In addition, the techniques used here have been extended to investigate mixing effects among the heavy quarkonium-like states~\cite{Chen_2013_a}.

\section{Published Article}

The scalar glueball and quark meson mixing paper was published in Nuclear Physics A in 2011. Links to the preprint and published journal versions are included below.

\begin{itemize}
 \item D.~Harnett, R.T.~Kleiv, K.~Moats and T.G.~Steele, Near-maximal mixing of scalar gluonium and quark mesons: a Gaussian sum-rule analysis, \href{http://www.sciencedirect.com/science/article/pii/S0375947410007554}{Nucl. Phys. A850 (2011) 110}, \href{http://arxiv.org/abs/0804.2195}{arXiv:0804.2195 [hep-ph]}.
\end{itemize}


\chapter{Conclusions}
\label{chapter_7_conclusions}
Heavy quarkonium spectroscopy is a rapidly changing field, both experimentally and theoretically. In recent years many heavy quarkonium-like states have been discovered by the Babar, Belle, BES-III, CDF, CLEO, D0, and LHCb experiments. It is entirely possible that more heavy quarkonium-like states will be discovered by these experiments, or by new experiments being planned such as Belle-II~\cite{Aushev_2010_a} and ${\rm \overline{P}ANDA}$~\cite{Lutz_2009_a}. The heavy quarkonium sector provides perhaps the most promising ``hunting ground'' for exotic hadrons. Firm theoretical predictions for the properties of exotic hadrons are needed in order to determine the true nature of the heavy quarkonium-like states.

The main theme of research presented in this thesis has been to utilize QSR techniques to determine mass predictions for exotic hadrons that could exist among the heavy quarkonium-like states. This work has direct implications for the XYZ states. In Chapter~\ref{chapter_3_hybrids} the mass of the $J^{PC}=0^{-+}$ charmonium hybrid was found to be $3.82\pm0.13\,{\rm GeV}$, which is compatible with the $Y(3940)$. In Ref.~\cite{Abe_2004_a} it was suggested that this particle could be a charmonium hybrid, and the mass prediction extracted in Chapter~\ref{chapter_3_hybrids} is compatible with this interpretation. More experimental work is needed to establish the $J^{PC}$ quantum numbers of this state. Similarly, the $1^{++}$ charmonium hybrid mass was predicted to be $5.13\pm0.25\,{\rm GeV}$ in Chapter~\ref{chapter_3_hybrids}. The LHCb collaboration has confirmed that the $X(3872)$ has $J^{PC}=1^{++}$ ~\cite{Aaij_2013_a}, therefore the mass prediction in Chapter~\ref{chapter_3_hybrids} helps to rule out the pure 
charmonium hybrid interpretation of this state~\cite{Li_2004_a}. In Chapter~\ref{chapter_4_Qq_diquark} the $J^P=0^+$ and $1^+$ charm-light diquark masses were predicted to be $1.86\pm0.05\,{\rm GeV}$ and $1.87\pm0.10\,{\rm GeV}$. In Ref.~\cite{Maiani_2004_a} the $X(3872)$ was interpreted as a tetraquark using a constituent diquark model. The $J^P=0^+$ and $1^+$ charm-light diquark masses were determined to be $1.93\,{\rm GeV}$, which is compatible with the mass predictions extracted in Chapter~\ref{chapter_4_Qq_diquark}. This agreement provides QCD support for the predictions of the constituent diquark model developed in Ref.~\cite{Maiani_2004_a}. In particular, this agreement provides indirect support for the tetraquark interpretation of the $Z_c^\pm\left(3895\right)$. The $J^P=0^+$ and $1^+$ bottom-light diquark masses were also extracted in Chapter~\ref{chapter_4_Qq_diquark}, finding a common mass of $5.08\pm0.04\,{\rm GeV}$. This is in reasonable agreement with the constituent bottom-light diquark mass of $5.20\,{\rm GeV}$ determined from a constituent diquark model of the $Y_b\left(10890\right)$ in Ref.~\cite{Ali_2011_a}. Therefore the bottom-light diquark mass prediction extracted in Chapter~\ref{chapter_4_Qq_diquark} supports the tetraquark interpretations of the $Y_b\left(10890\right)$, $Z_b^{\pm}(10610)$ and $Z_b^{\pm}(10650)$. 

A secondary theme in this research has been renormalization methodology. QSR calculations involve correlation functions of composite local operators. The renormalization of these composite operators can significantly complicate QSR calculations. In Chapter~\ref{chapter_4_Qq_diquark}, next-to-leading order perturbative contributions to the heavy-light diquark correlation function were calculated. In order to renormalize these contributions the heavy quark mass and diquark current must be renormalized. The scalar diquark operator renormalization factor was determined in Chapter~\ref{chapter_5_diquark_renorm}. A QSR analysis of mixing between scalar mesons and gluonium was performed in Chapter~\ref{chapter_6_mixing}. The leading order perturbative contribution to the bare non-diagonal correlation function was found to contain a non-local divergence. This problem was resolved through the use of the renormalized scalar glueball operator, which mixes under renormalization with the scalar meson operator. The 
renormalization induced contributions of the scalar meson operator served to cancel the divergence in the bare non-diagonal correlation function. The research in Chapters~\ref{chapter_4_Qq_diquark} and \ref{chapter_6_mixing} illustrates the two distinct ways in which composite operator renormalization can complicate QSR calculations: it can be required in leading order contributions due to operator mixing or in higher order contributions due to the multiplicative renormalization of the current being used. 

The QSR calculations in this thesis have largely been concerned with heavy quarkonium-like states that contain heavy quarks. Unlike calculations that involve only light quarks whose masses can be neglected, the heavy quark mass cannot be ignored. In practice this means that loop integrals that involve heavy quarks are much more complicated than those that involve only light quarks. The loop integration techniques developed in Chapter~\ref{chapter_2_loops} are crucial to the QSR analyses in Chapters~\ref{chapter_3_hybrids} and \ref{chapter_4_Qq_diquark}. In addition, the renormalization methodology used in Chapters~\ref{chapter_4_Qq_diquark} and \ref{chapter_6_mixing} is dependent upon the loop integration methods discussed in Chapter~\ref{chapter_2_loops}. 

The research presented in this thesis can be extended in several ways. First, the QSR studies of $J^{PC}=1^{++}$ and $0^{-+}$ heavy quarkonium hybrids in Chapter~\ref{chapter_3_hybrids} have been extended to additional $J^{PC}$ channels in Ref.~\cite{Chen_2013_b}. This will provide useful information regarding the spectrum of heavy quarkonium hybrids, enabling a comparison between QSR and lattice QCD predictions~\cite{Liu_2012_a}. Second, the heavy-light diquark analysis in Chapter~\ref{chapter_4_Qq_diquark} can be generalized to doubly-heavy diquarks, which could be used to study heavy baryons as well as doubly-charmed or doubly-bottomed tetraquarks. Third, the scalar diquark operator renormalization factor determined in Chapter~\ref{chapter_5_diquark_renorm} could be used to extend existing QSR studies of scalar diquarks to higher orders. Finally, the renormalization methodology applied to the mixing between scalar mesons and gluonia in Chapter~\ref{chapter_6_mixing} can be applied to study possible mixing 
among the heavy quarkonium-like states. These methods have been used to study mixing between heavy quarkonium hybrids and four-quark states in Ref.~\cite{Chen_2013_a}.

The unanticipated XYZ states have heralded a golden age in hadron spectroscopy. In order to determine if any of these states are exotic hadrons, theoretical calculations are needed to clearly establish the expected properties of exotic hadrons that may coexist with heavy quarkonia. The QSR method is a powerful, QCD-based technique that can be used to perform these calculations. To date, there have been many QSR studies of heavy quarkonium-like states. However most of these have focused on four-quark states and have only included leading order perturbative contributions in the OPE. It is desirable to extend QSR calculations to higher order so that more complete and accurate predictions for the properties of exotic hadrons can be obtained. In order to do so, the renormalization methodology and loop integration techniques discussed in this thesis are essential. The techniques used in this thesis can be used to extract more accurate QSR predictions of the properties of exotic hadrons and therefore aid in efforts 
to determine the true natures of the heavy quarkonium-like states.




\uofsbibliography[amsplain_no_dash]{refs_no_titles}








\providecommand{\bysame}{\leavevmode\hbox to3em{\hrulefill}\thinspace}
\providecommand{\MR}{\relax\ifhmode\unskip\space\fi MR }
\providecommand{\MRhref}[2]{%
  \href{http://www.ams.org/mathscinet-getitem?mr=#1}{#2}
}
\providecommand{\href}[2]{#2}
\begin{thebibliography}{100}

\bibitem{Aad_2012_a}
G.~Aad et~al., Phys. Lett. \textbf{B716} (2012), 1.

\bibitem{Aaij_2013_a}
R~Aaij et~al., {arXiv:1302.6269 [hep-ex]}.

\bibitem{Aaij_2011_a}
R.~Aaij et~al., Eur. Phys. J. \textbf{C72} (2012), 1972.

\bibitem{Abazov_2004_a}
V.M. Abazov et~al., Phys. Rev. Lett. \textbf{93} (2004), 162002.

\bibitem{Abe_2004_a}
K.~Abe et~al., Phys. Rev. Lett. \textbf{94} (2005), 182002.

\bibitem{Ablikim_2013_a}
M.~Ablikim et~al., Phys. Rev. Lett. \textbf{110} (2013), 252001.

\bibitem{Abramowitz_1964_a}
M.~Abramowitz and I.A. Stegun, \emph{{Handbook of Mathematical Functions}},
  Dover, 1964.

\bibitem{Acosta_2003_a}
D.~Acosta et~al., Phys. Rev. Lett. \textbf{93} (2004), 072001.

\bibitem{AlFiky_2005_a}
M.T. AlFiky, F.~Gabbiani, and A.A. Petrov, Phys. Lett. \textbf{B640} (2006),
  238.

\bibitem{Ali_2011_a}
A.~Ali, C.~Hambrock, and W.~Wang, Phys. Rev. \textbf{D85} (2012), 054011.

\bibitem{Anselmino_1992_a}
M.~Anselmino, E.~Predazzi, S.~Ekelin, S.~Fredriksson, and D.~B. Lichtenberg,
  Rev. Mod. Phys. \textbf{65} (1993), 1199--1234.

\bibitem{Appelquist_1975_a}
T.~Appelquist and J.~Carazzone, Phys. Rev. \textbf{D11} (1975), 2856.

\bibitem{Aubert_2004_a}
B.~Aubert et~al., Phys. Rev. Lett. \textbf{93} (2004), 041801.

\bibitem{Aubert_2007_a}
B.~Aubert et~al., Phys. Rev. Lett. \textbf{101} (2008), 082001.

\bibitem{Aushev_2010_a}
T.~Aushev, W.~Bartel, A.~Bondar, J.~Brodzicka, T.E. Browder, et~al.,
  {arXiv:1002.5012 [hep-ex]}.

\bibitem{Bagan_1992_a}
E.~Bagan, M.R. Ahmady, V.~Elias, and T.G. Steele, Z. Phys. \textbf{C61} (1994),
  157.

\bibitem{Barnes_1995_a}
T.~Barnes, F.E. Close, and E.S. Swanson, Phys. Rev. \textbf{D52} (1995),
  5242--5256.

\bibitem{Becchi_1976_a}
C.~Becchi, A.~Rouet, and R.~Stora, Ann. Phys. \textbf{98} (1976), 287.

\bibitem{Berg_2012_a}
R.~Berg, D.~Harnett, R.T. Kleiv, and T.G. Steele, Phys. Rev. \textbf{D86}
  (2012), 034002.

\bibitem{Beringer_2012_a}
J.~Beringer et~al., Phys. Rev. \textbf{D86} (2012), 010001.

\bibitem{Bertlmann_1984_a}
R.A. Bertlmann, G.~Launer, and E.~de~Rafael, Nucl. Phys. \textbf{B250} (1985),
  61.

\bibitem{Bethke_2009_a}
S.~Bethke, Eur. Phys. J. \textbf{C64} (2009), 689.

\bibitem{Binosi_2003_a}
D.~Binosi and L.~Theussl, Comput. Phys. Commun. \textbf{161} (2004), 76.

\bibitem{Bjorken_1964_a}
J.D. Bjorken and S.D. Drell, \emph{{Relativistic Quantum Mechanics}},
  McGraw-Hill, 1964.

\bibitem{Bollini_1972_a}
C.G. Bollini and J.J. Giambiagi, Phys. Lett. \textbf{B40} (1972), 566.

\bibitem{Bondar_2011_a}
A.~Bondar et~al., Phys. Rev. Lett. \textbf{108} (2012), 122001.

\bibitem{Boos_1990_a}
E.E. Boos and A.I. Davydychev, Theor. Math. Phys. \textbf{89} (1991), 1052.

\bibitem{Brambilla_2010_a}
N.~Brambilla, S.~Eidelman, B.K. Heltsley, R.~Vogt, G.T. Bodwin, et~al., Eur.
  Phys. J. \textbf{C71} (2011), 1534.

\bibitem{Broadhurst_1993_a}
D.J. Broadhurst, J.~Fleischer, and O.V. Tarasov, Z. Phys. \textbf{C60} (1993),
  287.

\bibitem{Chatrchyan_2012_a}
S.~Chatrchyan et~al., Phys. Lett. \textbf{B716} (2012), 30.

\bibitem{Abe_2007_a}
K.F. Chen et~al., Phys. Rev. Lett. \textbf{100} (2008), 112001.

\bibitem{Chen_2013_a}
W.~Chen, H.Y. Jin, R.T. Kleiv, T.G. Steele, M.~Wang, et~al., {arXiv:1305.0244
  [hep-ph]}.

\bibitem{Chen_2013_b}
W.~Chen, R.T. Kleiv, T.G. Steele, B.~Bulthuis, D.~Harnett, J.~Ho, T.~Richards,
  and S.L. Zhu, {arXiv:1304.4522 [hep-ph]}.

\bibitem{Chetyrkin_1980_a}
K.G. Chetyrkin, A.L. Kataev, and F.V. Tkachov, Nucl. Phys. \textbf{B174}
  (1980), 345.

\bibitem{Chetyrkin_1981_a}
K.G. Chetyrkin and F.V. Tkachov, Nucl. Phys. \textbf{B192} (1981), 159.

\bibitem{Choi_2003_a}
S.K. Choi et~al., Phys. Rev. Lett. \textbf{91} (2003), 262001.

\bibitem{Close_2003_a}
F.E. Close and P.R. Page, Phys. Lett. \textbf{B578} (2004), 119--123.

\bibitem{Colangelo_2000_a}
P.~Colangelo and A.~Khodjamirian, {arXiv:0010175 [hep-ph]}.

\bibitem{Collins_1984_a}
John~C. Collins, \emph{{Renormalization: an Introduction to Renormalization,
  the Renormalization Group and the Operator-Product Expansion}}, Cambridge
  University Press, 1984.

\bibitem{Davydychev_1990_a}
A.I. Davydychev, J. Math. Phys. \textbf{33} (1992), 358.

\bibitem{Dosch_1988_b}
H.G. Dosch, M.~Jamin, and S.~Narison, Phys. Lett. \textbf{B220} (1989), 251.

\bibitem{Dosch_1988_a}
H.G. Dosch, M.~Jamin, and B.~Stech, Z. Phys. \textbf{C42} (1989), 167.

\bibitem{Dubnicka_2010_a}
S.~Dubnicka, A.Z. Dubnickova, M.A. Ivanov, and J.G. Korner, Phys. Rev.
  \textbf{D81} (2010), 114007.

\bibitem{Ebert_2005_a}
D.~Ebert, R.N. Faustov, and V.O. Galkin, Phys. Lett. \textbf{B634} (2006), 214.

\bibitem{Ecker_1994_a}
G.~Ecker, Prog. Part. Nucl. Phys. \textbf{35} (1995), 1.

\bibitem{Elias_1998_a}
V.~Elias, A.H. Fariborz, F.~Shi, and T.G. Steele, Nucl. Phys. \textbf{A633}
  (1998), 279.

\bibitem{Erdelyi_1953_a}
A.~Erdelyi, \emph{{Higher Transcendental Functions (Bateman Manuscript
  Project)}}, vol.~1, McGraw-Hill Book Company, Inc., 1953.

\bibitem{Faccini_2013_a}
R.~Faccini, L.~Maiani, F.~Piccinini, A.~Pilloni, A.D. Polosa, et~al.,
  {arXiv:1303.6857 [hep-ph]}.

\bibitem{Faddeev_1967_a}
L.~D. Faddeev and V.~N. Popov, Phys. Lett. \textbf{25B} (1967), 19.

\bibitem{Feynman_1965_a}
R.P Feynman and A.R. Hibbs, \emph{{Quantum Mechanics and Path Integrals}},
  McGraw-Hill, 1965.

\bibitem{Gell_Mann_1964_a}
M.~Gell-Mann, Phys. Lett. \textbf{8} (1964), 214.

\bibitem{Gell_Mann_1968_a}
M.~Gell-Mann, R.J Oakes, and B.~Renner, Phys. Rev. \textbf{175} (1968), 2195.

\bibitem{Godfrey_2009_a}
S.~Godfrey, {arXiv:0910.3409 [hep-ph]}.

\bibitem{Govaerts_1986_a}
J.~Govaerts, L.J. Reinders, P.~Francken, X.~Gonze, and J.~Weyers, Nucl. Phys.
  \textbf{B284} (1987), 674.

\bibitem{Govaerts_1984_a}
J.~Govaerts, L.J. Reinders, H.R. Rubinstein, and J.~Weyers, Nucl. Phys.
  \textbf{B258} (1985), 215.

\bibitem{Govaerts_1985_a}
J.~Govaerts, L.J. Reinders, and J.~Weyers, Nucl. Phys. \textbf{B262} (1985),
  575.

\bibitem{Greenberg_1964_a}
O.W. Greenberg, Phys. Rev. Lett. \textbf{13} (1964), 598--602.

\bibitem{Griffiths_1987_a}
D.J. Griffiths, \emph{{Introduction to Elementary Particles}}, John Wiley \&
  Sons, Inc., 1987.

\bibitem{Harnett_2008_a}
D.~Harnett, R.T. Kleiv, K.~Moats, and T.G. Steele, Nucl. Phys. \textbf{A850}
  (2011), 110.

\bibitem{Harnett_2012_a}
D.~Harnett, R.T. Kleiv, T.G. Steele, and H.Y. Jin, J. Phys. \textbf{G39}
  (2012), 125003.

\bibitem{Harris_2009_a}
D.~Harris and K.~Riesselmann, \emph{{Deconstruction: Standard Model
  Discoveries}}, Symmetry Magazine \textbf{6} (2009), 30.

\bibitem{Huber_2005_a}
T.~Huber and D.~Maitre, Comput. Phys. Commun. \textbf{175} (2006), 122.

\bibitem{Huber_2007_a}
Tobias Huber and D.~Maitre, Comput. Phys. Commun. \textbf{178} (2008), 755.

\bibitem{Iofa_1976_a}
M.Z. Iofa and I.V. Tyutin, Theor. Math. Phys. \textbf{27} (1976), 316.

\bibitem{Jamin_1985_a}
M.~Jamin and M.~Kremer, Nucl. Phys. \textbf{B277} (1986), 349.

\bibitem{Jamin_2001_a}
M.~Jamin and B.O. Lange, Phys. Rev. \textbf{D65} (2002), 056005.

\bibitem{Jamin_1989_a}
M.~Jamin and M.~Neubert, Phys. Lett. \textbf{B238} (1990), 387.

\bibitem{Kalmykov_2006_a}
M.Y. Kalmykov, JHEP \textbf{0604} (2006), 056.

\bibitem{Kim_2012_a}
Y.~Kim, I.J. Shin, and T.~Tsukioka, Prog. Part. Nucl. Phys. \textbf{68} (2013),
  55.

\bibitem{Kleiv_2010_a}
R.T. Kleiv and T.G. Steele, J. Phys. \textbf{G38} (2011), 025001.

\bibitem{Kleiv_2013_a}
R.T. Kleiv, T.G. Steele, A.~Zhang, and I.~Blokland, Phys. Rev. \textbf{D87}
  (2013), 125018.

\bibitem{Klempt_2007_a}
E.~Klempt and A.~Zaitsev, Phys. Rept. \textbf{454} (2007), 1.

\bibitem{Kronfeld_2012_a}
A.S. Kronfeld, Ann. Rev. Nucl. Part. Sci. \textbf{62} (2012), 265.

\bibitem{Kwong_1987_a}
W.~Kwong, J.L. Rosner, and C.~Quigg, Ann. Rev. Nucl. Part. Sci. \textbf{37}
  (1987), 325.

\bibitem{Lee_2009_a}
I.W. Lee, A.~Faessler, T.~Gutsche, and V.E. Lyubovitskij, Phys. Rev.
  \textbf{D80} (2009), 094005.

\bibitem{Lewin_1981_a}
L.~Lewin, \emph{{Polylogarithms and Associated Functions}}, North-Holland,
  1981.

\bibitem{Li_2004_a}
B.A. Li, Phys. Lett. \textbf{B605} (2005), 306--310.

\bibitem{Liu_2012_a}
L.~Liu et~al., JHEP \textbf{1207} (2012), 126.

\bibitem{Liu_2011_a}
L.~Liu, S.M. Ryan, M.~Peardon, G.~Moir, and P.~Vilaseca, arXiv:1112.1358
  [hep-lat].

\bibitem{Liu_2008_a}
X.~Liu, Z.G. Luo, Y.R. Liu, and S.L. Zhu, Eur. Phys. J. \textbf{C61} (2009),
  411.

\bibitem{Liu_2013_a}
Z.Q. Liu et~al., Phys. Rev. Lett. \textbf{110} (2013), 252002.

\bibitem{Luke_1969_a}
Y.L. Luke, \emph{{The Special Functions and their Approximations}}, vol.~1,
  Academic Press, 1969.

\bibitem{Lutz_2009_a}
M.F.M. Lutz et~al., {arXiv:0903.3905 [hep-ex]}.

\bibitem{Maiani_2004_a}
L.~Maiani, F.~Piccinini, A.D. Polosa, and V.~Riquer, Phys. Rev. \textbf{D71}
  (2005), 014028.

\bibitem{Maris_2003_a}
P.~Maris and C.D. Roberts, Int. J. Mod. Phys. \textbf{E12} (2003), 297.

\bibitem{Matheus_2006_a}
R.~Matheus, S.~Narison, M.~Nielsen, and J.M. Richard, Phys. Rev. \textbf{D75}
  (2007), 014005.

\bibitem{Mathieu_2008_a}
V.~Mathieu, N.~Kochelev, and V.~Vento, Int. J. Mod. Phys. \textbf{E18} (2009),
  1.

\bibitem{Mertig_1998_a}
R.~Mertig and R.~Scharf, Comput. Phys. Commun. \textbf{111} (1998), 265.

\bibitem{Narison_2007_a}
S.~Narison, \emph{{QCD as a Theory of Hadrons: From Partons to Confinement}},
  Cambridge University Press, 2007.

\bibitem{Narison_2010_a}
S.~Narison, Phys. Lett. \textbf{B693} (2010), 559--566.

\bibitem{Narison_2011_a}
S.~Narison, Phys. Lett. \textbf{B706} (2012), 412.

\bibitem{Narison_1983_a}
S.~Narison and R.~Tarrach, Phys. Lett. \textbf{B125} (1983), 217.

\bibitem{Neubert_1993_a}
M.~Neubert, Phys. Rept. \textbf{245} (1994), 259.

\bibitem{Nielsen_2009_a}
M.~Nielsen, F.S. Navarra, and S.H. Lee, Phys. Rept. \textbf{497} (2010), 41.

\bibitem{Novikov_1983_a}
V.A. Novikov, M.A. Shifman, A.I. Vainshtein, and V.I. Zakharov, Fortsch. Phys.
  \textbf{32} (1984), 585.

\bibitem{Ochs_2013_a}
W.~Ochs, J. Phys. G \textbf{40} (2013), 043001.

\bibitem{Pascual_1984_a}
P.~Pascual and R.~Tarrach, \emph{{QCD: Renormalization for the Practitioner}},
  Springer-Verlag, 1984.

\bibitem{Passarino_1978_a}
G.~Passarino and M.J.G. Veltman, Nucl. Phys. \textbf{B160} (1979), 151.

\bibitem{Perantonis_1990_a}
S.~Perantonis and C.~Michael, Nucl. Phys. \textbf{B347} (1990), 854--868.

\bibitem{Peskin_1995_a}
M.E. Peskin and D.V. Schroeder, \emph{{An Introduction to Quantum Field
  Theory}}, Westview Press, 1995.

\bibitem{Polya_1974_a}
G.~P\'{o}lya and G.~Latta, \emph{{Complex Variables}}, Wiley, 1974.

\bibitem{Qiao_2010_a}
C.F. Qiao, L.~Tang, G.~Hao, and X.Q. Li, J. Phys. \textbf{G39} (2012), 015005.

\bibitem{Rainville_1960_a}
E.D. Rainville, \emph{{Special Functions}}, The Macmillan Company, 1960.

\bibitem{Reinders_1984_a}
L.J. Reinders, H.~Rubinstein, and S.~Yazaki, Phys. Rept. \textbf{127} (1985),
  1.

\bibitem{Richard_2012_a}
J.M. Richard, {arXiv:1205.4326 [hep-ph]}.

\bibitem{Shifman_1978_a}
M.A. Shifman, A.I. Vainshtein, and V.I. Zakharov, Nucl. Phys. \textbf{B147}
  (1979), 385.

\bibitem{Shifman_1978_b}
M.A. Shifman, A.I. Vainshtein, and V.I. Zakharov, Nucl. Phys. \textbf{B147}
  (1979), 448.

\bibitem{Slater_1966_a}
L.J. Slater, \emph{{Generalized Hypergeometric Functions}}, Cambridge
  University Press, 1966.

\bibitem{Slavnov_1975_a}
A.A. Slavnov, Sov. Jour. Part. Nucl. \textbf{5} (1975), 303.

\bibitem{Smirnov_2004_a}
V.A. Smirnov, \emph{{Evaluating Feynman Integrals}}, vol. 211, Springer, 2004.

\bibitem{Srednicki_2007_a}
Mark Srednicki, \emph{{Quantum Field Theory}}, Cambridge University Press,
  2007.

\bibitem{Steinhauser_2002_a}
M.~Steinhauser, Phys. Rept. \textbf{364} (2002), 247--357.

\bibitem{Swanson_2003_a}
E.S. Swanson, Phys. Lett. \textbf{B588} (2004), 189.

\bibitem{Swanson_2006_a}
E.S. Swanson, Phys. Rept. \textbf{429} (2006), 243.

\bibitem{tHooft_1972_a}
G.~'t~Hooft and M.J.G. Veltman, Nucl. Phys. \textbf{B44} (1972), 189.

\bibitem{tHooft_1972_b}
G.~'t~Hooft and M.J.G. Veltman, Nucl. Phys. \textbf{B50} (1972), 318.

\bibitem{Tarasov_1996_a}
O.V. Tarasov, Phys. Rev. \textbf{D54} (1996), 6479--6490.

\bibitem{Tarasov_1997_a}
O.V. Tarasov, Nucl. Phys. \textbf{B502} (1997), 455--482.

\bibitem{Tarasov_1998_a}
O.V. Tarasov, Acta Phys. Polon. \textbf{B29} (1998), 2655.

\bibitem{Taylor_1971_a}
J.C. Taylor, Nucl. Phys. \textbf{B33} (1971), 436.

\bibitem{Terasaki_2007_a}
K.~Terasaki, Prog. Theor. Phys. \textbf{118} (2007), 821--826.

\bibitem{Thomas_2008_a}
C.E. Thomas and F.E. Close, Phys. Rev. \textbf{D78} (2008), 034007.

\bibitem{Tornqvist_2004_a}
N.A. Tornqvist, Phys. Lett. \textbf{B590} (2004), 209--215.

\bibitem{Voloshin_2003_a}
M.B. Voloshin, Phys. Lett. \textbf{B579} (2004), 316.

\bibitem{Wang_2010_a}
Z.G. Wang, Eur. Phys. J. \textbf{C71} (2011), 1524.

\bibitem{Wang_2011_a}
Z.G. Wang, Commun. Theor. Phys. \textbf{59} (2013), 451.

\bibitem{Wilson_1969_a}
K.G. Wilson, Phys. Rev. \textbf{179} (1969), 1499.

\bibitem{Xiao_2013_a}
T.~Xiao, S.~Dobbs, A.~Tomaradze, and K.K. Seth, {arXiv:1304.3036 [hep-ex]}.

\bibitem{Zhang_2006_a}
A.~Zhang, T.~Huang, and T.G. Steele, Phys. Rev. \textbf{D76} (2007), 036004.

\bibitem{Zhu_2005_a}
S.L. Zhu, Phys. Lett. \textbf{B625} (2005), 212.

\bibitem{Zweig_1964_a}
G.~Zweig, CERN-TH \textbf{412} (1964), 80.

\end{thebibliography}

%

\uofsappendix

\chapter{Conventions}
\label{appendix_a_conventions}
\doublespacing

For brevity four-vectors are often written without a Lorentz index, that is, it is to be understood that $x=\left(x^0\,,x^1\,,x^2\,,x^3\right)$. When used, three-vectors are denoted as $\mathbf{x}=\left(x^1\,,x^2\,,x^3\right)$. The following convention is used for the four-dimensional Minkowski space metric:
\begin{equation}
g_{\mu\nu}=\left[
\begin{array}{rrrr}
1 &  0 &  0 &  0 \\
0 & -1 &  0 &  0 \\
0 &  0 & -1 &  0 \\
0 &  0 &  0 & -1 
\end{array}
\right] \,,
\quad
g_{\mu\nu}\,p^{\mu}k^{\nu}=p \cdot k = p^0 k^0 - \mathbf{p}\cdot\mathbf{k} \,. 
\end{equation}
In $d$-dimensions, the metric is defined such that $g^{\mu\nu}g_{\mu\nu}=d$. The Einstein summation convention is assumed on all indices, that is, a product containing repeated spinor, colour, Lorentz or SU(3) indices is summed over the full range of the indices. Apart from Lorentz indices, no distinction is made between raised and lowered indices. That is, $A^a_\mu = A_{\mu \, a}$, for instance.

We use the conventions of Ref.~\cite{Bjorken_1964_a} for the Dirac gamma matrices. In what follows each matrix element is itself a two by two matrix (\textit{i.e.} the two by two identity matrix is denoted as $1$). The specific forms are
\begin{equation}
\gamma^0=\left[
\begin{array}{rr}
1 &  0 \\
0 &  -1 
\end{array}
\right] 
\,, \quad 
\gamma^i = 
\left[
\begin{array}{rr}
0 & \sigma^i  \\
-\sigma^i & 0 
\end{array}
\right] \,, \quad i=\left\{1\,,2\,,3\right\} \,,
\end{equation}
where $\sigma^i$ is a Pauli matrix.  In these conventions,
\begin{equation}
\gamma^5=i\gamma^0\gamma^1\gamma^2\gamma^3=
\left[
\begin{array}{rr}
0 &  1 \\
1 &  0 
\end{array}
\right] 
\,, 
\end{equation}
from which it follows that $\left(\gamma_5\right)^2=1$. Ref.~\cite{Collins_1984_a} discusses various approaches to defining $\gamma_5$ in $d$-dimensions. In QSR calculations using dimensional regularization it is conventional to define $\gamma_5$ such that $\left\{\gamma_5\,,\gamma_\mu\right\}=0$~\cite{Narison_2007_a}. The charge conjugation operator is defined as
\begin{gather}
C = i \gamma^2 \gamma^0 \,.
\label{definition_of_charge_conjugation_operator}
\end{gather}
The following properties are useful in Chapters~\ref{chapter_4_Qq_diquark} and \ref{chapter_5_diquark_renorm}:
\begin{gather}
C^{-1} = C^T = -C \,, \quad C^2=-1\,, \quad C\gamma_\mu^T C = \gamma_\mu \,, \quad \left[C\,,\gamma_5\right]=0 \,,
\end{gather}
where $T$ denotes the transpose.

Natural units are used, where $\hbar=c=1$. Using the relativistic invariant it can be shown that
\begin{gather}
E^2 = \left(\mathbf{p}c\right)^2 + \left(mc^2\right)^2 \,, \quad \xrightarrow{c = 1} \quad  \left[E\right]=\left[\mathbf{p}\right]=\left[m\right] \,,
\label{relativistic_invariant}
\end{gather}
so that energy, momentum and mass have identical dimensions in this system of units. It is conventional to chose energy units as the base unit for all quantities. For instance, in natural units the masses of the electron and proton are approximately $0.511\,{\rm MeV}$ and $938\,{\rm MeV}$, respectively.

Dimensional analysis in natural units is straightforward. We define $\left[m\right]=1$, from which it follows that $\left[E\right]=\left[\mathbf{p}\right]=\left[p_\mu\right]=1$. Because momenta and derivatives are related through Fourier transforms, $\left[\partial_\mu\right]=\left[p_\mu\right]=1$. From the expression for a plane wave it can be shown that
\begin{gather}
e^{i p \cdot x} \quad \rightarrow \quad \left[ p \cdot x \right] = 0 \,, \quad \rightarrow \quad \left[x\right] = -1 \,,
\label{plane_wave_exponential}
\end{gather}
and consequently $\left[d^d x\right]=-d$. By a similar argument it can be shown that 
\begin{gather}
\exp{\left[i\int \, d^d x \, \mathcal{L} \right]} \quad \rightarrow \quad \left[ \int \, d^d x \, \mathcal{L} \right] = 0 \quad \rightarrow \quad \left[\mathcal{L}\right] = d \,.
\label{action_exponential}
\end{gather}
This can be used to determine the dimensions of the fields and parameters appearing in the QCD Lagrangian~\eqref{bare_qcd_lagrangian}. For instance, from the quark term we can show that
\begin{gather}
\left[\bar{Q}\gamma^\mu\partial_\mu Q\right] = d \quad \rightarrow \quad \left[\bar{Q}\right]=\left[Q\right]=\frac{d-1}{2} \,,
\end{gather}
because the Dirac Gamma matrix is dimensionless. From the gluon term we find that
\begin{gather}
\left[G^{\mu\nu}_a G^a_{\mu\nu} \right] = d \quad \rightarrow \quad \left[G^a_{\mu\nu}\right]=\left[\partial_\mu A^a_\mu \right] =\frac{d}{2} 
\quad \rightarrow \quad \left[ A^a_\mu \right] = \frac{d-2}{2} \,.
\end{gather}
The units of the coupling can be determined from the quark-gluon interaction term:
\begin{gather}
\left[g \bar{Q} \frac{\lambda^a}{2} \gamma^\mu A^a_\mu Q \right] = d \,, \quad \rightarrow \quad \left[g\right] = \frac{4-d}{2} \,, 
\end{gather}
because the Gell-Mann matrix is dimensionless. The renormalization scale has $\left[\mu\right]=1$, therefore
\begin{gather}
 \left[ \alpha \right]  = \left[ g^2 \mu^{d-4} \right] = 0 \,,
\end{gather}
and hence $\alpha$ is an appropriate expansion parameter in $d$-dimensions.


\chapter{Mathematical Functions}
\label{appendix_b_math}
\doublespacing

This appendix briefly summarizes the relevant properties of the special functions that are used in Chapters~\ref{chapter_2_loops}, \ref{chapter_3_hybrids}, \ref{chapter_4_Qq_diquark}, \ref{chapter_5_diquark_renorm} and \ref{chapter_6_mixing}. The material in Section~\ref{GammaFunction} is taken from Refs.~\cite{Polya_1974_a,Abramowitz_1964_a}, that of Section~\ref{HypergeometricFunctions} can be found in Refs.~\cite{Erdelyi_1953_a,Slater_1966_a,Luke_1969_a,Rainville_1960_a}, and Ref.~\cite{Lewin_1981_a} contains the material in Section~\ref{Polylogarithms}.

\section{The Gamma Function}
\label{GammaFunction}

\begin{figure}[htb]
\centering
\includegraphics[scale=1]{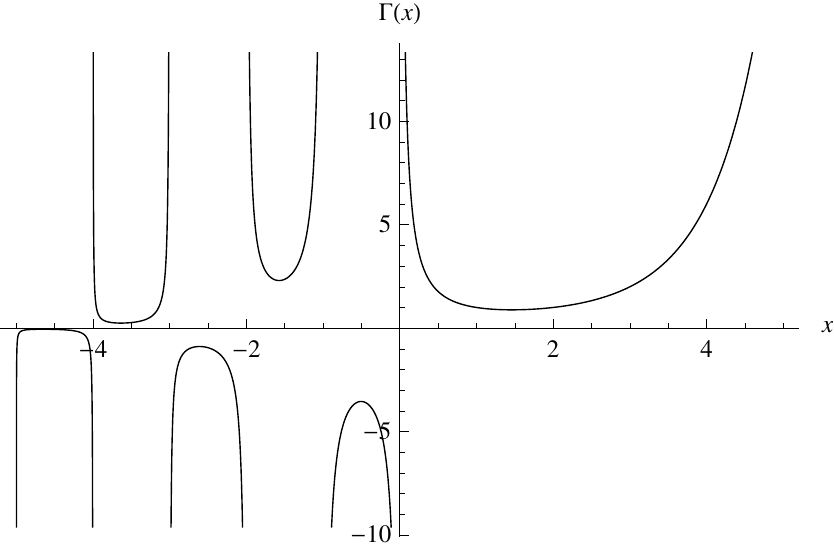}
\caption{The Gamma function.}
\label{Gamma_Function}
\end{figure}

As we have seen, the Gamma function arises frequently in dimensional regularization. A plot of the Gamma function is shown in Fig.~\ref{Gamma_Function}.

\noindent The Gamma function provides an extension of the factorial to non-integers:
\begin{gather}
 \Gamma\left(n\right) = \left(n-1\right)! \,, \quad \Gamma\left(z+1\right) = z \Gamma\left(z\right) \,.
\label{defintion_of_gamma_function}
\end{gather}
$\Gamma\left(z\right)$ has a simple pole when its argument is zero or a negative integer, the residue of which is
\begin{gather}
\left. \rm Res \, \Gamma\left( z\right) \right|_{z\to-n}   = \frac{\left(-1\right)^n}{n!} \,.
\label{residue_of_gamma_function}
\end{gather}
Note that there is no value of $z$ for which $\Gamma\left(z\right)=0$, hence $\left[\Gamma\left(z\right)\right]^{-1}$ is an entire function. An integral representation is given by
\begin{gather}
\Gamma\left(z\right) = \int_0^\infty dt \, t^{z-1} e^{-t} \,, \quad {\rm Re}(z) > 0\,.
\label{integral_representation_of_gamma_function}
\end{gather}
Using~\eqref{integral_representation_of_gamma_function}, it can be shown that $\Gamma\left(\frac{1}{2}\right)=\sqrt{\pi}$. The argument of the Gamma function can be simplified using the identity
\begin{gather}
\Gamma\left(2z\right) = \frac{2^{2x-\frac{1}{2}}}{\sqrt{2\pi}} \Gamma\left(z\right) \Gamma\left(z+\frac{1}{2}\right) \,.
\label{gamma_function_duplication_identity}
\end{gather}
In order to construct series expansions of the Gamma function, it is helpful to introduce the Digamma function
\begin{gather}
\psi\left(z\right) = \frac{1}{\Gamma\left(z\right)} \frac{d}{dz} \Gamma\left(z\right) \,,
\label{digamma_function}
\end{gather}
and the closely related Polygamma function
\begin{gather}
\psi^{\left(n\right)} \left(z\right) = \frac{d^n}{dz^n} \psi\left(z\right) \,.
\label{polygamma_function}
\end{gather}
Numerical values of these functions at $z=1$ are 
\begin{gather}
\psi\left(1\right) = -\gamma_E \,, \quad \psi^{\left(n\right)}\left(1\right) = \left(-1\right)^n \, n! \, \zeta\left(n+1\right) \,,
\label{polygamma_functions_at_z=1} 
\end{gather}
where $\gamma_E \simeq 0.577$ is the Euler-Mascheroni constant and $\zeta\left(z\right)$ is the Riemann Zeta function. Using these results, it is easy to show that
\begin{gather}
\Gamma\left(1+z\right) = 1 - \gamma_{E} z +\frac{1}{2}\left[\gamma_{E}^2+\zeta\left(2\right)\right]z^2 + \mathcal{O}\left(z^3\right) \,.
\label{expansion_of_Gamma_z+1} 
\end{gather}
Using~\eqref{defintion_of_gamma_function} and~\eqref{expansion_of_Gamma_z+1}, it can be shown that
\begin{gather}
\Gamma\left(z\right) = \frac{1}{z} - \gamma_{E} +\frac{1}{2}\left[\gamma_{E}^2+\zeta\left(2\right)\right]z + \mathcal{O}\left(z^2\right) \,.
\label{expansion_of_Gamma_z} 
\end{gather}
The Beta function is defined in terms of the Gamma function as 
\begin{gather}
B\left(a\,,b\right) = \frac{\Gamma\left(a\right) \Gamma\left(b\right)}{ \Gamma\left(a+b\right) } \,.
\label{definition_of_Beta_function}
\end{gather}
Integral representations of the Beta function are 
\begin{gather}
B\left(a\,,b\right) = \int_0^1 dx \, x^{a-1} \left(1-x\right)^{b-1} = \int_0^\infty dx \, \frac{x^{a-1}}{\left(1+x\right)^{a+b}} \,.
\label{integral_representation_of_beta_function}
\end{gather}

\section{Hypergeometric Functions}
\label{HypergeometricFunctions}

The generalized hypergeometric function is defined as 
\begin{gather}
\begin{split}
&\phantom{}_p F_q \left[ a_1 \,, a_2 \,, \ldots a_p \,; b_1 \,, b_2 \,, \ldots b_q \,; z \right] = \,
\phantom{}_p F_q
\left[
\begin{array}{c|}
a_1 \,,  a_2 \,, \ldots a_p \\
b_1 \,,  b_2 \,, \ldots b_q 
\end{array} \, z\right] 
\\
&=\frac{\Gamma\left(b_1\right)\Gamma\left(b_2\right)\ldots\Gamma\left(b_q\right)}{\Gamma\left(a_1\right)\Gamma\left(a_2\right)\ldots\Gamma\left(a_p\right)}
 \sum_{n=0}^\infty 
 \frac{\Gamma\left(a_1+n\right)\Gamma\left(a_2+n\right)\ldots\Gamma\left(a_p+n\right)}{\Gamma\left(b_1+n\right)\Gamma\left(b_2+n\right)\ldots\Gamma\left(b_q+n\right)}
 \frac{z^n}{n!} \,,
\end{split} 
\label{pFq_series_representation} 
\end{gather}
where the constants $a_i$, $b_i$ are called indices and uniquely define each generalized hypergeometric function. From this definition it is clear that the ordering of the indices is irrelevant and that if any $a_i=b_j$ ($1\leq i\leq p\,, \, 1\leq j\leq q$), the generalized hypergeometric function $\phantom{}_p F_q$ reduces to $\phantom{}_{p-1} F_{q-1}$. When $p=q+1$, the generalized hypergeometric function has a branch cut on the interval $z\in\left[1\,,\infty\right)$. Generalized hypergeometric functions can also be represented in terms of Mellin-Barnes contour integrals. For instance, the following contour integral representation
\begin{gather}
\begin{split}
\phantom{}_p F_q & \left[ a_1 \,, a_2 \,, \ldots a_p \,;  b_1 \,, b_2 \,, \ldots b_q \,; z \right] 
\\
&=\frac{\Gamma\left(b_1\right)\Gamma\left(b_2\right)\ldots\Gamma\left(b_q\right)}{\Gamma\left(a_1\right)\Gamma\left(a_2\right)\ldots\Gamma\left(a_p\right)}
\int_{-i\infty}^{i\infty} \frac{ds}{2\pi i} \frac{\Gamma\left(a_1+s\right)\Gamma\left(a_2+s\right)\ldots\Gamma\left(a_p+s\right)}{\Gamma\left(b_1+s\right)\Gamma\left(b_2+s\right)\ldots\Gamma\left(b_q+s\right)}
\Gamma\left(-s\right)\left(-z\right)^s 
\end{split}
\label{pFq_Mellin_Barnes_representation}
\end{gather}
is completely equivalent to the series representation~\eqref{pFq_series_representation}. The integral can be evaluated using the residue theorem, and the integration along the imaginary axis can be shifted to avoid poles if needed. Note that the Gamma functions of the form $\Gamma\left(c+s\right)$ have poles in the left half plane, while the Gamma function $\Gamma\left(-s\right)$ has poles in the right half plane. In order to reproduce~\eqref{pFq_series_representation}, the contour should be closed in the right half plane, and the residues can be calculated using~\eqref{residue_of_gamma_function}. If the integration contour is closed appropriately, it can be shown that all contributions apart from the integration along the imaginary axis are zero. The proof of this is rather delicate and is not given here (see Refs.~\cite{Slater_1966_a,Rainville_1960_a}). Using~\eqref{pFq_Mellin_Barnes_representation}, we can write
\begin{gather}
\phantom{}_1 F_0 \left[n\,; \phantom{} \,; z\right] = \frac{1}{\left(1-z\right)^n} = \frac{1}{\Gamma\left(n\right)} \int_{-\infty}^{i\infty} \frac{ds}{2\pi i} \Gamma\left(n+s\right) \Gamma\left(-s\right) \left(-z\right)^s \,.
\label{1F0_Mellin_Barnes_representation}
\end{gather}
This identity permits massive propagators to be expressed as contour integrals of massless propagators and is the foundation of the Mellin-Barnes techniques used in Chapter~\ref{chapter_2_loops} to calculate loop integrals with two massive propagators. An important identity for contour integrals of the form~\eqref{pFq_Mellin_Barnes_representation} is Barnes' Lemma:
\begin{gather}
\int_{-\infty}^{i\infty} \frac{ds}{2\pi i} \Gamma\left(a+s\right) \Gamma\left(b+s\right) \Gamma\left(c+s\right) \Gamma\left(d+s\right)
= \frac{\Gamma\left(a+c\right)\Gamma\left(a+d\right)\Gamma\left(b+c\right)\Gamma\left(b+d\right)}{\Gamma\left(a+b+c+d\right)} \,.
\label{Barnes_Lemma}
\end{gather}

Most special functions encountered in Mathematical Physics can be expressed in terms of generalized hypergeometric functions (see Ref.~\cite{Luke_1969_a} for a partial list). The most commonly known hypergeometric function is the Gauss hypergeometric function, $\phantom{}_2 F_1\left[a\,,b\,; c\,; z\right]$, of which the Chebyshev, Gegenbauer, Jacobi and Legendre polynomials are special cases. The Gauss hypergeometric function has the integral representation
\begin{gather}
\phantom{}_2 F_1\left[a\,,b\,; c\,; z\right] = \frac{\Gamma\left(c\right)}{\Gamma\left(a\right)\Gamma\left(c-b\right)} \int_0^1 dt \, t^{b-1} \left(1-t\right)^{c-b-1} \left(1-tz\right)^{-a} \,, \quad {\rm Re}\, (c) > {\rm Re}\, (b) > 0 \,.
\label{2F1_hypergeometric_function_integral_representation}
\end{gather}
At $z=1$, the Gauss hypergeometric function reduces to
\begin{gather}
\phantom{}_2 F_1\left[a\,,b\,; c\,; 1\right] = \frac{\Gamma\left(c\right)\Gamma\left(c-a-b\right)}{\Gamma\left(c-a\right)\Gamma\left(c-b\right)} \,, \quad {\rm Re}\left(c-a-b\right) > 0 \,.
\label{2F1_at_z=1}
\end{gather}
It is interesting to note that there are recurrence relations relating $\phantom{}_2 F_1\left[a\,,b\,; c\,; z\right]$ and the contiguous functions $\phantom{}_2 F_1\left[a\pm1\,,b\pm1\,; c\pm1\,; z\right]$ (See Ref.~\cite{Abramowitz_1964_a}). In Ref.~\cite{Tarasov_1998_a} it is pointed out that these are closely related to the generalized recurrence relations discussed in Chapter~\ref{chapter_2_loops}. 

\section{Polylogarithms}
\label{Polylogarithms}

In Chapter~\ref{chapter_2_loops} it was pointed out that loop integrals that include an external momentum and a mass often lead to generalized hypergeometric functions whose indices of are $d$-dependent. Higher order terms in the epsilon expansion of these often involve Polylogarithm functions~\cite{Lewin_1981_a}. The simplest Polylogarithm is the Dilogarithm, which is defined as 
\begin{gather}
{\rm Li}_2\left(z\right) = - \int_0^z ds \, \frac{\log{\left(1-s\right)}}{s} \,,
\label{definition_of_dilogarithm_function}
\end{gather}
which has the same branch cut as the generalized hypergeometric function~\eqref{pFq_series_representation}. The numerical value at $z=1$ is given by
\begin{gather}
 {\rm Li}_2\left(1\right) = \frac{\pi^2}{6} \,.
\end{gather}
In general, the Polylogarithm is defined recursively:
\begin{gather}
{\rm Li}_n\left(z\right) =  \int_0^z ds \, \frac{{\rm Li}_{n-1}\left(s\right)}{s} \,,
\label{definition_of_polylogarithm_function}
\end{gather}
which for $n=3$ is called the Trilogarithm. The Dilogarithm satisfies the identity
\begin{gather}
{\rm Li}_2 \left(1-z\right) = \frac{\pi^2}{6} - \log{\left(z\right)}\log{\left(1-z\right)} - {\rm Li}_2 \left(z\right) \,.
\label{dilogarithm_identity}
\end{gather}
The Trilogarithm satisfies a similar identity,
\begin{gather}
\begin{split}
{\rm Li}_3 \left(1-z\right) &= \frac{\pi^2}{6}\log{\left(1-z\right)} + \frac{1}{6}\log^3{\left(1-z\right)}  - \frac{1}{2} \log^2{\left(1-z\right)}
\log{\left(z\right)} - {\rm Li}_3 \left(z\right)
\\
&- {\rm Li}_3 \left(\frac{z}{z-1}\right) + \zeta\left(3\right) \,.
\end{split}
\label{trilogarithm_identity}
\end{gather}
As mentioned previously, the Mathematica package HypExp can perform epsilon expansions of some generalized hypergeometric functions. This was used in the heavy-light diquark calculation in Chapter~\ref{chapter_4_Qq_diquark}. However, when this package was used some functions with an inappropriate branch cut structure were generated. Using the identities \eqref{dilogarithm_identity} and \eqref{trilogarithm_identity}, the functions with this branch structure can be canceled identically, and those that remain have the appropriate branch cut structure.

\end{document}